\documentclass[sn-mathphys-num]{sn-jnl}

\usepackage{adjustbox}
\usepackage{graphicx}%
\usepackage{multirow}%
\usepackage{amsmath,amssymb,amsfonts}%
\usepackage{amsthm}%
\usepackage{mathrsfs}%
\usepackage[title]{appendix}%
\usepackage{xcolor,colortbl}
\usepackage[most]{tcolorbox}
\usepackage{textcomp}%
\usepackage{manyfoot}%
\usepackage[utf8]{inputenc}
\usepackage{booktabs}%
\usepackage[ruled,linesnumbered]{algorithm2e}
\usepackage{listings}%
\usepackage{subfigure}
\usepackage{subcaption}
\usepackage{pgfplots}
\pgfplotsset{compat=1.18}
\usepackage{geometry}

\usepackage{subcaption}
\usepackage{tabularx}
\usepackage[dvipsnames]{xcolor}
\usepackage[normalem]{ulem}
\usepackage{enumitem}
\usepackage{pifont}
\usepackage{array}
\usepackage{lipsum}
\usepackage[utf8]{inputenc}
\usepackage{hyperref}
\usepackage{tabularray} 
\usepackage{graphicx}
\usepackage{caption}

\usepackage{caption}

\usepackage{minted}
\setminted{
    fontsize=\footnotesize,
    bgcolor=gray!10,
    frame=single,
    linenos,
    breaklines
}

\SetKwInput{KwInput}{Input}                
\SetKwInput{KwOutput}{Output} 

\definecolor{mycolor_box}{HTML}{F5FFFA} 
\definecolor{mycolor_title}{HTML}{FFEBCD} 

\theoremstyle{thmstyleone}%
\theoremstyle{thmstyletwo}%

\theoremstyle{thmstylethree}%

\begin{document}

\title{Model Compression vs. Adversarial Robustness: An Empirical Study on Language Models for Code}


\author*[1]{\fnm{Md. Abdul} \sur{Awal}}\email{abdul.awal@usask.ca}
\author[1]{\fnm{Mrigank} \sur{Rochan}}\email{mrochan@cs.usask.ca}
\author[1]{\fnm{Chanchal K.} \sur{Roy}}\email{chanchal.roy@usask.ca}

\affil[1]{
  \orgdiv{Department of Computer Science}, 
  \orgname{University of Saskatchewan}, 
  \orgaddress{\city{Saskatoon}, \state{Saskatchewan}, \country{Canada}}
}


\abstract{Transformer-based language models for code have shown remarkable performance in various software analytics tasks, but their adoption is hindered by high computational costs, slow inference speeds, and substantial environmental impact. Model compression techniques such as pruning, quantization, and knowledge distillation have gained traction in addressing these challenges. However, the impact of these strategies on the robustness of compressed language models for code in adversarial scenarios remains poorly understood. Understanding how these compressed models behave under adversarial attacks is essential for their safe and effective deployment in real-world applications. To bridge this knowledge gap, we conduct a comprehensive evaluation of how commonly used compression strategies affect the adversarial robustness of compressed code models. We assess the robustness of compressed versions of six widely used language models for code across three software analytics tasks, using six evaluation metrics and five state-of-the-art adversarial attack techniques. Our findings indicate that compressed models generally maintain comparable performance to their uncompressed counterparts. However, when subjected to adversarial attacks, compressed models exhibit significantly reduced robustness. This vulnerability is consistent across all three compression techniques, with knowledge-distilled models exhibiting the most pronounced performance degradation. These results reveal a trade-off between model size reduction and adversarial robustness, underscoring the need for careful consideration when deploying compressed models in security-critical software applications. Our study highlights the need for further research into compression strategies that balance computational efficiency and adversarial robustness, which is essential for deploying reliable language models for code in real-world software applications.}


\maketitle

\section{Introduction}
\label{sec1}
Transformer-based \cite{vaswani2017attention} language models for code\footnote{Throughout this study, language models for code and models denote the same concept and are used interchangeably unless stated otherwise.} have achieved remarkable success across a range of software analytics tasks, including clone detection, code summarization, code completion, vulnerability detection, and code search \cite{roy2007survey, feng2020codebert, guo2020graphcodebert, lu2021codexglue, ahmad2021unified}. However, their widespread adoption is often hindered by practical constraints, notably high computational costs, low inference speeds, and significant carbon footprint \cite{shi2022compressing, shi2024greening, shi2024efficient, schwartz2020green, d2024compression, saad2024alpine, chen2025smaller}. For example, CodeBERT \cite{feng2020codebert}, a state-of-the-art (SOTA) code language model with 125 million parameters and a size of 476 MB, can experience response latencies of up to 1.5 seconds, consume 0.32 kilowatt-hours (kWh) of energy, and generate approximately 0.14 kilograms of CO\textsubscript{2} emission on consumer-grade laptops\footnote{A CodeBERT model integrated into IDEs may be invoked thousands of times daily by a developer, reflecting typical usage patterns \cite{hellendoorn2019code}}.

To address these challenges, model compression techniques such as pruning \cite{sanh2020movement}, quantization \cite{zafrir2019q8bert}, and knowledge distillation \cite{hinton2015distilling} have recently attracted significant attention in software engineering. Researchers and practitioners are actively exploring these methods to optimize the deployment of language models for code in real-world applications \cite{shi2022compressing, shi2024greening, wei2023towards, d2024compression}. For example, Shi et al. \cite{shi2022compressing} proposed \textit{Compressor}, a knowledge distillation-based approach that compresses CodeBERT \cite{feng2020codebert} and GraphCodeBERT \cite{guo2020graphcodebert} into compact 3 MB models, significantly reducing inference latency with minimal impact on model performance. In a follow-up study, Shi et al. \cite{shi2024greening} extended their approach to address issues related to energy consumption and carbon emission. Similarly, Wei et al. \cite{wei2023towards} conducted an empirical study on quantized models for code generation tasks, analyzing their resource usage, carbon footprint, and computational efficiency, with a focus on reducing inference latency and memory footprint. However, these efforts to improve efficiency, energy consumption, and performance have largely overlooked the robustness of compressed models, especially under adversarial attacks. This oversight could lead to vulnerabilities, such as adversarial inputs manipulating a compressed model in security-critical applications like automated code review \cite{li2022codereviewer, shi2019automatic} or vulnerability detection \cite{zhou2019devign, ding2024vulnerability, risse2025top}, resulting in missed flaws or incorrect assessments. Moreover, the lack of robustness to adversarial attacks in compressed models can propagate bugs and increase technical debt during software maintenance and evolution. For instance, failing to detect code clones under adversarial attacks can lead to increased technical debt, potentially resulting in substantial costs, estimated at \$3.61 per line of code \cite{guo2016exploring, technicaldebtcost, mondal2019empirical}. While prior studies have investigated adversarial attacks on uncompressed models \cite{yang2022natural, du2023extensive, tian2023code, zhang2022towards}, they have not examined adversarial attacks in the context of compressed models. Therefore, it is crucial to gain a deeper understanding of how compressed models are impacted by adversarial threats to ensure their safe and effective deployment in software analytics tasks \cite{shi2022compressing, panichella2025metamorphic}.

In this study, we systematically investigate the impact of three common model compression strategies: pruning \cite{sanh2020movement}, quantization \cite{zafrir2019q8bert}, and knowledge distillation \cite{hinton2015distilling} on the robustness of compressed models under adversarial attacks. Our work is partly inspired by recent research in computer vision that evaluates the robustness of compressed large pre-trained models under adversarial attacks \cite{wei2023towards, xu2021beyond, ye2019adversarial}. We employ five SOTA adversarial attack techniques: ALERT \cite{yang2022natural}, BeamAttack \cite{du2023extensive}, Metropolis-Hastings Modifier (MHM) \cite{zhang2020generating}, WIR-Random \cite{zeng2022extensive}, and CODA \cite{tian2023code} to assess the performance of compressed versions of six widely used code language models: CodeBERT \cite{feng2020codebert}, GraphCodeBERT \cite{guo2020graphcodebert}, CodeGPT \cite{lu2021codexglue}, CodeT5 \cite{wang2021codet5}, UniXcoder \cite{guo2022unixcoder}, and PLBART \cite{ahmad2021unified}. Our investigation spans three software analytics tasks: clone detection, code summarization, and vulnerability detection. Following prior works \cite{du2023extensive, yang2022natural}, we evaluate the robustness of compressed models using several standard metrics, including Attack Success Rate (\%ASR), Average Model Query (AMQ), Identifier Change Rate (ICR), Token Change Rate (TCR), Average Edit Distance (AED), and Average Code Similarity (ACS).

Our study reveals that compressed models perform comparably to their uncompressed counterparts across clone detection, code summarization, and vulnerability detection tasks in the absence of adversarial attacks. However, under adversarial attacks, compressed models exhibit reduced robustness, with knowledge-distilled models experiencing the most significant performance drop. Pruning and quantization tend to be more robust than knowledge distillation, but all compression methods show vulnerabilities to attacks, and no single technique consistently outperforms others across tasks. Notably, our results confirm a trade-off between model compactness and adversarial robustness, where more intensive compression techniques, such as knowledge distillation, yield smaller models at the cost of increased vulnerability. These findings imply that practitioners should carefully weigh the trade-offs between computational efficiency and adversarial robustness when deploying compressed models in security-critical applications. Ultimately, our study highlights the need for further research into compression strategies that balance computational efficiency and adversarial robustness of compressed language models for code in software analytics.

In summary, this study makes the following major contributions:

\begin{itemize}[left=0pt]
    \item \textbf{Originality.} To the best of our knowledge, we present the first comprehensive study to investigate the impact of different model compression strategies on the robustness of compressed language models for code under adversarial attacks across well-studied software analytics tasks.

    \item \textbf{Extensive study.} We conduct a thorough assessment of the robustness of compressed versions of six widely used code language models: CodeBERT, GraphCodeBERT, CodeGPT, CodeT5, UniXcoder, and PLBART, across three software analytics tasks: clone detection, code summarization, and vulnerability detection. To ensure a comprehensive analysis, we employ five SOTA adversarial attack methods: ALERT, BeamAttack, MHM, WIR-Random, and CODA. We then evaluate the robustness of compressed models using six standard metrics.
    
    \item \textbf{Open science}. The code and the corresponding datasets used in our empirical study are publicly available\footnote{https://doi.org/10.5281/zenodo.15272413}

\end{itemize}

\section{Background}
\label{back}
This section briefly introduces the key concepts and terminology relevant to this study.

\subsection{Language Models for Code}
\label{PTMC}
Transformer-based \cite{vaswani2017attention} pre-trained language models for code have been widely employed and have achieved SOTA performance on numerous downstream software analytics tasks. The model training paradigm comprises two key phases: \textit{pre-training} and \textit{fine-tuning}. During pre-training, the model is exposed to a large dataset and learns general language representations through self-supervised learning, eliminating the need for explicitly labeled data \cite{feng2020codebert}. In the fine-tuning phase, the pre-trained model is adapted to a specific task using labeled data, where only a subset of parameters is updated through supervised learning techniques \cite{lu2021codexglue}. As a result, fine-tuning simplifies the training process and significantly lowers the training costs associated with various downstream tasks in software analytics \cite{devlin2018bert}.

Transformer-based language models for code are typically classified into three architectural types: encoder-only, decoder-only, and encoder-decoder models \cite{zeng2022extensive}. Encoder-only models, such as CodeBERT \cite{feng2020codebert}, process the entire input sequence simultaneously and are primarily designed for program understanding tasks, such as vulnerability detection and clone identification. Decoder-only models, such as CodeGPT \cite{lu2021codexglue}, generate tokens sequentially and are widely used for program generation tasks, including code completion, code generation, and code summarization. Encoder-decoder models, including PLBART \cite{ahmad2021unified}, integrate an encoding stage that learns a representation of the input with a decoding stage that produces the output sequence. This architecture is particularly suitable for sequence-to-sequence tasks, including code summarization and programming language translation.

\subsection{Model Compression}
A key challenge in the practical adoption of large models lies in their substantial computational costs, high memory requirements, and environmental impact (e.g., carbon footprint) \cite{schwartz2020green, xu2023survey}. To tackle these problems and enhance its sustainability, the AI community has proposed various model compression approaches to minimize these models' memory requirement, energy consumption, and inference latency \cite{zafrir2019q8bert, hinton2015distilling, xu2023survey, zhu2024survey, ye2019adversarial, castano2023exploring, hort2023exploratory, wang2024model, xu2024surveyKD, xu2024survey}. The fundamental model compression strategies are pruning \cite{sanh2020movement}, quantization \cite{zafrir2019q8bert}, and knowledge distillation \cite{hinton2015distilling}, and each targets a different aspect of model optimization. \textbf{Knowledge distillation} transfers the knowledge from a larger, more complex model (the \textit{teacher}) to a smaller, more efficient model (the \textit{student}), which is trained to approximate the teacher's behavior. This compression strategy creates a model that requires less memory and delivers faster inference. \textbf{Quantization} reduces the precision of the model's weights and activations, typically converting them from floating-point (e.g., \texttt{float32}) representations to lower-bit (e.g., \texttt{int8} or \texttt{float8}). This helps reduce the memory footprint and speeds up inference with minimal impact on performance. \textbf{Pruning} involves removing less important weights or neurons from a model to reduce its size while maintaining accuracy. This strategy can result in sparse models that are more computationally efficient. In Section \ref{MCS}, we provide a detailed description of each compression strategy within the context of our empirical study.

\subsection{Adversarial Attack}
\label{Attack}
An adversarial attack refers to the process of generating adversarial examples by introducing subtle perturbations to the original input. From a human perspective, an adversarial example appears nearly identical to its original counterpart, yet it causes the model to produce incorrect or inconsistent results compared to the original input. Zegedy et al. \cite{szegedy2013intriguing} were pioneers in introducing the concept of adversarial attacks in computer vision, demonstrating that SOTA image classifiers can be misled by carefully crafted pixel-level perturbations that are imperceptible to human eyes. For instance, we expect an image classifier $f: X \rightarrow Y$ to predict the test instance correctly (e.g., $y_{truth} \in Y$) for an image $x \in X$. However, an attack introduces minor perturbations to $x$, resulting in an adversarial example $x'$ that misleads the model $f$.

In domains such as images, where inputs are composed of numeric pixel values that can be smoothly perturbed within a valid range, generating adversarial examples that are imperceptible to humans is relatively straightforward. In contrast, software artifacts such as source code are composed of structured tokens governed by strict syntactic and semantic rules. This rigid and highly constrained structure presents greater challenges for generating valid adversarial examples, as even minor perturbations may disrupt compilation or alter program semantics \cite{zhang2022towards, zhang2020generating}. Therefore, adversarial examples generated to attack models must satisfy three requirements \cite{du2023extensive}:

\begin{enumerate}
    \item Adversarial examples should effectively mislead models, causing them to produce incorrect outputs and thereby degrading overall performance (e.g.,  $f(x') \neq f(x) = y_{truth}$).
    \item Adversarial perturbations must preserve syntactic correctness by conforming to the programming language's syntax rules. For instance, in Python, an identifier cannot start with a number and must consist only of letters, digits, and underscores after perturbations
    \item $x'$ must remain semantically equivalent to $x$ after perturbations, preserving the same functionality and yielding identical outputs for the same inputs.
\end{enumerate}

\begin{figure}[htbp]
  \centering
  \subfigure[Model output before adversarial attacks, Target class = 1, Predicted class = 1]{\label{OrgExam}\includegraphics[width=6.5cm, height=3cm]{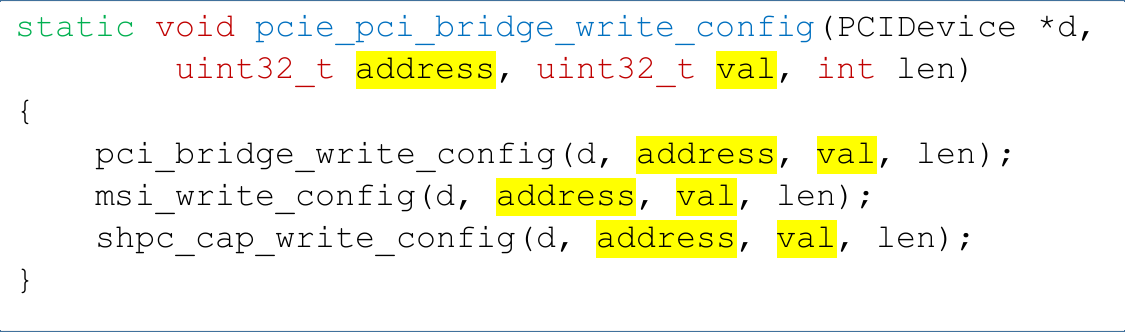}}
  \subfigure[Model output after adversarial attacks, Target class = 1, Predicted class = 0]{\label{AdvExam}\includegraphics[width=6.5cm, height=3cm]{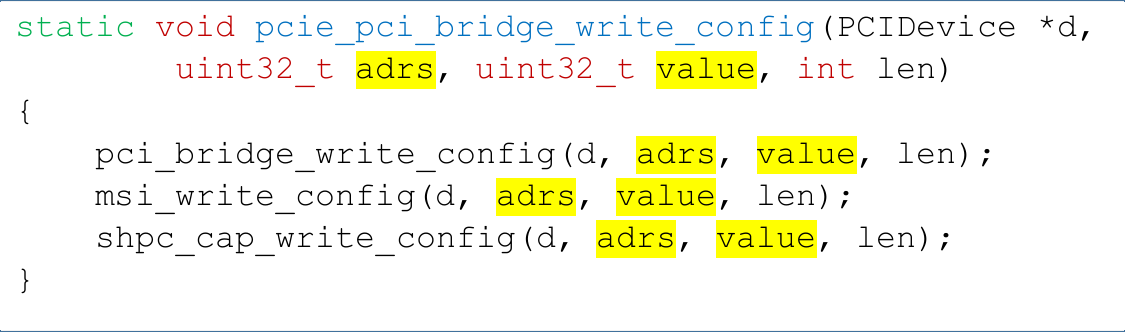}}
  \caption{A code snippet before (Figure \ref{OrgExam}) and after (Figure \ref{AdvExam}) an adversarial attack performed using ALERT \cite{yang2022natural} on the CodeBERT model fine-tuned for vulnerability detection task on the Devign \cite{zhou2019devign} dataset. The adversarial example preserves program semantics and functional correctness while introducing minimal identifier-level perturbations that alter the model's prediction, demonstrating the fine-tuned model's susceptibility to semantics-preserving natural adversarial attacks.}
  \label{AttackExample}
\end{figure}

Adversarial attacks targeting clone detection and vulnerability detection must satisfy all three of those requirements. Notably, the first requirement for the code summarization task alone cannot determine the success of the attack. Therefore, following the methodology outlined in existing studies \cite{du2023extensive, zeng2022extensive}, we define an attack as successful when the BLEU-4 score between the adversarial and reference summaries is 0, indicating that the adversarial summary does not correspond to the reference summary in any meaningful way. For example, when the reference summary is ``\texttt{returns the maximum value of the array}" and the summary produced by the model on adversarially modified code is ``\texttt{initializes the database connection,}" the BLEU-4 score is zero because there are no overlapping n-grams, which indicates a complete semantic mismatch. When an input is subtly modified in accordance with the aforementioned constraints to produce a perturbed version, the resulting sample is referred to as an adversarial example. If this modification causes models to make an incorrect prediction, it is considered an adversarial attack. For instance, Figure \ref{AttackExample} illustrates how a simple identifier renaming can create a successful adversarial example that misleads the model’s prediction. In this example, a SOTA adversarial attack technique ALERT \cite{yang2022natural} replaces the identifiers \textit{`address'} with \textit{`adrs'} and \textit{`val'} with \textit{`value'}. While these changes are trivial for humans and preserve the code’s syntax and semantics, they cause the model to make an incorrect prediction on the transformed input.

Adversarial attacks are divided into two categories based on the attacker’s level of access to the target model: 1) black-box attacks \cite{ilyas2018black}, where attackers only have access to model outputs, and 2) white-box attacks \cite{ebrahimi2017hotflip}, where attackers have full access to the subject models, including outputs, gradients, and hyperparameters. Beyond the access-based taxonomy, adversarial attacks are frequently classified by the attacker's objective as either targeted or untargeted. Targeted attacks involve manipulating the inputs to force the model to produce a specific incorrect output predetermined by the adversary, thereby directing the prediction toward a chosen target label or sequence \cite{szegedy2013intriguing}. Conversely, untargeted attacks aim to induce any form of misprediction without specifying a target outcome, which generally makes them easier to construct but less precise in their effects \cite{carlini2017adversarial}. In software analytics tasks, targeted attacks aim to force the model toward a specific incorrect prediction, while untargeted attacks aim to cause general prediction errors that undermine model robustness.

\subsection{Adversarial Robustness}
It refers to a model's ability to maintain its performance and generate consistent outputs, even when confronted with adversarial examples \cite{zhang2022towards, szegedy2013intriguing}. For instance, if a model can retain high accuracy and perform effectively under adversarial attacks, it is considered robust against such attacks. Adversarial robustness is typically evaluated by measuring the frequency with which adversarial examples successfully mislead the model \cite{yang2022natural}. A higher success rate of adversarial attacks indicates reduced robustness, whereas a lower success rate indicates greater resilience to such attacks.

In software analytics tasks, adversarial robustness is particularly critical because adversarial perturbations often involve subtle code transformations, such as identifier renaming or statement modifications, that preserve program semantics while misleading models. Consequently, evaluating the robustness of language models for code under adversarial attacks is essential to ensure their reliable deployment in practical software engineering environments \cite{du2023extensive}.

\section{Study Design}
\label{SD}
\begin{figure}[htbp]
\centering
\includegraphics[width=15cm, height = 4cm]{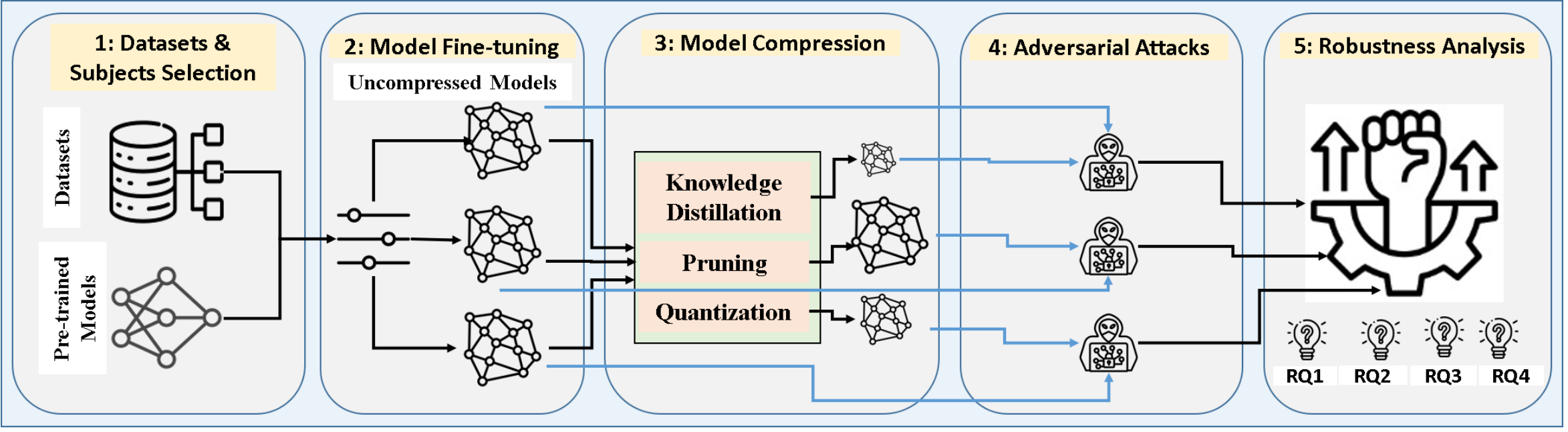}
\caption{Our study methodology for investigating the impact of compression strategies on the robustness of language models for code under adversarial attacks. Step 1 involves selecting benchmark datasets and code language models to define the experimental subjects. Step 2 fine-tunes the selected models to obtain uncompressed baseline models. Step 3 applies three compression techniques: knowledge distillation, pruning, and quantization, to generate compressed model variants. Step 4 subjects both uncompressed and compressed models to adversarial attacks to evaluate their vulnerability. Step 5 conducts a comprehensive robustness analysis to compare models across the research questions and assess the impact of compression on model reliability and robustness.} 
\label{workflow}
\end{figure}

The primary objective of this study is to investigate the impact of compression strategies on the robustness of compressed language models for code under adversarial attacks. This section outlines the design of our study, as illustrated in Figure \ref{workflow}, which includes the selection of models, the compression and adversarial attack strategies employed, and the benchmark datasets utilized. Additionally, we describe the attack settings and standard evaluation metrics applied to address the following research questions (RQs):

\begin{itemize}

    \item \textbf{RQ1: To what extent do different adversarial attack techniques preserve the syntactic and semantic quality of adversarial examples generated for uncompressed and compressed models?} This question investigates how different adversarial attack techniques influence the syntactic and semantic quality of adversarial examples, where quality reflects the characteristics defined in Section \ref{Attack}. By evaluating these characteristics, we ensure that any observed performance degradation is attributable to the models themselves, thereby reflecting the effectiveness of adversarial examples in assessing model robustness before and after compression.

    \item \textbf{RQ2: Does model compression degrade the adversarial robustness of language models for code compared to their uncompressed counterparts?} This research question aims to assess whether compression compromises the resilience of models against adversarial attacks. To investigate the impact, we directly compare the robustness of compressed and uncompressed models under adversarial attacks.

    \item \textbf{RQ3: Among different compression strategies, which one best preserves adversarial robustness?} In RQ3, we systematically assess which compression strategy most effectively preserves the adversarial robustness of models. To address this research question, we evaluate and compare the performance of different compression strategies on models when subjected to adversarial attacks.

    \item \textbf{RQ4: Does reducing model size through compression lead to a trade-off with adversarial robustness?} This research question examines whether increasing compression efficiency, that is, reducing model size, comes at the cost of decreased adversarial robustness. We evaluate this trade-off using \%ASR and the AMQ metrics.

\end{itemize}

\subsection{Datasets and Subjects}
This Section outlines the rationale for our selection of models and datasets, highlighting their role in enhancing the relevance and impact of our study.

\subsubsection{Datasets}
\label{DS}
To ensure a comprehensive understanding of the impact of model compression strategies on the robustness of models, we examine three software analytics tasks across two categories: (1) program understanding (e.g., \textit{clone detection} (\textbf{CD}), \textit{vulnerability detection} (\textbf{VD})) and (2) generation (\textit{code summarization} (\textbf{CS})). As the representative benchmark for clone detection, we select \textit{BigCloneBench} \cite{svajlenko2014towards}. Furthermore, BigCloneBench is widely used in model compression research for code language models \cite{shi2022compressing, shi2024greening, panichella2025metamorphic, shi2024efficient}. To better understand the impact of model compression strategies and simplify the fine-tuning process, we use the filtered dataset proposed by Wang et al. \cite{wang2020detecting}. Their data filtering process includes removing unlabeled examples, balancing the number of clone and non-clone instances, and scaling the dataset to ensure computational efficiency. As a result, the final filtered dataset comprises 90,102 training examples and 4,000 examples each for validation and testing.

For vulnerability detection, we select the \textit{Open Web Application Security Project} (OWASP) benchmark\footnote{https://owasp.org/www-project-benchmark}, a widely used Java test suite designed to assess the effectiveness of vulnerability detection tools \cite{hough2020revealing, sayar2023depth}. Following Du et al. \cite{du2023extensive}, we use \texttt{version 1.1} of this benchmark, which provides a larger dataset better suited for model training. This version includes 13,041 training examples (5206 non-vulnerable and 7835 vulnerable examples) and 4,000 examples each for validation and testing. For code summarization, we use the \textit{CodeSearchNet} dataset \cite{husain2019codesearchnet}, which includes data from six programming languages and is widely adopted for evaluating code summarization models. Consistent with previous studies \cite{lu2021codexglue, du2023extensive, zeng2022extensive, awal2025largelanguagemodelsrobust}, we focus exclusively on the Java sub-dataset, which contains 164,923 examples for training, 5,183 for validation, and 10,955 for testing.

\subsubsection{Target Language Models for Code}
\label{TPTMCs}
Transformer-based \cite{vaswani2017attention} language models for code have been effectively applied to a wide range of software analytics tasks, including clone detection \cite{feng2020codebert, lu2021codexglue, ahmad2021unified}, vulnerability detection \cite{hough2020revealing, sayar2023depth}, and code summarization \cite{husain2019codesearchnet, ahmed2024automatic}. Language models are generally classified into three categories: \textit{encoder-only}, \textit{decoder-only}, and \textit{encoder-decoder} models \cite{zeng2022extensive}. Aloisio et al. \cite{d2024compression} empirically investigated the impact of various model compression strategies on the efficiency and effectiveness of the \textit{encoder-only} CodeBERT \cite{feng2020codebert} model. However, to better understand how compression strategies affect the robustness of models under adversarial attacks, we select two models from each category of architecture. Specifically, we select CodeBERT \cite{feng2020codebert} and GraphCodeBERT \cite{guo2020graphcodebert} from the \textit{encoder-only} category, CodeGPT \cite{lu2021codexglue} and UniXcoder\footnote{UniXcoder can be used as encoder-only, decoder-only, and encoder-decoder model based on how its modality-aware attention masks are configured.} \cite{guo2022unixcoder} from the \textit{decoder-only} category, and PLBART \cite{ahmad2021unified} and CodeT5 \cite{wang2021codet5} from the \textit{encoder-decoder} category.

\subsection{Model Fine-tuning}
\label{ModFineTune}
To obtain robust task-specific baselines, the standard supervised fine-tuning paradigm is employed. For each task, the pre-trained checkpoint is initialized with publicly available weights and subsequently optimized on the corresponding training split. To ensure fairness and comparability, experimental settings, including learning rate, number of epochs, batch size, and data splits, are adopted from prior studies \cite{shi2022compressing, shi2024greening}, and the official training pipeline provided by the \textit{CodeXGLUE} \cite{lu2021codexglue} benchmark is reused. This approach facilitates reproducibility and alignment with established evaluation protocols.

During fine-tuning, task-specific heads are attached to each backbone model. For classification tasks such as clone detection and vulnerability detection, a fully connected classification layer is applied over the contextual representation. For the code summarization task, models are fine-tuned in a sequence-to-sequence generation setting using teacher forcing. Beam search is employed during decoding, consistent with common practice in code summarization, and stochastic sampling is not utilized. Model selection is based on validation performance, and the best checkpoint is retained as the uncompressed baseline for subsequent compression and robustness evaluation. This fine-tuning procedure ensures that all models achieve competitive task performance before applying compression techniques.

\subsection{Model Compression Approaches}
\label{MCS}
Model compression strategies have recently attracted increasing attention in software engineering research \cite{shi2022compressing, shi2024greening, wei2023towards}. Shi et al. \cite{shi2022compressing} introduced \textit{Compressor}, a knowledge distillation-based technique to compress CodeBERT and GraphCodeBERT into 3 MB. The literature offers various model compression strategies, including pruning \cite{lecun1989optimal}, quantization \cite{gray1998quantization}, knowledge distillation \cite{hinton2015distilling}, low-rank factorization \cite{sainath2013low}, and embedding decomposition \cite{lan2019albert}. In this study, we focus on pruning, quantization, and knowledge distillation because, to the best of our knowledge, these techniques are the most extensively investigated and adopted compression approaches in prior software analytics research \cite{shi2022compressing, shi2024greening, panichella2025metamorphic, d2024compression}. Additionally, these compression methods constitute complementary paradigms that facilitate a comprehensive robustness assessment across distinct compression mechanisms.

\subsubsection{Knowledge Distillation}
\label{KD}
Knowledge Distillation (\textbf{KD}) \cite{hinton2015distilling} is a method for transferring knowledge from a large, complex model (the \textit{teacher}) to a smaller, more efficient model (the \textit{student}). It is computationally intensive, as it usually requires retraining the student model from scratch. Several KD-based methods have been introduced, including \textit{Compressor} \cite{shi2022compressing}, \textit{AVATAR} \cite{shi2024greening}, and \textit{MORPH} \cite{panichella2025metamorphic}. These approaches employ a similar teacher-to-student knowledge transfer paradigm based on logit alignment in classification contexts. Due to this methodological similarity, \textit{Compressor} is selected as the representative KD method in this study, as it is specifically tailored for encoder-only transformer models and is well-suited to the program understanding tasks examined in our empirical analysis.

\textit{Compressor} initially employs a genetic algorithm-based approach to search for a simplified model architecture that resembles the teacher model, aiming to maximize \textit{Giga floating-point operations} (GFLOPs) while minimizing the model size within a fixed memory budget (measured in megabytes, MB). Subsequently, knowledge distillation is employed wherein the teacher model processes unlabeled data to generate soft targets (e.g., logits), which are then used to supervise the training of the student model. \textit{Compressor} uses the Kullback-Leibler (KL) divergence \cite{kullback1951information} loss function to minimize the loss between the output of the teacher and student model. KL divergence loss is defined as follows: 

\begin{equation}
    \mathcal{L} = -\frac{1}{n} \sum_{i}^{n} softmax(\frac{p_i}{T})\text{log} \Bigl( softmax(\frac{q_i}{T}) \Bigl) T^2
    \label{KLD}
\end{equation}

In ``\eqref{KLD}'', $n$ is the number of examples in the training set. $p_i$ and $q_i$ represent the logits of the teacher and student models, respectively. $T$ is the temperature index used to control the softmax function, as introduced by Hinton et al. \cite{hinton2015distilling}. It is essential to note that the logits (e.g., $p_i$) generated by the teacher model are fixed, whereas the logits (e.g., $q_i$) produced by the student model change as retraining progresses. Finally, to simplify our implementation, we keep the vocabulary size unchanged when using \textit{Compressor} to compress language models for code.

It is important to note that in our study, \textit{Compressor} is applied only to the CodeBERT and GraphCodeBERT models, as it is specifically designed for encoder-only models. The method distills knowledge by aligning the logits of the teacher and student models for classification tasks. Extending \textit{Compressor} to generation tasks, such as code summarization, would require sequence-level distillation, which is not supported by the current implementation. Although recent work such as \textit{SODA} \cite{chen2025smaller} targets decoder-based code models and program generation tasks, a complete and reproducible implementation was not publicly available at the time of our study\footnote{https://shorturl.at/OEdFD}. Including this method without full reproducibility would compromise experimental fairness and rigor. Therefore, \textit{SODA} is not included in our experimental analysis.

\subsubsection{Pruning}
\label{prun}
Pruning (\textbf{PR}) \cite{sanh2020movement} is a compression technique that eliminates redundant or less significant weights from a model, enhancing storage and memory efficiency. It can also improve computational throughput and inference latency without substantially compromising model performance. The pruning technique can be categorized into \textit{structured}, \textit{unstructured}, and \textit{semi-structured} pruning. Structured pruning eliminates entire components, such as neurons, attention heads, or layers, based on defined criteria while maintaining the model's overall architecture. In contrast, unstructured pruning removes individual weights, leading to irregular sparsity. Semi-structured pruning strikes a balance, offering both fine-grained sparsity and structural regularity. Based on the work of Aloisio et al. \cite{d2024compression}, we use the \textit{unstructured global pruning}\footnote{https://pytorch.org/tutorials/intermediate/pruning\_tutorial.html} technique implemented in the PyTorch library, as it preserves the original model architecture. Furthermore, we apply pruning to the weights of all linear layers in the network, using the $ L1 \ norm$ as the criterion for selection \cite{gordon2020compressing}. Similar to Aloisio et al. \cite{d2024compression}, we analyze various pruning configurations, ranging from 20\% to 60\% of the weights in 5\% intervals (e.g. [20\%, 25\%, 30\%, 35\%, 40\%, 45\%, 50\%, 55\%, 60\%]). We begin by fine-tuning the models for each task, followed by pruning. Finally, we remove the internal copy of the original weights created during pruning and store the model for further experiments.

\subsubsection{Quantization}
\label{quant}
Quantization (\textbf{QT}) \cite{zafrir2019q8bert} reduces a model's size by decreasing its weights' bit-width (i.e., \texttt{float32} to \texttt{int8}), thereby enhancing storage, memory demand, computational cost, and inference speed. Broadly, quantization can be divided into two main categories: \textit{post-training quantization} (PTQ) and \textit{quantization-aware training} (QAT). Based on the work of Aloisio et al. \cite{d2024compression}, we use the PTQ method implemented in Hugging Face’s Optimum-Quanto library\footnote{https://github.com/huggingface/optimum-quanto}. We consider two quantization configurations by converting full precision (\texttt{float32}) weights to \texttt{int8} and \texttt{float8}. We perform quantization in three steps: (1) fine-tune the models for each task; (2) quantize the fine-tuned models and calibrate their activation functions using the validation dataset; (3) freeze the quantized models and store them for further experiments.  


\subsubsection{Selection of Compressed Models}
While \textit{Compressor} produces a single compressed model, pruning and quantization generate multiple compressed models depending on the configurations used. Therefore, selecting a single compressed model to evaluate its robustness to adversarial attacks is crucial when pruning and quantization are applied. The intuitive approach is to choose the model with the least performance degradation. Thus, the final compressed model, obtained via pruning and quantization, is selected based on the best trade-off between compression rate and a performance drop of no more than 5\% on the test dataset relative to the uncompressed model for each software analytics task. This design choice reflects real-world scenarios in which compressed models with significant performance degradation are unlikely to be deployed in practice \cite{dong2025can}. Following the studies by Shi et al. \cite{shi2022compressing, shi2024efficient}, we use only the \textit{accuracy} metric to compare the performance of compressed versus uncompressed models for selecting the best models obtained through pruning and quantization. Thus, after selecting the compressed model variants, they are subsequently used to evaluate the adversarial robustness.

\subsection{Adversarial Attacks}
In this section, we outline the adversarial attack strategies employed in our study to assess the robustness of compressed and uncompressed models.

\subsubsection{Attack Approaches}

Adversarial attacks provide a systematic approach to evaluating the robustness of machine learning models by introducing carefully crafted perturbations that mislead model predictions while preserving the original input semantics. In this empirical study, adversarial attacks play a key role in assessing the impact of compression strategies on the robustness of compressed models.

Following the recommendation of Du et al. \cite{du2023extensive}, this study employs black-box and untargeted adversarial attacks against models. We select black-box attacks over white-box techniques because, in real-world deployments, the internal parameters and gradients of compressed models are typically inaccessible to adversaries. Black-box attacks, therefore, better represent realistic threat scenarios where only model inputs and outputs can be queried. Additionally, these methods are model-agnostic and can be applied uniformly across various compression strategies without modifying the model architecture or requiring access to training data. We adopt untargeted attacks rather than targeted ones because they represent a more general and practical threat model, in which the adversary aims to induce incorrect predictions rather than enforce a specific output. Additionally, untargeted attacks are less constrained and more broadly applicable across software analytics tasks, which makes them suitable for evaluating the overall robustness of compressed models. Furthermore, adversarial perturbations are expected to be subtly small, ensuring the preservation of the original input's syntactic and semantic properties while effectively inducing prediction errors in the target models \cite{yang2022natural, du2023extensive, zhang2020generating, awal2024large}.

To this end, we employ four\footnote{Four attack approaches: ALERT: \url{https://github.com/soarsmu/attack-pretrain-models-of-code/}, BEAM: \url{https://github.com/CGCL-codes/Attack_PTMC/tree/main}, MHM: \url{https://github.com/SEKE-Adversary/MHM}, and Wir-Random: \url{https://github.com/ZZR0/ISSTA22-CodeStudy}} SOTA black-box adversarial attack techniques that focus only on identifier renaming in source code: ALERT \cite{yang2022natural}, BeamAttack \cite{du2023extensive}, Metropolis-Hastings Modiﬁer (MHM) \cite{zhang2020generating}, and WIR-Random \cite{zeng2022extensive}. These techniques are designed to introduce minimal yet effective perturbations that do not alter the code's functionality but are sufficient to challenge model robustness. In addition to identifier renaming-based attacks, the more advanced Code Difference guided Adversarial example generation technique, CODA\footnote{CODA: \url{https://github.com/tianzhaotju/CODA}} \cite{tian2023code}, is utilized. CODA generates adversarial examples through semantic-preserving code transformations guided by structural and identifier differences between target and reference inputs. Below, we provide a brief introduction to the selected black-box adversarial attack approaches:

\begin{itemize}

    \item \textbf{ALERT}. Yang et al. \cite{yang2022natural} introduced ALERT, a black-box attack technique that takes into account both the operational and natural semantics of source code when creating adversarial examples. The method only considers local identifiers for substitution based on their context in the code. ALERT employs two selection methods for substituting identifiers: \textit{greedy} and \textit{genetic} algorithms. We set the exact same hyperparameter setting as used in ALERT to generate adversarial examples. For example, we consider 30 candidates for each substitute identifier.

    \item \textbf{WIR-Random}. Zeng et al. \cite{zeng2022extensive} proposed WIR-Random, a black-box attack technique that works on identifier substitution in the code. It employs \textit{Word Importance Rank (WIR)} to find the candidate substitution sequence of each identifier and ranks them according to their importance in the prediction probabilities of the model. Finally, WIR-Random replaces the sorted identifiers one by one based on the randomly selected candidates.

    \item \textbf{MHM}. Zhang et al. \cite{zhang2020generating} introduced a black-box attack method that involves renaming identifiers in the code using the Metropolis-Hastings (M-H) sampling concept. The MHM technique replaces each identifier with candidate substitutions through iterative M-H sampling. MHM is governed by two hyperparameters: the maximum number of iterations and the number of candidate identifiers in each iteration. Consistent with the original article \cite{zhang2020generating}, we set these hyperparameters at 100 and 30, respectively.

    \item \textbf{BeamAttack}. Du et al. \cite{du2023extensive} proposed BeamAttack, an efficient and effective method for adversarial code attacks. BeamAttack's efficiency lies in its ability to organize identifiers into groups based on the statements they are part of and to iteratively prioritize the selection of identifiers most likely to lead to successful attacks. This approach significantly reduces the attack cost compared to exploring the entire identifier space. Furthermore, BeamAttack mitigates the risk of being trapped in local optima, a drawback of sequential identifier examination methods such as WIR-Random \cite{zeng2022extensive}.

    \item \textbf{CODA}. The Code Difference guided Adversarial example generation (CODA) technique is adopted to generate adversarial examples for code language models. CODA leverages code differences between a target input and reference inputs to guide the generation process, which significantly reduces the search space of candidate perturbations. The method extracts transformation ingredients from two types of differences: \textit{code structure differences} and \textit{identifier differences}, both of which enable semantic-preserving transformations. Using these ingredients, CODA applies equivalent-structure and identifier-renaming transformations to iteratively modify the original code while preserving its functionality. The generated variants are evaluated by the target model, and those that cause prediction changes are considered adversarial examples.
    
    
\end{itemize}

\subsubsection{Attack Settings}
\label{AS}
We use the fine-tuned models described in Section \ref{TPTMCs} as targets for our attacks and adapt the original implementations of the five attack approaches used in this study. To achieve this, we retain the default hyperparameter settings of these attack methods. For instance, ALERT evaluates 30 candidate substitutions per identifier, whereas MHM uses at most 100 iterations and considers 30 candidate identifiers per iteration for substitution. We consider the entire test set, consisting of $4,000$ instances, as the target for attacks in both vulnerability detection and clone detection tasks. For the code summarization task, we randomly select $4,000$ examples from the test set as attack instances, consistent with the number of target instances used in the other two tasks \cite{du2023extensive}. Since the four identifier-renaming attack approaches selected in this study focus solely on identifier renaming when generating adversarial examples, we exclude test code snippets that do not contain identifiers. Additionally, we concentrate only on instances that models correctly predict before adversarial attacks, following the standard approach used in existing studies \cite{yang2022natural, zhang2020generating}. If a model misclassifies a sample under standard conditions, further modification of that sample does not yield meaningful insights into the model’s vulnerability to adversarial perturbations. Consequently, incorrectly predicted samples are excluded when generating adversarial examples to ensure that reported attack success rates accurately reflect the capacity of perturbations to convert correct predictions into incorrect ones. Finally, it is essential to emphasize that we compare the impact of model compression strategies on the robustness of compressed models compared to their uncompressed counterparts using consistent experimental settings across all target models and subject systems in this study.

\subsection{Robustness Analysis}
\label{RDAnalysis}

We use two sets of evaluation metrics in our experiment. The first set quantitatively compares the impact of compression strategies on the robustness of models under adversarial attacks. The second set quantitatively evaluates the quality of adversarial examples, ensuring that the changes are minimal yet effective in misleading the models.

\subsubsection{Quantitative Analysis}
\label{RDQuant}
We adopt \textit{Attack Success Rate} (\%ASR) \cite{yang2022natural} and \textit{Average Model Query} (AMQ) \cite{du2023extensive} for the quantitative analysis. \textbf{\%ASR} represents the proportion of test samples where an attack method successfully generates adversarial examples. It is defined as below:

\begin{equation}
    \%\text{ASR} = \frac{|\{x \in X \wedge M(x') \neq M(x)\}|}{\left|\left\{x \in X \mid M(x)=y\right\}\right|}
    \label{ASREQN}
\end{equation}

Where $x$ is an original input, $x'$ is a perturbed input. A higher \%ASR indicates that the models are more vulnerable to adversarial attacks. \textbf{AMQ} refers to the number of queries made to the target model during the generation of adversarial examples. A lower AMQ suggests that the model is more vulnerable to adversarial attacks.

\subsubsection{Adversarial Example Quality Analysis}
\label{RDQual}
We conduct an adversarial example quality analysis to assess the effectiveness of the generated adversarial samples and to verify that the applied modifications remain minimal yet sufficient to mislead the models. Excessive modifications to the original test samples may produce out-of-distribution inputs, in which case the resulting performance degradation may not accurately indicate a genuine lack of model robustness. Therefore, adversarial perturbations should remain minimal and must preserve the original semantics of the input. Since the inputs to models in the selected software analytics tasks consist of code, we adopt established quantitative quality metrics from Du et al. \cite{du2023extensive}, \textit{Identifier Change Rate} (ICR), \textit{Token Change Rate} (TCR), \textit{Average Edit Distance} (AED), and \textit{Average Code Similarity} (ACS) to evaluate the quality of the generated adversarial examples. For a set of $m$ adversarial examples, let each $i$-th code snippet contain $k_i$ identifiers, of which $n_i$ are modified to generate the adversarial example. The \textbf{ICR} is then computed as below:

\begin{equation}
    \text{ICR} = \frac{\sum_{1}^{m} n_i}{\sum_{1}^{m} k_i}
    \label{ICREQN}
\end{equation}

\textbf{TCR} measures the proportion of tokens altered in the adversarial example relative to the total number of tokens in the original code. \textbf{AED} measures the differences between tokens at the character level, indicating the number of edits needed to transform one token into another. \textbf{ACS} is calculated using cosine similarity between the embeddings generated by CodeBERT before and after the adversarial attacks are applied. \textit{High-quality adversarial examples typically exhibit lower ICR, AED, and TCR values, along with higher ACS scores}.

Nonparametric statistical tests are employed to assess whether observed differences in the quality of adversarial examples generated by various attack techniques are statistically significant. The Friedman test \cite{friedman1937use} is utilized to compare the quality of adversarial examples across multiple attack techniques using the evaluated metrics. Additionally, the Wilcoxon signed-rank test \cite{woolson2007wilcoxon} is applied to assess statistically significant differences between adversarial examples generated for compressed and uncompressed models using the same attack techniques. These tests are chosen because they do not require the evaluation metrics to be normally distributed. A $p$-value less than 0.05 indicates that the observed differences in adversarial example quality are statistically significant. The corresponding statistical results and $p$-values are reported in the experimental results section.

\subsection{Experimental Settings}
\label{ExpSetting}
All experiments were conducted on a Linux server running Ubuntu 22.04.5 LTS, equipped with an Intel Xeon Platinum 8356H CPU, 3.0 TB RAM, and four NVIDIA A100 GPUs with 80 GB memory each. To ensure reproducibility, we fixed the random seed to 42 across all experiments. 
The selected models were fine-tuned following the training pipeline and hyperparameter configurations defined in the CodeXGLUE benchmark \cite{lu2021codexglue} and prior adversarial attack studies \cite{yang2022natural, du2023extensive}. Since task-specific hyperparameters (e.g., learning rate, batch size, tokenizer code length, and epoch) were adopted for clone detection, code summarization, and vulnerability prediction, we provide complete fine-tuning configurations for all models and tasks in our publicly available replication package to ensure full reproducibility.

\section{Empirical Results}
\label{EMPR}
In this section, we present results from our extensive experiments evaluating the impact of model compression techniques on the adversarial robustness of language models for code relative to their uncompressed counterparts. 

\subsection{Selection of Compressed Models Based on Accuracy Preservation}

\begin{figure}[htbp]
  \centering
  \subfigure[Clone Detection]{\label{CDPrune}\includegraphics[width=6cm, height=4.5cm]{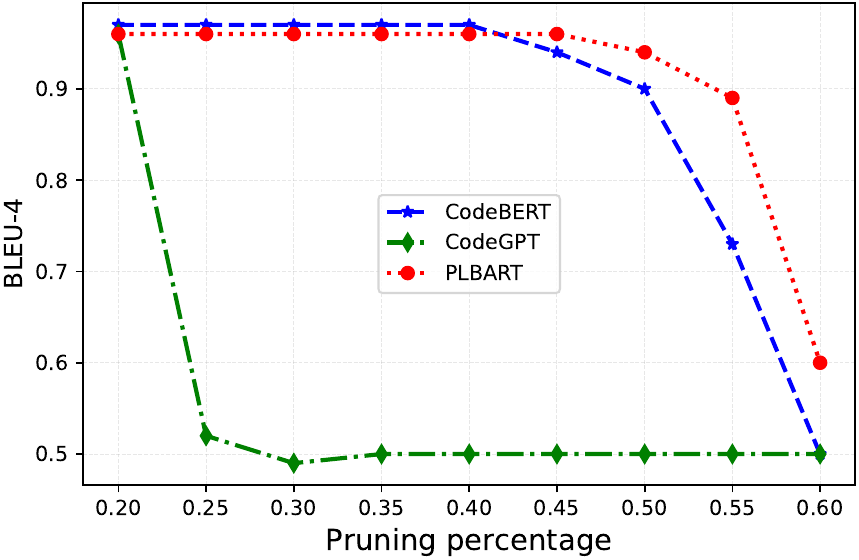}}
  \subfigure[Vulnerability Detection]{\label{VDPrune}\includegraphics[width=6cm, height=4.5cm]{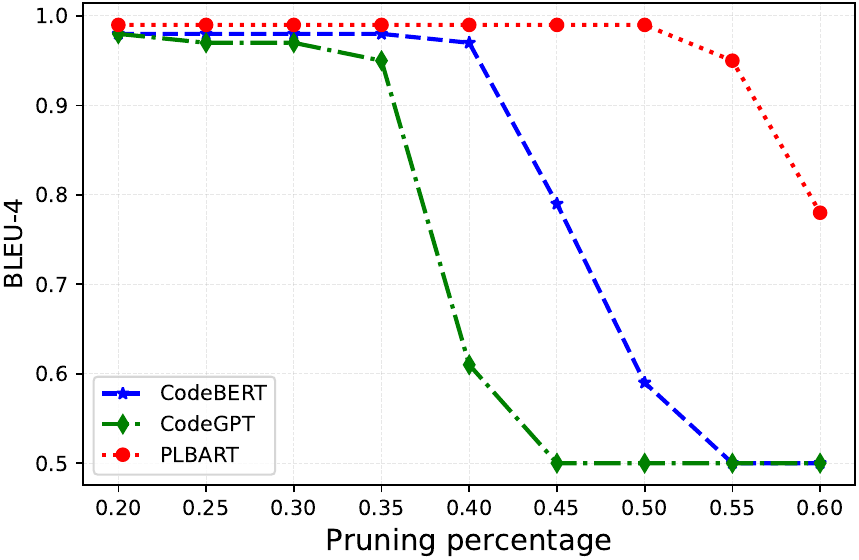}}
  \subfigure[Code Summarization]{\label{CSPrune}\includegraphics[width=6cm, height=4.5cm]{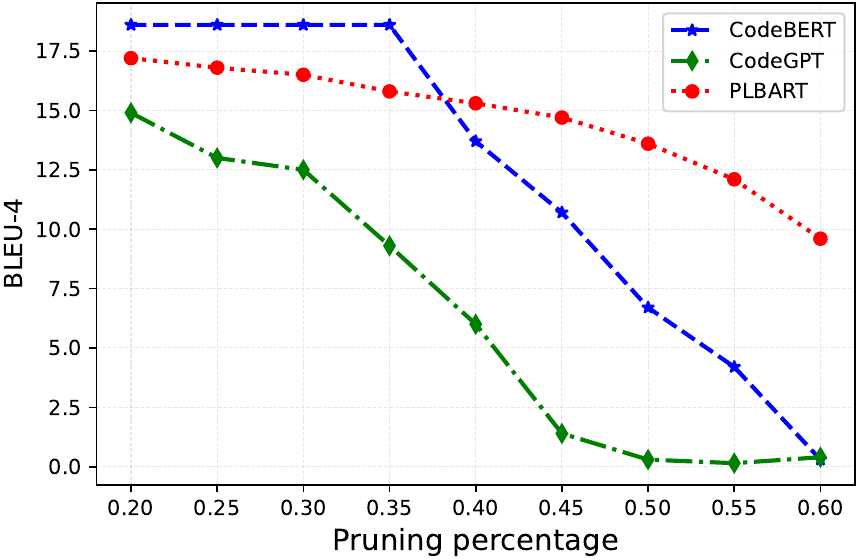}}
  \caption{Impact of different pruning ratios on model accuracy across software analytics tasks, highlighting the trade-off between model compression ratio and accuracy preservation.}
  \label{PruningVsAccuracy}
\end{figure}

To evaluate the impact of compression strategies on the robustness of code language models under adversarial attacks, we explore different levels of pruning and quantization, as described in Section \ref{MCS}. Figure \ref{PruningVsAccuracy} illustrates how compressed models perform across various weight pruning ratios. For clarity, we present results for representative models from each architectural category. The full set of experimental results is available in our replication package.

As shown in Figures \ref{VDPrune} and \ref{CSPrune}, up to 40\% of the weights can be pruned from the CodeBERT model in the vulnerability detection and code summarization tasks without exceeding a 5\% drop in accuracy. For clone detection, as shown in \ref{CDPrune}, a slightly higher pruning ratio of up to 45\% is acceptable. In contrast, CodeGPT demonstrates a significantly steeper decline in performance as the pruning ratio increases, indicating that this decoder-only architecture is more sensitive to aggressive weight removal. PLBART exhibits moderate degradation relative to CodeGPT but still experiences performance drops at lower pruning thresholds than CodeBERT. These findings indicate that encoder-based models such as CodeBERT tolerate pruning more effectively, whereas generative architectures, particularly CodeGPT, are more vulnerable to performance degradation under compression.

A similar selection strategy is applied for quantized models. Two quantization configurations (int8 and float8) are evaluated, and the variant that retains the highest task performance while limiting the accuracy drop to no more than 5\% relative to the uncompressed model is selected. According to this criterion, the selected configurations are as follows: CodeBERT uses \texttt{float8} for clone detection, \texttt{int8} for vulnerability detection, and \texttt{float8} for code summarization; CodeGPT uses \texttt{int8} for clone detection and vulnerability detection, and \texttt{float8} for code summarization; PLBART uses \texttt{float8} for all three tasks: clone detection, vulnerability detection, and code summarization. Since the knowledge distillation approach, \textit{Compressor}, produces only one compressed variant, we use that version in our subsequent analysis. This selection strategy ensures that the most effective compressed models, in terms of both compression and accuracy, are included in our evaluation. Their performance is reported in Table \ref{TablePerformCDVD} for program understanding tasks and in Table \ref{TablePerformCS} for program generation tasks, along with the corresponding uncompressed models.

\begin{table*}[htbp]
\centering
\caption{Pre-attack predictive performance (Accuracy) of uncompressed and compressed models across program understanding tasks: clone detection and vulnerability detection. Results are reported on the test dataset before adversarial attacks. Note: \textit{Compressor} is applicable only to encoder-only models under classification settings. UC: Uncompressed, PR: Pruned, QT: Quantized, KD: Knowledge Distillation (\textit{Compressor}).}
\begin{tabular}{l|l|cccc|cccc}
\toprule
\textbf{Architecture} & \textbf{Model}
& \multicolumn{4}{c|}{\textbf{Clone Detection}}
& \multicolumn{4}{c}{\textbf{Vulnerability Detection}} \\
\cmidrule(lr){3-6} \cmidrule(lr){7-10}
& & \textbf{UC} & \textbf{PR} & \textbf{QT} & \textbf{KD}
& \textbf{UC} & \textbf{PR} & \textbf{QT} & \textbf{KD} \\
\midrule

\multirow{2}{*}{Encoder-only}
& CodeBERT      & 0.96 & 0.93 & 0.96 & 0.96 & 0.98 & 0.98 & 0.98 & 0.98 \\
& GraphCodeBERT & 0.97 & 0.93 & 0.96 & 0.93 & 0.71 & 0.68 & 0.69 & 0.68 \\
\midrule

\multirow{2}{*}{Decoder-only}
& CodeGPT       & 0.96 & 0.95 & 0.96 & --   & 0.98 & 0.95 & 0.97 & --   \\
& UniXcoder     & 0.97 & 0.95 & 0.97 & --   & 0.99 & 0.98 & 0.99 & --   \\
\midrule

\multirow{2}{*}{Encoder-decoder}
& CodeT5        & 0.98 & 0.95 & 0.97 & --   & 0.98 & 0.96 & 0.98 & --   \\
& PLBART        & 0.97 & 0.94 & 0.96 & --   & 0.99 & 0.96 & 0.99 & --   \\
\bottomrule
\end{tabular}
\label{TablePerformCDVD}
\end{table*}

Table \ref{TablePerformCDVD} presents the pre-attack predictive performance of uncompressed and compressed models for clone detection and vulnerability detection tasks. The results demonstrate that the selected compression strategies generally preserve task-level performance across most models. Among encoder-only models, CodeBERT maintains comparable accuracy under pruning, quantization, and knowledge distillation for both tasks, with negligible performance degradation. Similarly, GraphCodeBERT also exhibits relatively stable performance under compression, though a slight decrease is observed in the vulnerability detection task, where accuracy declines from 0.71 to 0.68, suggesting modest sensitivity to compression.

Among decoder-only models, both CodeGPT and UniXcoder demonstrate robust predictive performance, with minimal accuracy differences between uncompressed and compressed variants. Similarly, the encoder–decoder models CodeT5 and PLBART retain high accuracy across both tasks under pruning and quantization, with only slight reductions observed. All selected compressed models satisfy the performance retention criterion of no more than a 5\% decrease relative to their uncompressed counterparts. This ensures that the compressed models remain practically viable and that subsequent adversarial robustness evaluations are not confounded by significant differences in baseline predictive performance.

Table \ref{TablePerformCS} presents the pre-attack predictive performance of uncompressed and compressed models on the code summarization task, as measured by BLEU-4. The results demonstrate that compression strategies typically preserve generation quality; however, the extent of this preservation varies across models and compression techniques.

\begin{table}[htbp]
\centering
\caption{Pre-attack predictive performance (BLEU-4) of uncompressed and compressed models on the code summarization task. Results are reported on the test dataset before adversarial attacks.}
\begin{tabular}{l|ccc}
\toprule
\textbf{Model} & \textbf{Uncompressed} & \textbf{Pruned} & \textbf{Quantized} \\
\midrule
CodeBERT   & 18.8 & 18.6 & 14.8 \\
CodeGPT    & 14.9 & 12.5 & 14.5 \\
PLBART     & 17.7 & 13.7 & 17.5 \\
UniXcoder  & 18.9 & 16.9 & 18.8 \\
\bottomrule
\end{tabular}
\label{TablePerformCS}
\end{table}

CodeBERT and UniXcoder maintain relatively stable performance under pruning, exhibiting only minor reductions in BLEU-4 scores. In contrast, quantization causes a more pronounced decrease for CodeBERT but has minimal effect on UniXcoder. For CodeGPT, pruning results in a greater performance reduction, whereas quantization largely preserves its original performance. Similarly, PLBART experiences a substantial decline under pruning, whereas quantization maintains performance comparable to that of the uncompressed model. Overall, the selected compressed models retain reasonably comparable generation quality, ensuring that subsequent adversarial robustness evaluation is not biased by substantial differences in baseline performance.

\subsection{Answering RQ1: Evaluating Adversarial Examples Quality} 
\begin{figure}[htbp]
  \centering
  \subfigure[ICR on CD]{\label{CDICR}\includegraphics[width=6cm, height=3.5cm]{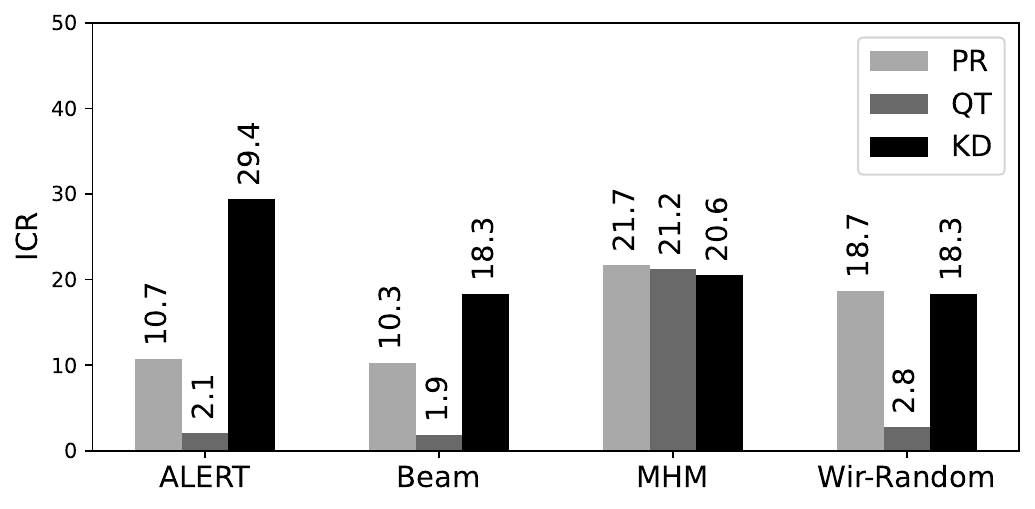}}
  \subfigure[ICR on VD]{\label{VDICR}\includegraphics[width=6cm, height=3.5cm]{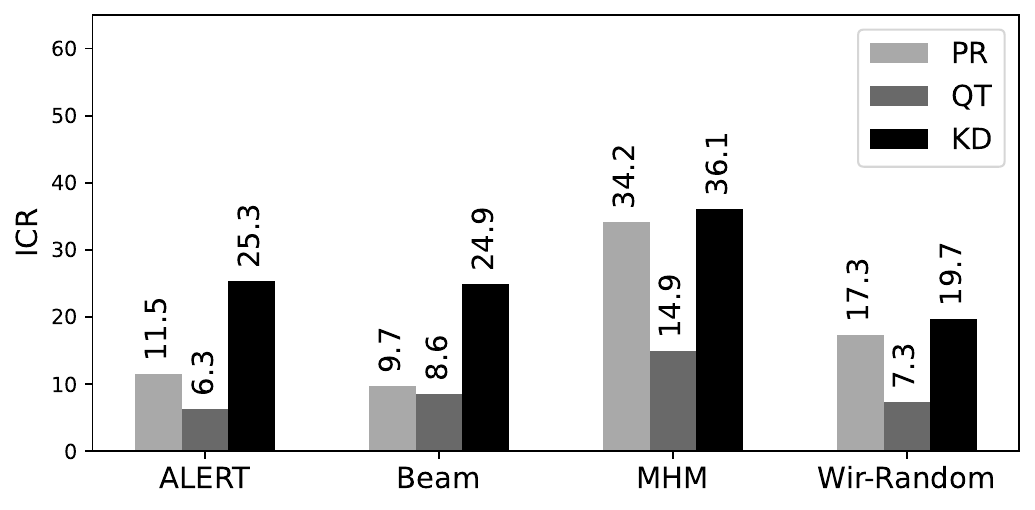}}
  \subfigure[TCR on CD]{\label{CDTCR}\includegraphics[width=6cm, height=3.5cm]{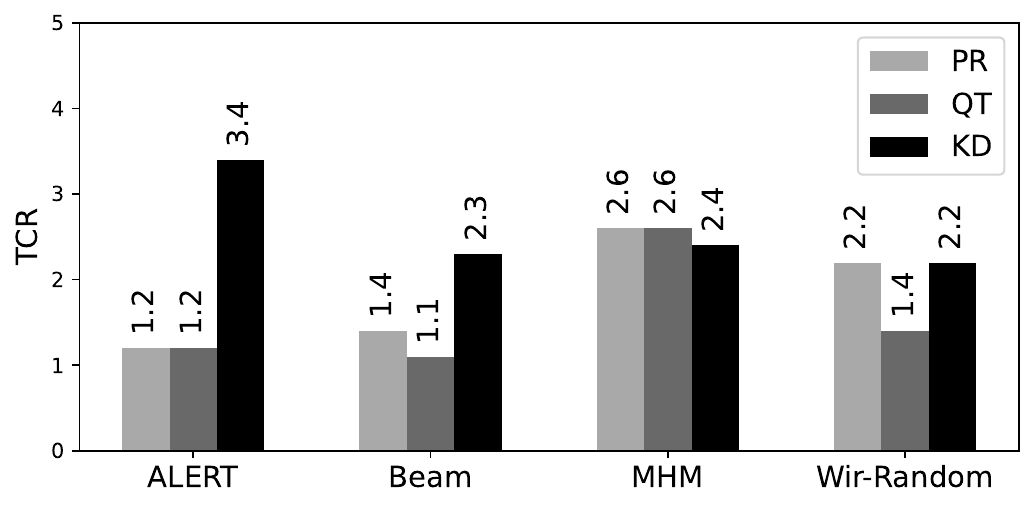}}
  \subfigure[TCR on VD]{\label{VDTCR}\includegraphics[width=6cm, height=3.5cm]{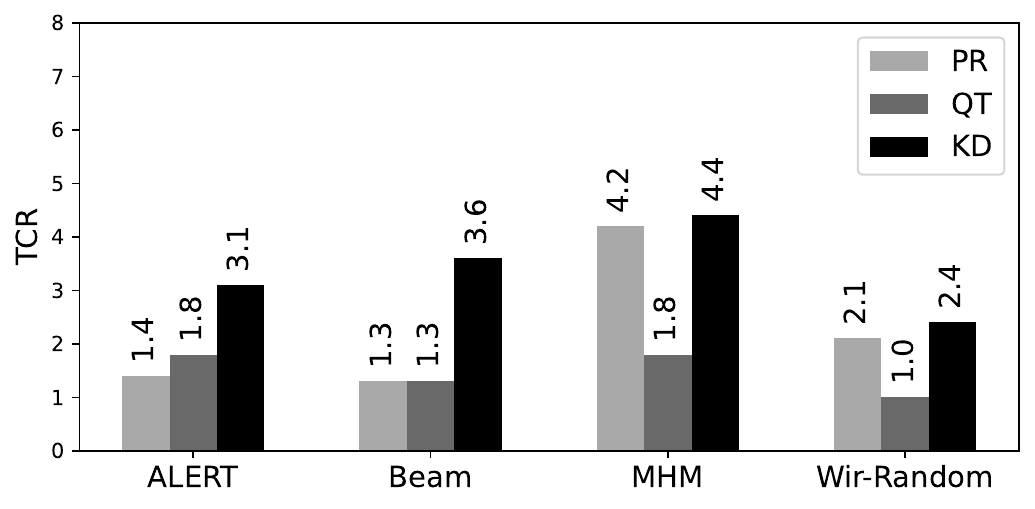}}
  \subfigure[ACS on CD]{\label{CDACS}\includegraphics[width=6cm, height=3.5cm]{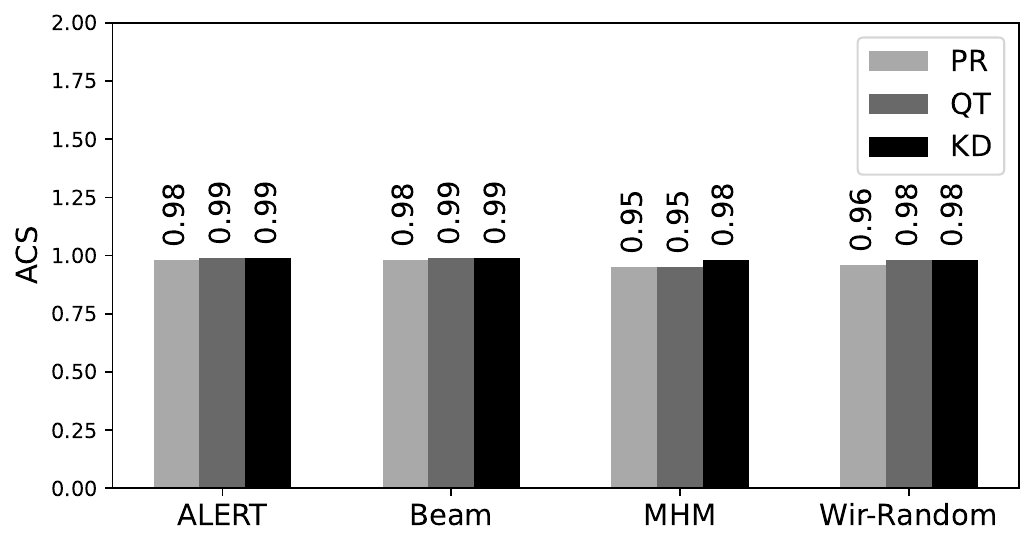}}
  \subfigure[ACS on VD]{\label{VDACS}\includegraphics[width=6cm, height=3.5cm]{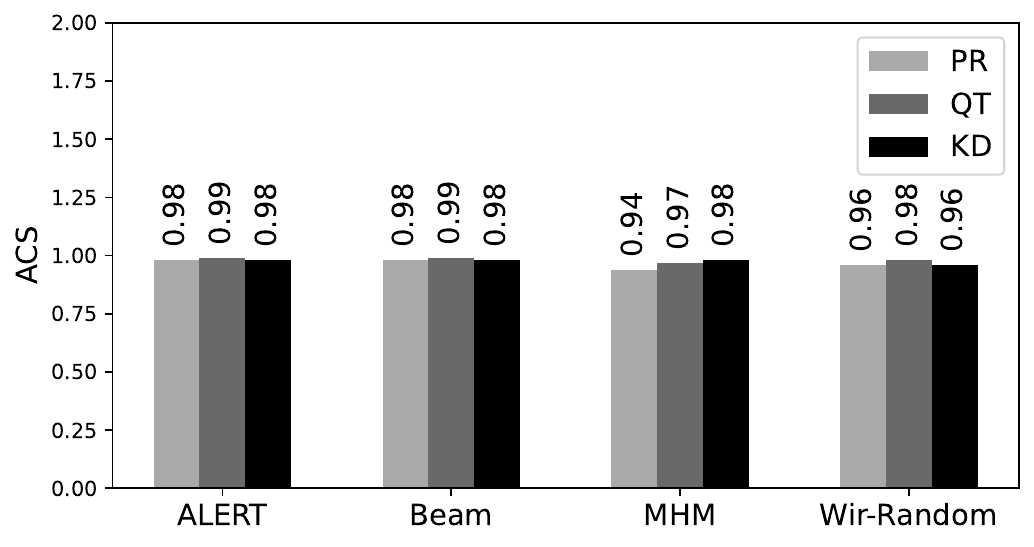}}
  \subfigure[AED on CD]{\label{CDAED}\includegraphics[width=6cm, height=3.5cm]{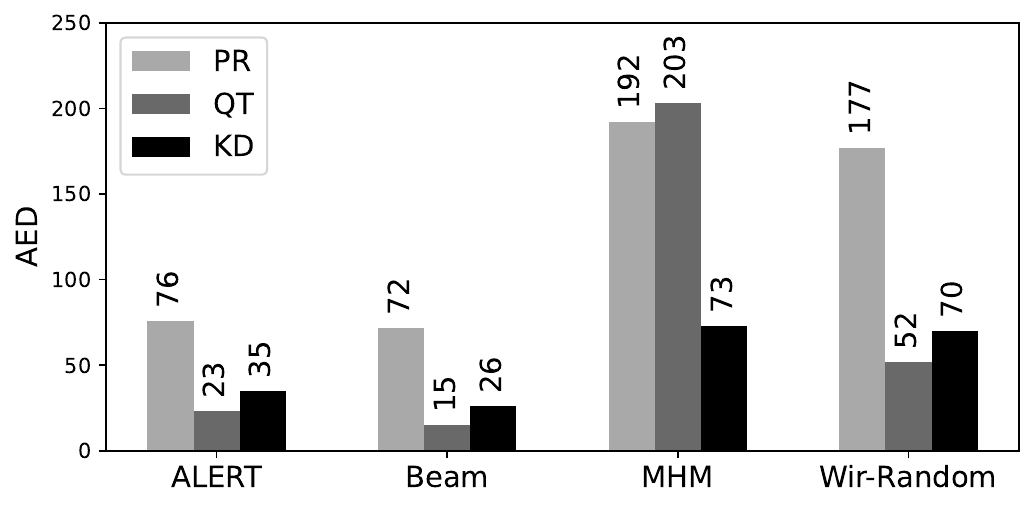}}
  \subfigure[AED on VD]{\label{VDAED}\includegraphics[width=6cm, height=3.5cm]{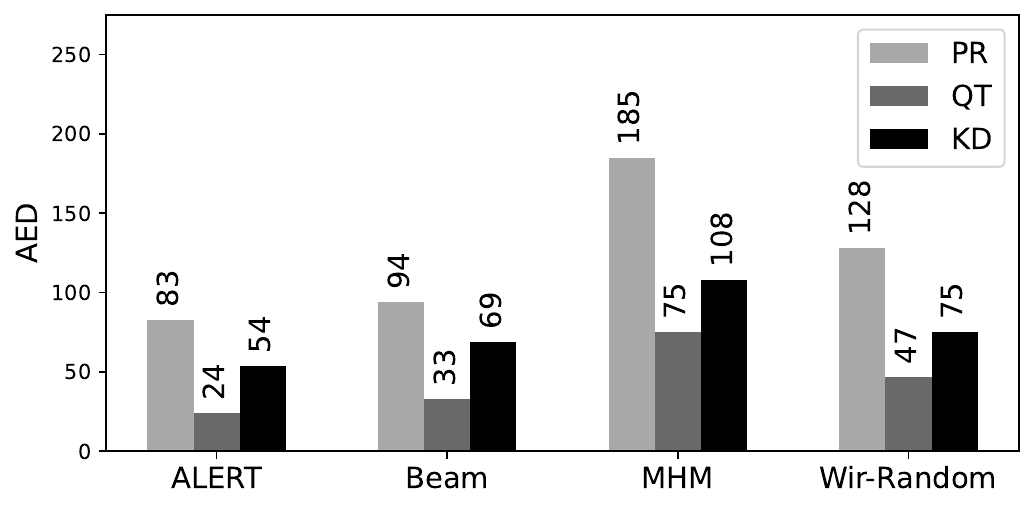}}
  \caption{Evaluating the quality of adversarial examples generated by four adversarial attack techniques (ALERT, BEAM, MHM, and WIR-Random) for pruned, quantized, and knowledge-distilled models for clone detection and vulnerability detection tasks.}
  \label{AdvExampleQuality}
\end{figure}

Intuitively, excessive modifications to the original inputs can lead to unrealistic adversarial examples, where the failure of uncompressed and compressed models may not accurately reflect an actual degradation in robustness. Therefore, this research question evaluates the syntactic and semantic quality of adversarial examples generated by adversarial attack techniques, ensuring that only minimal yet effective changes are made during the attacks. Through RQ1, we investigate how each attack technique impacts the quality of adversarial examples, focusing on their syntactic integrity and semantic coherence. Given the extensive nature of our experiments, we limit our analysis to the CodeBERT model fine-tuned on the clone detection and vulnerability detection tasks to enable a detailed and in-depth evaluation. The full set of experimental results for other models and tasks is provided in our replication package. Since all the adversarial attack techniques used in this study only rename \textit{identifiers} at the token-level when generating adversarial examples, the semantics of the adversarial examples remain the same as those of the original examples. Figure \ref{AdvExampleQuality} presents the quality evaluation results of adversarial examples generated by attack techniques targeting different compressed models, using four evaluation metrics.

\textit{High-quality adversarial examples typically exhibit lower ICR, AED, and TCR values and higher ACS scores}. Figure \ref{AdvExampleQuality} shows that, overall, the adversarial attack techniques successfully preserve the syntactic quality of the generated examples. In particular, the adversarial examples consistently achieve high ACS scores across both clone detection and vulnerability detection tasks for all compressed models. Adversarial examples also maintain strong overall quality in terms of ICR and TCR metrics for both pruned and quantized models. While some minor degradation is observed for examples targeting knowledge-distilled models, we conduct a Friedman test \cite{friedman1937use}, a non-parametric test suitable for assessing whether the quality of adversarial examples generated by the attack methods differs significantly across different compressed models. Prior to applying the test, the assumptions are verified: all compression methods are evaluated under the same experimental conditions, forming complete blocks. The evaluated metrics are continuous and rankable, and the evaluation settings are mutually independent. A $p$-value less than 0.05 indicates that the differences in the quality of adversarial examples generated across the different compressed models are statistically significant.

Table \ref{RQ3Stat} presents the $p$-values obtained from the Friedman test for the ICR, TCR, ACS, and AED metrics across the three compressed models. The results show that, apart from the two cases highlighted in bold ($p\text{-value} < 0.05$), there are no statistically significant differences in the quality of adversarial examples generated by the attack techniques for the pruned, quantized, and knowledge-distilled models. These findings suggest that the adversarial examples are generally of high quality and serve as effective inputs for evaluating the robustness of compressed language models for code.

\begin{table}[htbp]
\centering
\caption{Statistical comparison of adversarial example quality across pruned, quantized, and knowledge distilled models using the Friedman test. Reported $p$-values indicate whether differences among compression strategies are statistically significant.}
\begin{tabular}{c|c|c|c|c}
\hline
\textbf{Task} & \textbf{ICR} & \textbf{TCR} & \textbf{ACS} & \textbf{AED} \\
\hline
CD & 0.173 & 0.367 & 0.231 & 0.338 \\
\hline
VD & \textbf{0.018} & 0.126 & 0.092 & \textbf{0.017} \\
\hline
\end{tabular}
\label{RQ3Stat}
\end{table}

\begin{figure}[htbp]
  \centering
  \subfigure[CD]{\label{CD_Adv_Qual}\includegraphics[width=6.25cm, height=3.75cm]{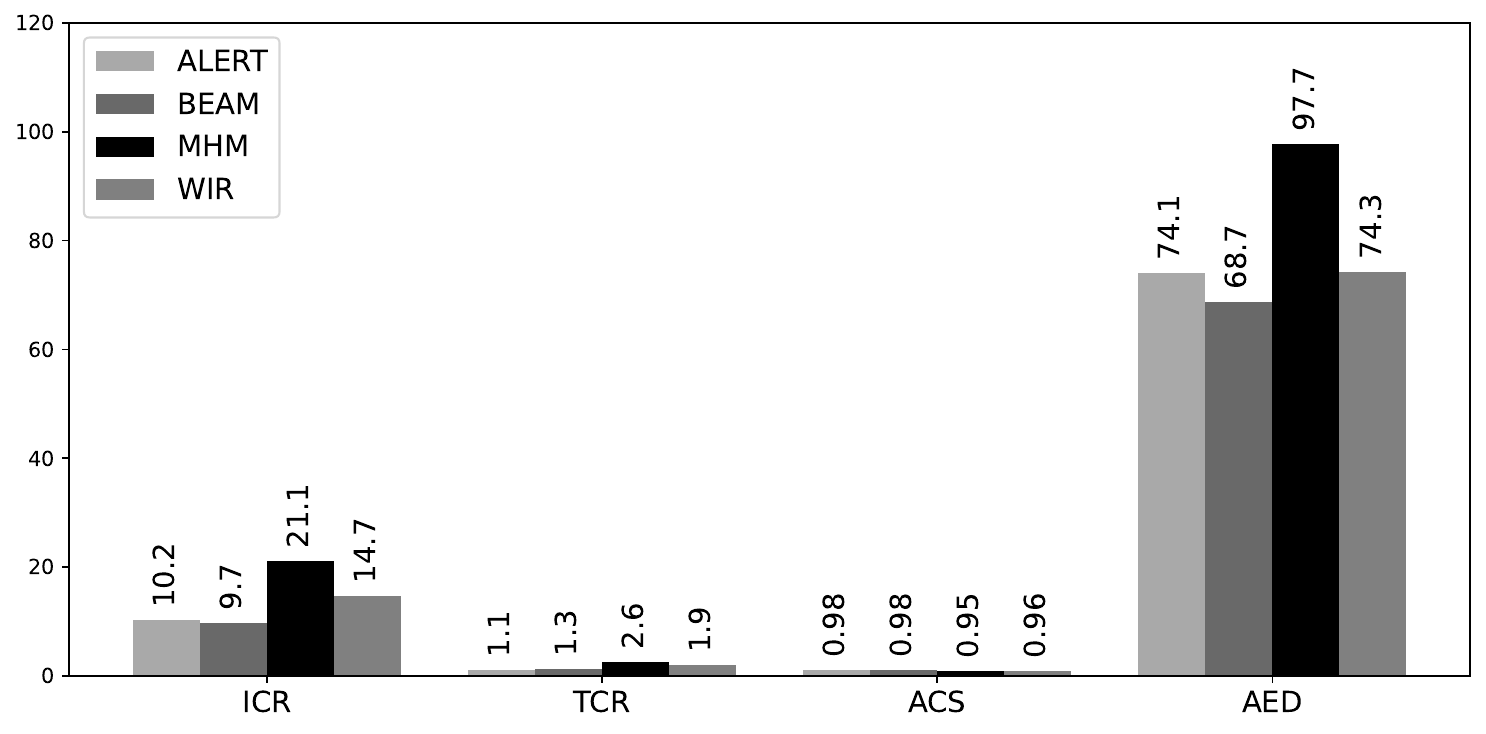}}
  \subfigure[VD]{\label{VD_Adv_Qual}\includegraphics[width=6.25cm, height=3.75cm]{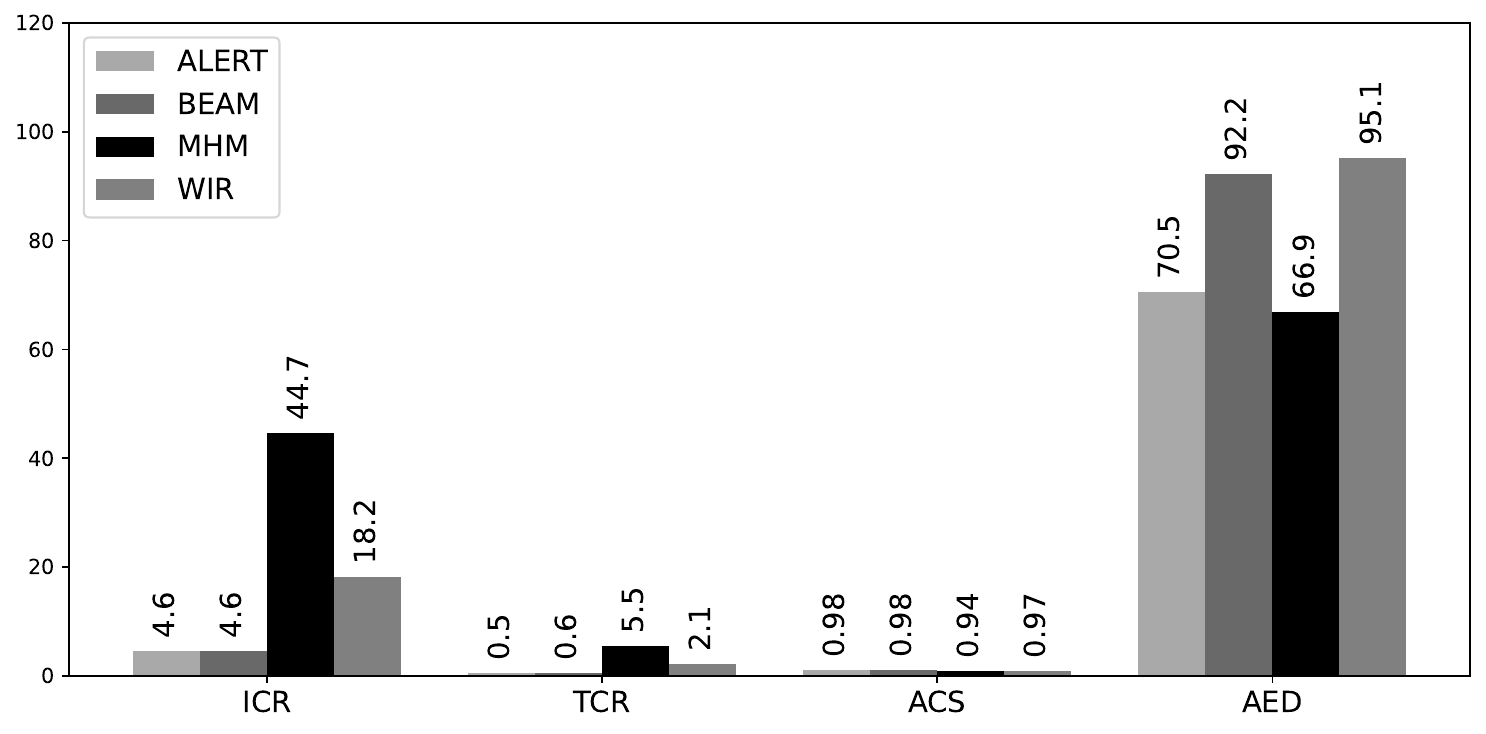}}
  \caption{Evaluating the quality of adversarial examples generated by four adversarial attack techniques (ALERT, BEAM, MHM, and WIR-Random) for uncompressed models for the clone detection and vulnerability detection tasks.}
  \label{AdvQualityOrg}
\end{figure}

Having established that the generated adversarial examples generally maintain high syntactic and semantic quality across compressed models, we next investigate whether similar patterns hold for uncompressed models. Conducting the same analysis on uncompressed models provides a fair basis for comparing the robustness of compressed and uncompressed models under adversarial attacks, enabling us to determine whether any significant changes in the robustness of compressed models compared to their uncompressed counterparts are attributable to the effect of the compression itself. Figure \ref{AdvQualityOrg} presents the quality evaluation results of adversarial examples generated by attack techniques targeting uncompressed models, using four evaluation metrics. We observe that the evaluation of syntactic and semantic quality for adversarial examples generated on uncompressed models shows a similar trend to that observed for compressed models, indicating that the attack techniques generally produce high-quality examples in both settings. Furthermore, the Friedman test (all the $p\text{-values are} >= 0.05$) confirms that there is no significant difference in the quality of adversarial examples generated by the different attack techniques for the uncompressed models.

While the above results indicate that adversarial examples generally maintain high syntactic and semantic quality for both compressed and uncompressed models, a direct comparison is required to assess whether any subtle differences in quality exist between the two model types. To this end, we conduct a pairwise comparison of adversarial example quality across compressed and uncompressed models for each attack technique and evaluation metric. We employ the Wilcoxon signed-rank test \cite{woolson2007wilcoxon} to evaluate the statistical significance of any observed differences. Table \ref{UncVSCmp} presents the results of this analysis, showing that all $p$-values exceed 0.05, indicating no statistically significant difference in the quality of adversarial examples between compressed and uncompressed models.

\begin{table}[htbp]
\centering
\caption{Wilcoxon signed-rank test $p$-values for assessing whether the quality of adversarial examples differs significantly between compressed (pruned, quantized, and knowledge-distilled) and uncompressed models on the clone detection (CD) and vulnerability detection (VD) task.}
\label{tab:wilcoxon_results}
\begin{tabular}{l|l|c|c}
\toprule
\textbf{Task} & \textbf{Comparison} & \textbf{$p$-value} & \textbf{Significant ($p<0.05$)?} \\
\midrule
\multirow{3}{*}{CD} 
  & Uncompressed vs. Pruning            & 0.083 & No  \\
  & Uncompressed vs. Quantization       & 0.099 & No   \\
  & Uncompressed vs. Knowledge Distill. & 0.252 & No  \\
\midrule
\multirow{3}{*}{VD} 
  & Uncompressed vs. Pruning            & 0.157 & No  \\
  & Uncompressed vs. Quantization       & 0.143 & No   \\
  & Uncompressed vs. Knowledge Distill. & 0.322 & No  \\
\bottomrule
\end{tabular}
\label{UncVSCmp}
\end{table}

\begin{center}
\begin{tcolorbox}[
    enhanced,
    attach boxed title to top left={yshift=-3mm,yshifttext=-1mm}, 
    colback=mycolor_box,                 
    colframe=black,                
    colbacktitle= mycolor_title,            
    coltitle=black,                
    title=Result RQ1,            
    fonttitle=\bfseries,           
    boxed title style={size=small},
    width=0.95\textwidth,            
    boxsep=1mm,                    
]

Adversarial examples generated by different attack techniques generally maintain high syntactic and semantic quality for both compressed and uncompressed models. Statistical analysis shows no significant difference in example quality between the two model types. Consequently, any performance degradation observed on these examples can be attributed to the vulnerability of the compressed and uncompressed models, enabling a fair comparison of their robustness under adversarial attacks.

\end{tcolorbox}
\label{RQ3Result}
\end{center}

\subsection{Answering RQ2: Comparing Adversarial Robustness of Compressed and Uncompressed Models}

\begin{figure}[htbp]
  \centering
  \subfigure[Clone Detection]{\label{CDPR}\includegraphics[width=7.5cm, height=6.25cm]{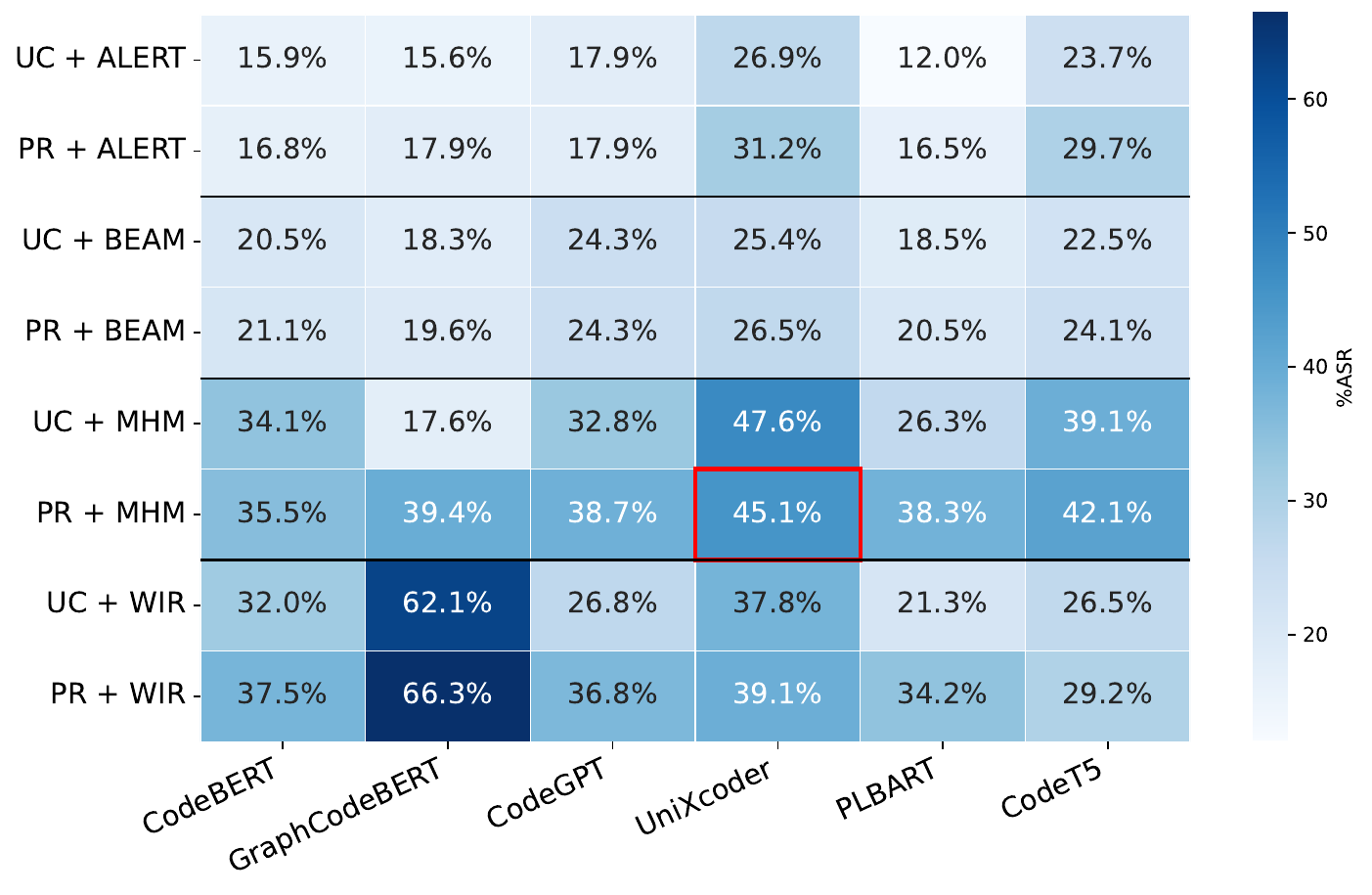}}
  \subfigure[Vulnerability Detection]{\label{VDPR}\includegraphics[width=7.5cm, height=6.25cm]{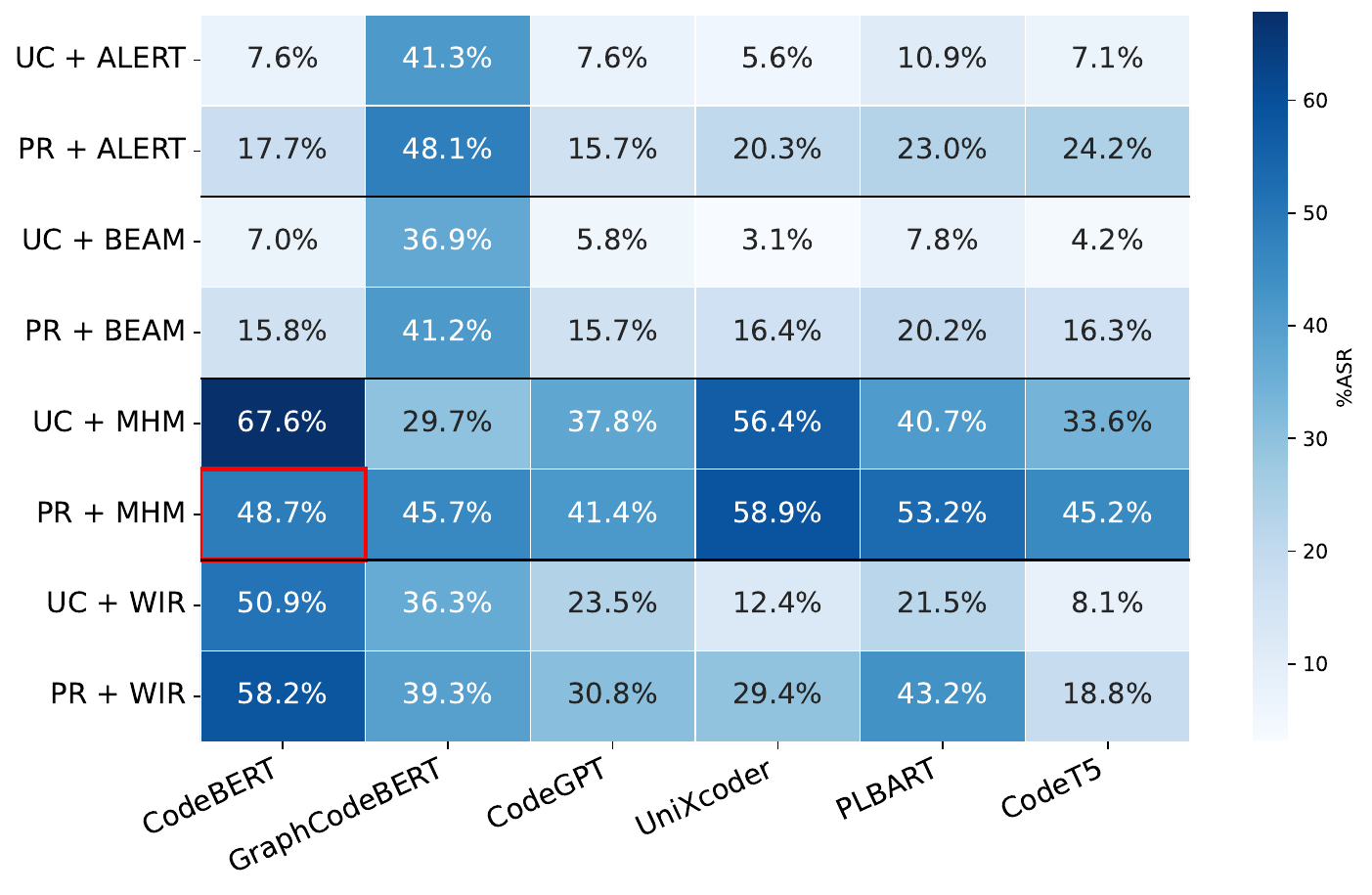}}
  \caption{Heatmap of \%ASR values from adversarial attacks on uncompressed (UC) and pruned (PR) models, illustrating the impact of pruning on model robustness across different code language models and attack techniques. Higher \%ASR indicates lower robustness.}
  \label{HeatmapAsrPR}
\end{figure}

Having identified the most effective compressed models based on their accuracy retention and rigorously examined the quality of the generated adversarial examples, we now investigate how well these models withstand adversarial attacks compared to their uncompressed counterparts. Figures \ref{HeatmapAsrPR}, \ref{HeatmapAsrQT}, \ref{HeatmapAsrKD}, and \ref{HeatmapAsrCS} illustrate how different compression strategies affect the performance of compressed models compared to their uncompressed counterparts under adversarial attacks. As shown in these figures, a general trend emerges across all tasks and models: compressed models exhibit lower robustness under adversarial attacks, as evidenced by higher \%ASR values than their uncompressed counterparts. This degradation is evident across all three compression strategies: pruning, quantization, and knowledge distillation, with only exceptions highlighted by the \textit{red} rectangle. Notably, across all attack techniques, no compressed models demonstrate superior robustness to their uncompressed versions across all tasks. Furthermore, models obtained via knowledge distillation show a pronounced decline in robustness, potentially due to the loss of fine-grained decision boundaries during the distillation process.

To further analyze these aggregate observations, we present a detailed examination of robustness outcomes across individual models, tasks, and compression techniques. We first examine the results for program understanding tasks, followed by the program generation task. This granular analysis reveals nuanced patterns and identifies areas where compression-induced vulnerabilities are most prominent.

\subsection*{Robustness Analysis of Uncompressed and Pruned Models for Program Understanding Tasks}
The impact of pruning on adversarial robustness is evaluated across six code language models for clone detection (CD) and vulnerability detection (VD) tasks. Figure~\ref{CDPR} and Figure~\ref{VDPR} display the corresponding \%ASR values, where higher \%ASR denotes lower robustness.

\textbf{CodeBERT:} In clone detection tasks, pruning consistently reduces robustness across all evaluated attack types except for two exceptions. For instance, under the ALERT attack, the \%ASR increases from 15.9\% (UCA) to 16.8\% (PRA), and under WIR-Random, it rises from 32.0\% to 37.5\% (Figure~\ref{CDPR}). A comparable pattern emerges in vulnerability detection (Figure~\ref{VDPR}), where pruning results in substantial increases in \%ASR, especially under more aggressive attacks such as WIR-Random. \textbf{GraphCodeBERT:} For clone detection, pruning results in substantial increases in \%ASR across most attack scenarios. For example, under MHM, \%ASR rises from 17.6\% to 39.4\% (Figure~\ref{CDPR}). In the context of vulnerability detection, pruning markedly amplifies vulnerability, with \%ASR increasing across all attacks, especially under ALERT and BEAM (Figure~\ref{VDPR}).

\textbf{CodeGPT: }In clone detection tasks, pruning results in substantial increases in \%ASR across most attack types. For example, under MHM, \%ASR rises from 32.8\% to 38.7\% (Figure~\ref{CDPR}). In vulnerability detection, pruning markedly amplifies vulnerability, with \%ASR increasing across all attacks, especially under ALERT and WIR-Random (Figure~\ref{VDPR}). These findings indicate that decoder-based architectures are more sensitive to pruning in adversarial settings. \textbf{UniXcoder:} It exhibits moderate reductions in robustness due to pruning. Although the increases in \%ASR are less pronounced than those observed for CodeGPT, consistent degradation occurs across both tasks. These findings suggest that, despite the advantages of hybrid representations, pruning continues to adversely affect UniXcoder's robustness against adversarial attacks.

\textbf{PLBART:} It exhibits significant robustness degradation under pruning, particularly in the context of vulnerability detection. As shown in Figures \ref{CDPR} and \ref{VDPR}, PLBART exhibits substantial, but task-dependent, robustness degradation. For clone detection, ALERT increases \%ASR from 12.0\% to 16.5\%, while WIR-Random increases \%ASR from 21.3\% to 34.2\%. In vulnerability detection, pruning results in substantial robustness degradation of compressed models compared to their uncompressed counterparts, such as an increase from 10.9\% to 23.0\% for ALERT and from 21.5\% to 43.2\% for WIR-Random attacks. These results indicate that encoder-decoder architectures are susceptible to pruning, especially in tasks that demand a deeper level of semantic understanding. \textbf{CodeT5:} It demonstrates consistent increases in \%ASR across both tasks when subjected to pruning. Although the extent of performance degradation varies by attack, pruning typically reduces robustness, especially under more aggressive adversarial strategies. These findings suggest that models with task-specific pretraining objectives remain vulnerable to pruning-induced biases.

\textbf{Overall Observations:} Across all six models and four attack techniques, pruning typically results in higher attack success rates, indicating reduced robustness. This effect is moderate in vulnerability detection and less significant in clone detection. Notably, no pruned model consistently surpasses its uncompressed counterpart across tasks and attacks, with only a few marginal cases demonstrating comparable performance. Overall, these findings suggest that pruning introduces systematic adversarial vulnerability across encoder-only, decoder-only, and encoder-decoder architectures, although the extent of degradation varies by model family and task.

\subsection*{Robustness Analysis of Uncompressed and Quantized Models for Program Understanding Tasks}
\begin{figure}[htbp]
  \centering
  \subfigure[Clone Detection]{\label{CDQT}\includegraphics[width=7.5cm, height=6.25cm]{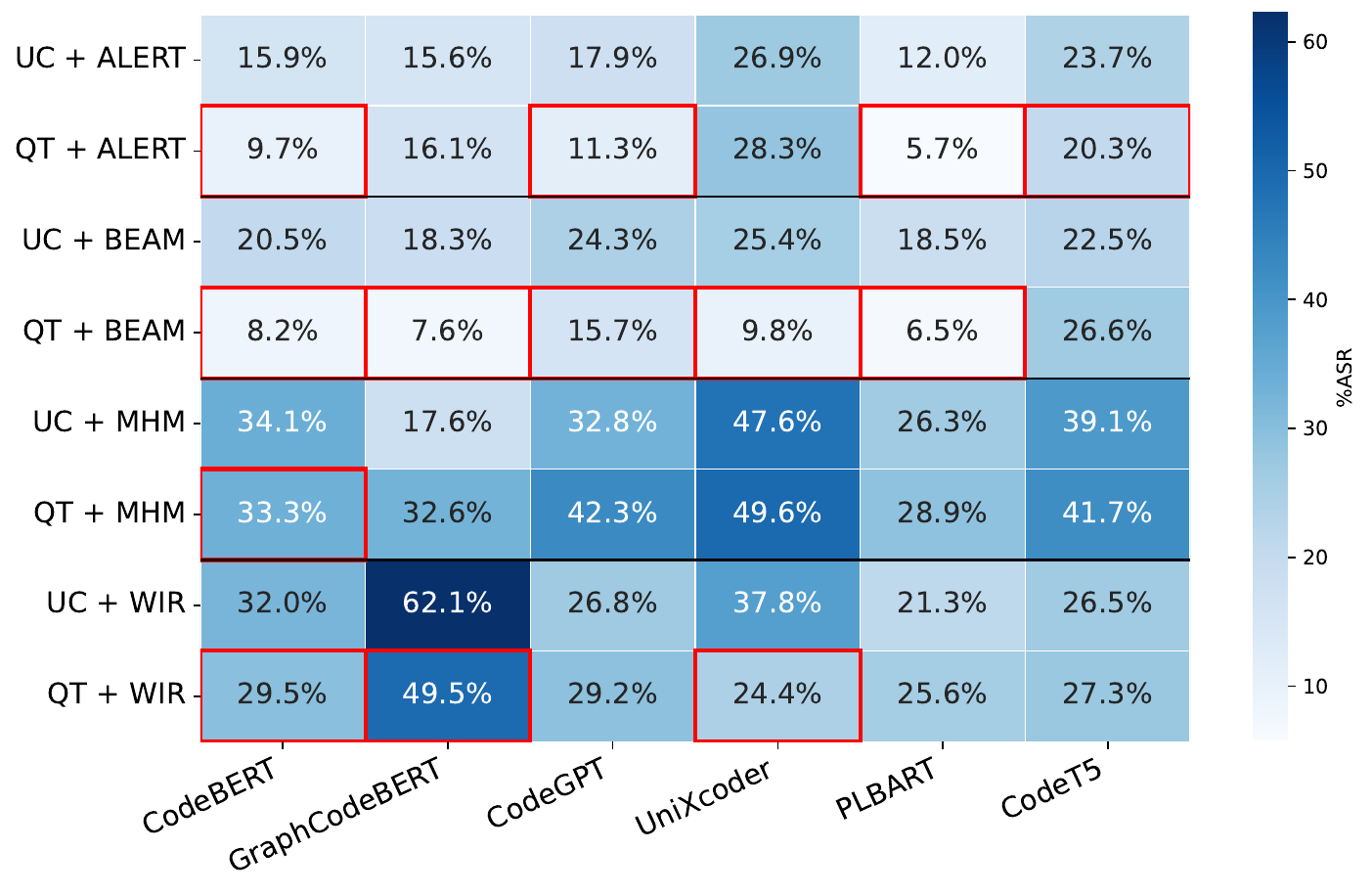}}
  \subfigure[Vulnerability Detection]{\label{VDQT}\includegraphics[width=7.5cm, height=6.25cm]{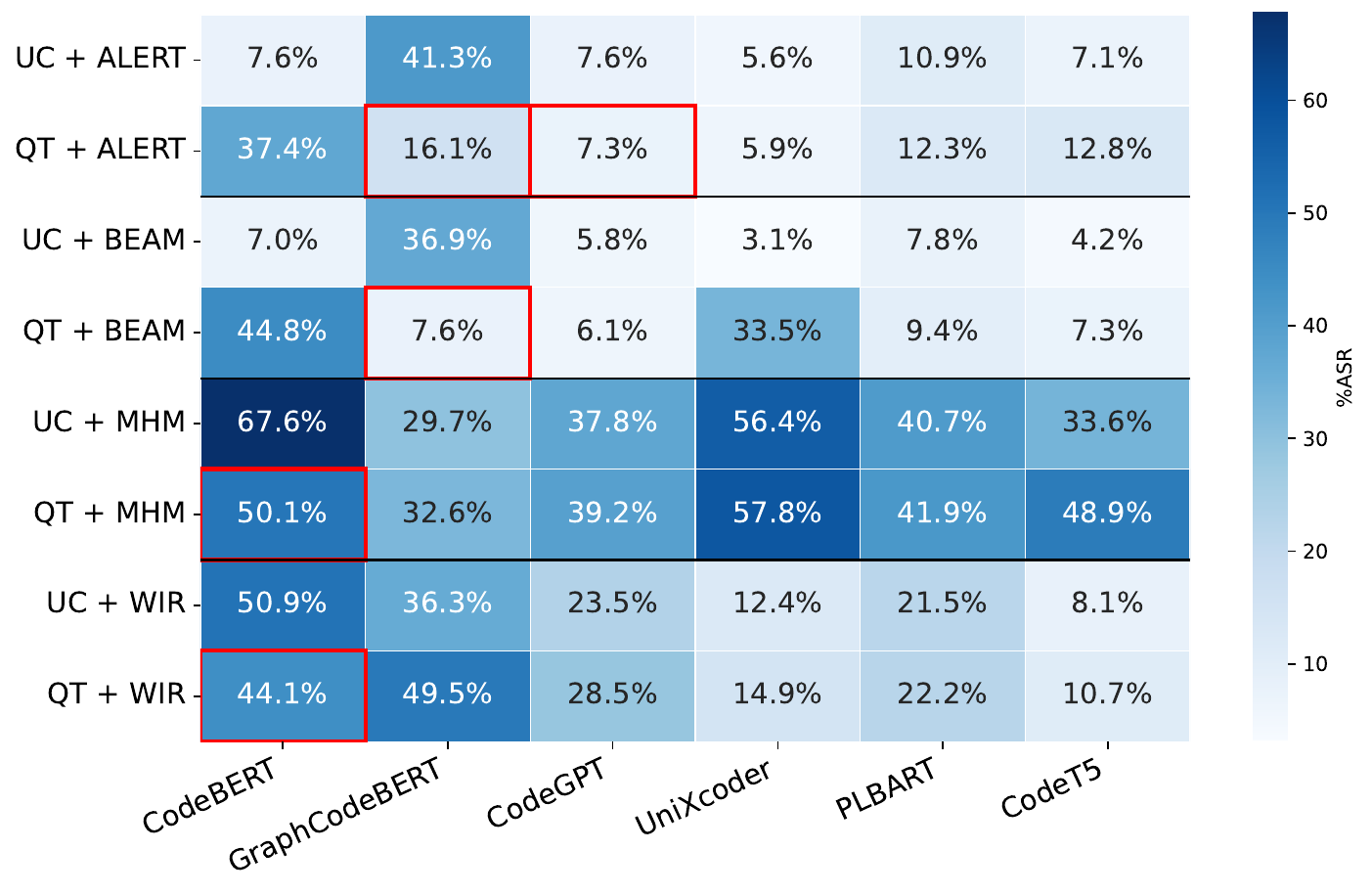}}
  \caption{Heatmap of \%ASR values from adversarial attacks on uncompressed (UC) and quantized (QT) models, illustrating the impact of quantization on model robustness across different code language models and attack techniques. Higher \%ASR indicates lower robustness.}
  \label{HeatmapAsrQT}
\end{figure}

The impact of quantization on adversarial robustness is analyzed across six code language models for clone detection (CD) and vulnerability detection (VD) tasks. Figure~\ref{CDQT} and Figure~\ref{VDQT} present the corresponding \%ASR values, where a higher \%ASR indicates lower robustness.

\textbf{CodeBERT:} Quantization exhibits a mixed but predominantly detrimental effect on CodeBERT. In clone detection, quantization enhances robustness to attacks; for instance, as shown in Figure \ref{CDQT}, ALERT decreases \%ASR from 15.9\% (UCA) to 9.7\% (QTA), and the BEAM attack drops from 20.5\% to 8.2\%. In vulnerability detection, as shown in Figure \ref{VDQT}, quantization significantly reduces robustness under ALERT, with rates increasing from 7.6\% to 37.4\%, while changes under other attacks remain comparatively moderate. Overall, CodeBERT demonstrates task and attack-dependent sensitivity to quantization. \textbf{GraphCodeBERT:} It exhibits variable performance when subjected to quantization. In the context of clone detection (Figure~\ref{CDQT}), robustness increases against certain attacks, such as BEAM, but declines against others, notably WIR-Random. For vulnerability detection (Figure~\ref{VDQT}), quantization markedly reduces robustness under MHM, with \%ASR rising from 29.7\% to 32.6\%. These findings indicate that, although GraphCodeBERT leverages structural representations, quantization impairs its robustness in a manner that depends on the attack type.

\textbf{CodeGPT:} It demonstrates relatively smaller, yet still significant, shifts in robustness. For clone detection, as shown in Figure \ref{CDQT}, quantization reduces \%ASR under ALERT from 17.9\% to 11.3\% and under BEAM from 24.3\% to 15.7\%, indicating enhanced robustness in these scenarios. In contrast, under MHM and WIR-Random, \%ASR increases from 32.8\% to 42.3\% and from 26.8\% to 29.2\%, respectively. In vulnerability detection, as shown in Figure \ref{VDQT}, the changes are modest; for example, ALERT decreases slightly from 7.6\% to 7.3\%, while WIR-Random increases from 23.5\% to 28.5\%. \textbf{UniXcoder:} It exhibits moderate variability when subjected to quantization. In the context of clone detection (Figure~\ref{CDQT}), quantization enhances robustness against specific attacks; however, performance deteriorates under attacks characterized by stronger perturbations. For vulnerability detection (Figure~\ref{VDQT}), quantization results in substantial increases in \%ASR under BEAM (e.g., from 3.1\% to 33.5\%), indicating reduced robustness. These findings suggest that hybrid architectures are still sensitive to reduced numerical precision in adversarial scenarios.


\textbf{PLBART:} It exhibits moderate performance degradation under quantization. For clone detection, as shown in Figure \ref{CDQT}, ALERT decreases \%ASR from 12.0\% to 5.7\%, while MHM and WIR-Random increase from 26.3\% to 28.9\% and from 21.3\% to 25.6\%, respectively. In vulnerability detection, as shown in Figure \ref{VDQT}, quantization results in modest increases, with ALERT rising from 10.9\% to 12.3\% and WIR-Random from 21.5\% to 22.2\%. \textbf{CodeT5:} It exhibits moderate yet consistent performance degradation when subjected to quantization. In the context of clone detection, quantization produces variable effects across different attack types. In contrast, for vulnerability detection (Figure~\ref{VDQT}), the attack success rate generally increases under stronger attacks, such as MHM (e.g., from 33.6\% to 48.9\%). These results suggest that although CodeT5 maintains relatively stable overall performance, quantization increases adversarial vulnerability.

\textbf{Observations across tasks:} Quantization demonstrates more variable behavior than pruning across different models and attack scenarios. Although some instances show enhanced robustness, the prevailing pattern, particularly in vulnerability detection tasks, is an increased attack success rate after quantization. Notably, no model consistently benefits from quantization across all evaluated tasks and attack types. Overall, these findings indicate that quantization yields robustness effects that depend on both the model architecture and the attack type. However, in most practical scenarios, quantization increases adversarial vulnerability.

\subsection*{Robustness Analysis of Uncompressed and Knowledge-distilled Models for Program Understanding Tasks}
\begin{figure}[htbp]
  \centering
  \subfigure[Clone Detection]{\label{CDKD}\includegraphics[width=4cm, height=6cm]{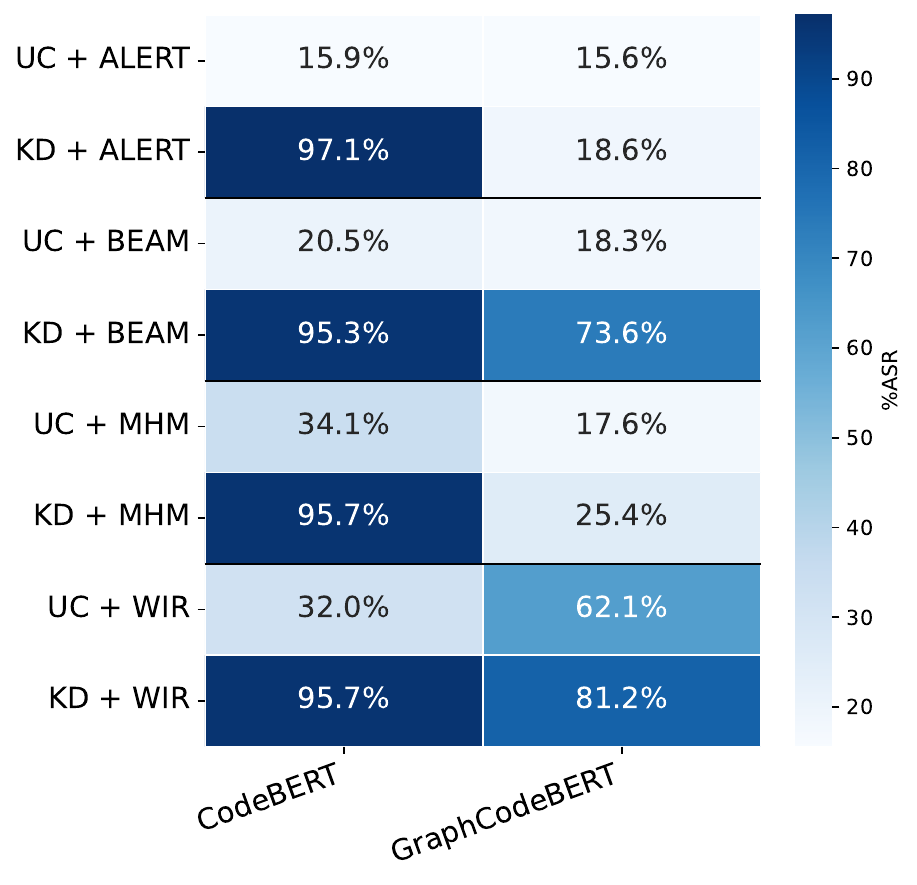}}
  \subfigure[Vulnerability Detection]{\label{VDKD}\includegraphics[width=4cm, height=6cm]{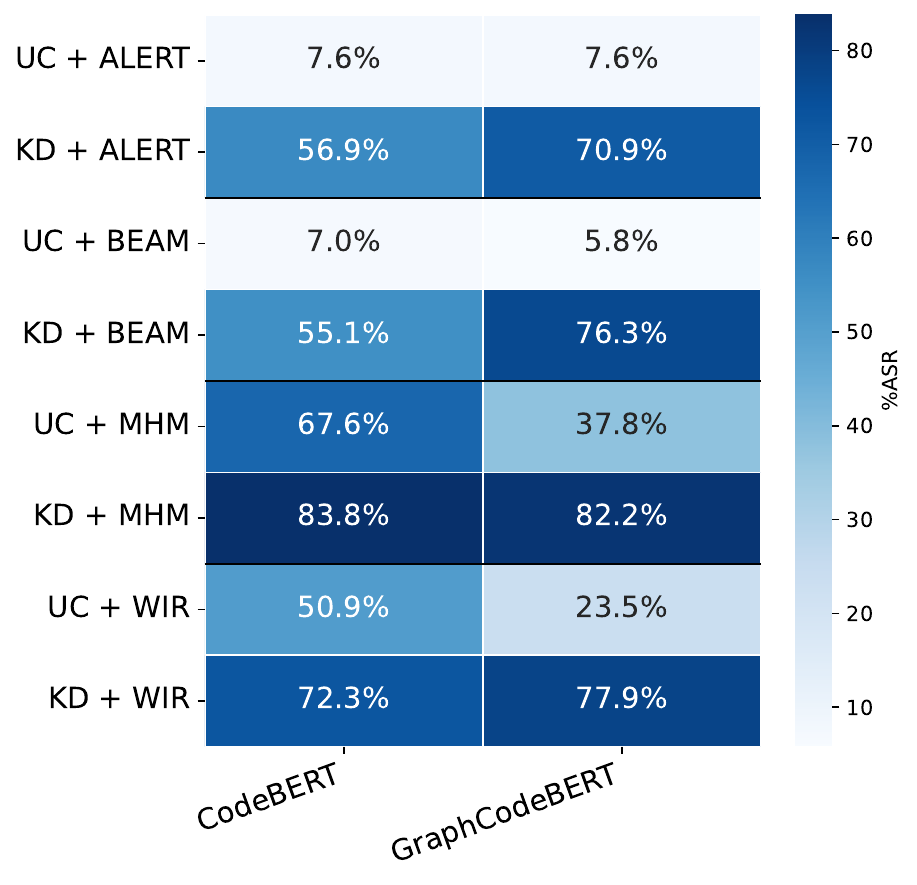}}
  \caption{Heatmap of \%ASR values from adversarial attacks on uncompressed (UC) and knowledge-distilled (KD) models, illustrating the impact of knowledge distillation on model robustness across different code language models and attack techniques. Higher \%ASR indicates lower robustness.}
  \label{HeatmapAsrKD}
\end{figure}

We next examine the impact of knowledge distillation on adversarial robustness of the CodeBERT and GraphCodeBERT models for the clone detection and vulnerability detection tasks. Figure~\ref{CDKD} and Figure~\ref{VDKD} present the corresponding \%ASR values, where a higher \%ASR indicates lower robustness.

\textbf{CodeBERT:} In clone detection tasks, knowledge distillation results in a substantial reduction in robustness across all evaluated attack techniques, as shown in Figure \ref{CDKD}. Under the ALERT attack, the \%ASR increases significantly from 15.9\% (UCA) to 97.1\% (KDA). A comparable trend is evident for the BEAM attack, with \%ASR rising from 20.5\% to 95.3\%. This decline in robustness is consistent for other attacks, as \%ASR for MHM increases from 34.1\% to 95.7\% and for WIR-Random from 32.0\% to 95.7\%. A consistent, though less pronounced, trend is evident in vulnerability detection, as shown in Figure \ref{VDKD}. Under ALERT, \%ASR increases from 7.6\% to 56.9\%, and under BEAM, it rises from 7.0\% to 55.1\%. The degradation is moderate under MHM and WIR-Random, where \%ASR increases from 67.6\% to 83.8\% and from 50.9\% to 72.3\%, respectively. Although the absolute values differ across attacks, the distilled model consistently demonstrates substantially greater vulnerability than the uncompressed baseline.

\textbf{GraphCodeBERT:} Similar to CodeBERT, GraphCodeBERT demonstrates significant degradation in adversarial robustness following knowledge distillation in both clone detection and vulnerability detection tasks. In clone detection, as illustrated in Figure~\ref{CDKD}, the degradation is particularly severe for BEAM attacks, with \%ASR increasing sharply from 18.3\% to 73.6\%. A similar pattern emerges in vulnerability detection, as shown in Figure~\ref{VDKD}, where knowledge distillation consistently increases \%ASR across all attack types. For instance, under the ALERT attack, \%ASR rises from 7.6\% to 70.9\%, and under BEAM, from 5.8\% to 76.3\%. For MHM, the increase is moderate, from 37.8\% to 82.2\%, while WIR-Random exhibits a substantial rise from 23.5\% to 77.9\%. These findings indicate that robustness degradation is more pronounced in vulnerability detection than in clone detection.

\textbf{Overall observation:} Across both tasks and all attack techniques, the knowledge-distilled CodeBERT and GraphCodeBERT models demonstrate the most significant robustness degradation among the evaluated compression strategies. In contrast to pruning and quantization, which show mixed or moderate effects, knowledge distillation consistently yields substantially higher attack success rates. These findings indicate that the distillation process likely smooths or compresses the student model's decision boundary, thereby diminishing its capacity to withstand small, adversarially crafted perturbations.

\subsection*{Robustness Analysis of Uncompressed, Pruned, and Quantized Models for Code Summarization}
\begin{figure}[htbp]
  \centering
  \subfigure[Uncompressed Vs. Pruning]{\label{CSPR}\includegraphics[width=6cm, height=6.25cm]{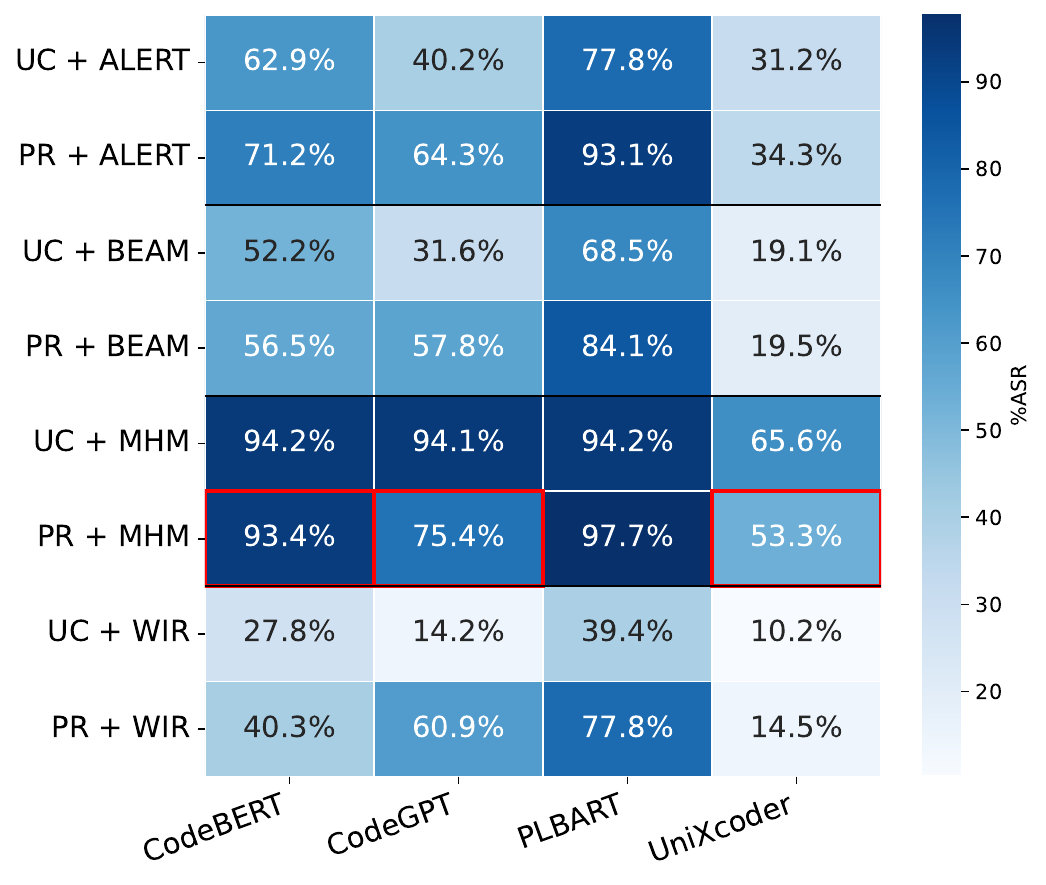}}
  \subfigure[Uncompressed Vs. Quantization]{\label{CSQT}\includegraphics[width=6cm, height=6.25cm]{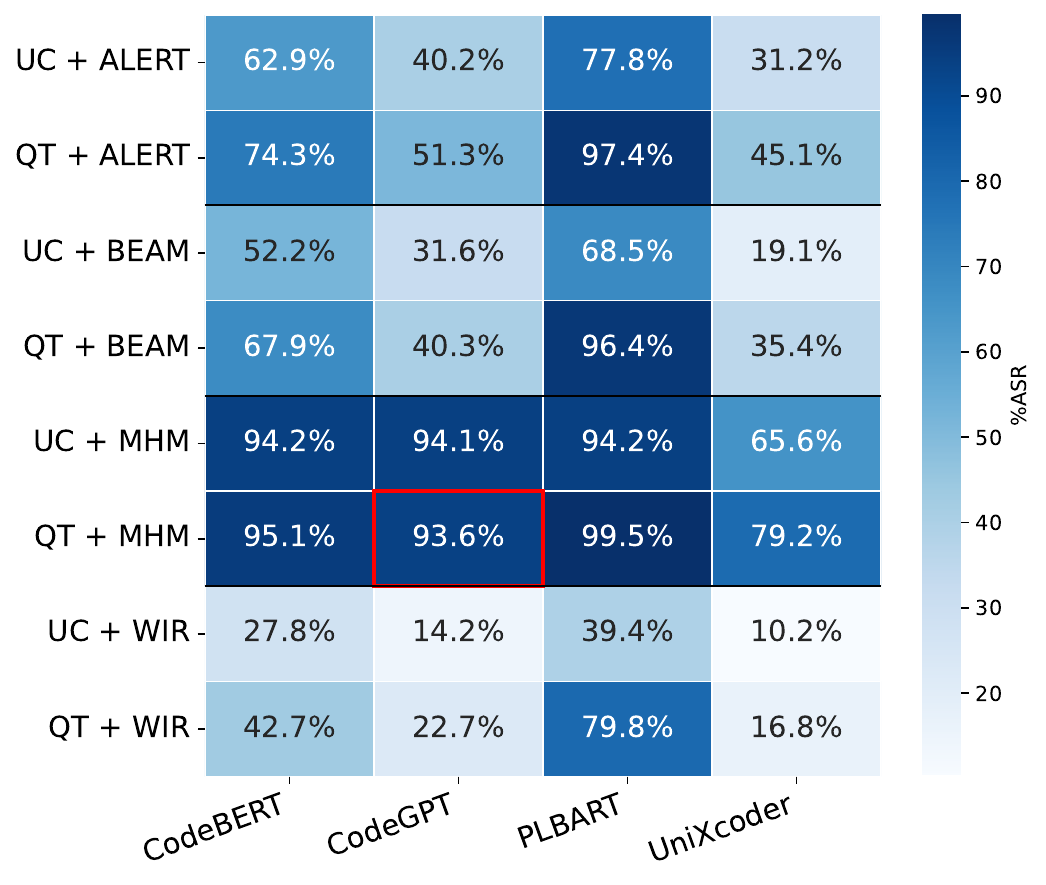}}
  \caption{Heatmap of \%ASR values from adversarial attacks on uncompressed (UC), pruned (PR), and quantized (QT) models for the code summarization task, illustrating the impact of compression on model robustness across different code language models and attack techniques. Higher \%ASR indicates lower robustness.}
  \label{HeatmapAsrCS}
\end{figure}

We analyze the impact of pruning and quantization on adversarial robustness in the code summarization (CS) task using four representative models: CodeBERT, CodeGPT, PLBART, and UniXcoder. The knowledge distillation approach, \textit{Compressor}, is excluded from consideration because it is applicable only to encoder-only models and classification-based tasks. To ensure comprehensive analysis across different model architectures, at least one representative model from each category (encoder-only, decoder-only, and encoder–decoder) is selected. Figure~\ref{CSPR} and Figure~\ref{CSQT} display the corresponding \%ASR values, where a higher \%ASR indicates lower robustness.

\textbf{CodeBERT:} In the case of CodeBERT, both pruning and quantization consistently diminish robustness across all evaluated attack types except the MHM attack. Under pruning, as shown in Figure~\ref{CSPR}, the \%ASR increases from 62.9\% (UC) to 71.2\% (PR) under ALERT and from 27.8\% to 40.3\% under WIR-Random. Similarly, quantization, as shown in Figure~\ref{CSQT}, further increases vulnerability, with \%ASR rising to 74.3\% under ALERT and 42.7\% under WIR-Random. Although the changes observed under MHM are relatively minor, the overall trend demonstrates that reducing model capacity or precision consistently undermines robustness in generative tasks.

\textbf{CodeGPT:} It demonstrates substantial performance degradation under both compression strategies except the MHM attack, with pruning exerting a particularly pronounced effect. Under pruning, as shown in Figure~\ref{CSPR}, \%ASR increases markedly from 40.2\% to 64.3\% under ALERT and from 14.2\% to 60.9\% under WIR-Random. Quantization, as shown in Figure~\ref{CSQT}, also results in consistent increases in \%ASR, although the effect is less severe than that of pruning (e.g., ALERT: 40.2\% to 51.3\%, WIR-Random: 14.2\% to 22.7\%). These results suggest that decoder-only architectures are highly susceptible to compression, especially under strong adversarial perturbations.

\textbf{PLBART:} It exhibits the most significant reduction in robustness among all evaluated models, particularly under quantization. During pruning, as shown in Figure~\ref{CSPR}, the \%ASR increases from 77.8\% to 93.1\% under ALERT and from 39.4\% to 77.8\% under WIR-Random. The effect of quantization is even more substantial, as shown in Figure~\ref{CSQT}, with \%ASR reaching very high levels across all attacks (e.g., ALERT: 77.8\% to 97.4\%, BEAM: 68.5\% to 96.4\%, MHM: 94.2\% to 99.5\%, WIR-Random: 39.4\% to 79.8\%). These findings indicate that encoder-decoder architectures are especially susceptible to reduced numerical precision and parameter removal in generative contexts.

\textbf{UniXcoder:} It exhibits moderate yet consistent performance degradation when subjected to both pruning and quantization. During pruning, as shown in Figure~\ref{CSPR}, the \%ASR increases from 31.2\% to 34.3\% under ALERT and from 10.2\% to 14.5\% under WIR-Random. In contrast, quantization, as shown in Figure~\ref{CSQT}, results in a more substantial degradation, with \%ASR rising to 45.1\% under ALERT and 35.4\% under BEAM. While UniXcoder demonstrates greater stability compared to other models, compression nonetheless introduces significant adversarial vulnerability.

\textbf{Overall Observations:} Across all four models, both pruning and quantization increase \%ASR, indicating reduced robustness in code summarization tasks. This degradation is consistently more severe than in classification tasks, which reflects the greater sensitivity of generative models to perturbations. Between the two compression strategies, quantization typically causes more pronounced degradation than pruning, especially for encoder-decoder models such as PLBART. Furthermore, stronger attacks such as WIR-Random consistently reveal larger robustness gaps. These results indicate that compression substantially undermines adversarial robustness in code summarization, with the extent of the degradation determined by both the model architecture and the compression technique.

\subsection*{Robustness Analysis for Program Understanding Tasks Under More Advanced Attack Technique: CODA}
After establishing that compressed models exhibit lower robustness than uncompressed counterparts under identifier renaming attacks, this study further examines their robustness against a more advanced adversarial attack technique, CODA \cite{tian2023code}, which performs both identifier renaming and semantic-preserving structural transformations, such as converting while loops to equivalent for loops. We focus on clone and vulnerability detection tasks to ensure a fair comparison, given that the knowledge distillation approach (\textit{Compressor}) is specifically designed for encoder-only models: CodeBERT and GraphCodeBERT, and classification-based program understanding tasks. Additionally, as the original implementation of CODA currently supports classification tasks, we apply it to clone detection and vulnerability detection tasks in this study.

\begin{figure}[htbp]
  \centering
  \subfigure[Clone Detection]{\label{CODACD}\includegraphics[width=6cm, height=5cm]{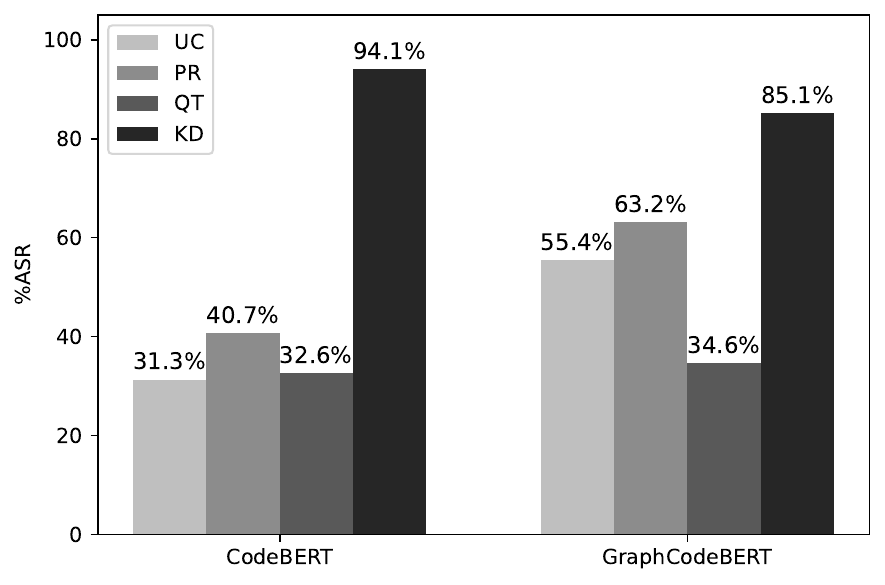}}
  \subfigure[Vulnerability Detection]{\label{CODAVD}\includegraphics[width=6cm, height=5cm]{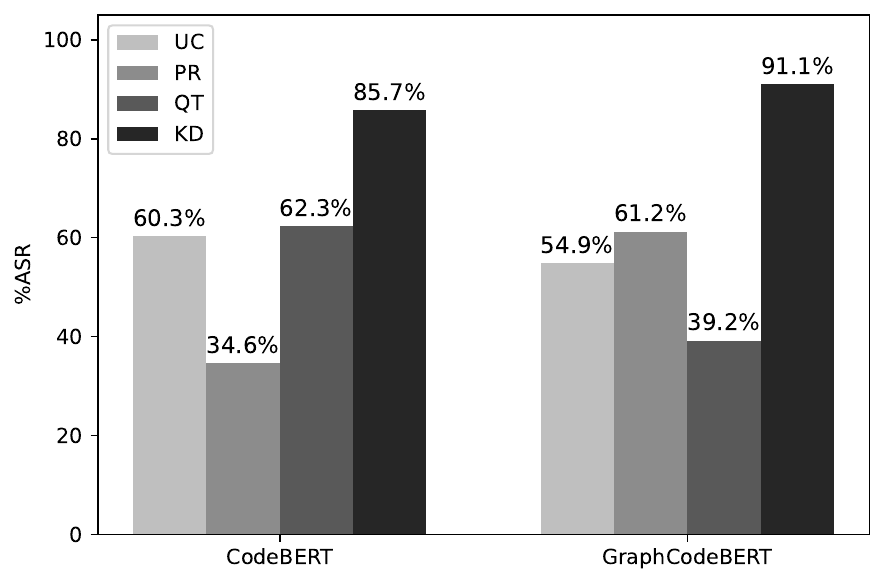}}
  \caption{Adversarial robustness comparison of uncompressed and compressed CodeBERT and GraphCodeBERT models under the advanced CODA attack for clone detection and vulnerability detection tasks. UC: Uncompressed, PR: Pruning, QT: Quantization, KD: Knowledge Distillation.}
  \label{CODAAttack}
\end{figure}

Figure \ref{CODAAttack} presents the \%ASR of the advanced CODA attack on both uncompressed and compressed CodeBERT and GraphCodeBERT models. These results reinforce previous findings from identifier renaming-based attacks: compressed models are generally more vulnerable to adversarial attacks than their uncompressed counterparts. In the clone detection task, the uncompressed CodeBERT model achieves a \%ASR of 31.3\%, while the pruned, quantized, and knowledge distilled variants reach 40.7\%, 32.6\%, and 94.1\%, respectively. GraphCodeBERT exhibits a similar yet more nuanced pattern. In the clone detection task, the uncompressed model achieves a \%ASR of 55.4\%. Pruning increases this value to 63.2\%, indicating reduced robustness. In contrast, quantization substantially improves robustness, reducing \%ASR to 34.6\%. Knowledge distillation, however, significantly degrades robustness, increasing \%ASR to 85.1\%.

A comparable pattern emerges in the vulnerability detection task, with a few exceptions. The uncompressed CodeBERT model records a \%ASR of 60.3\%, while the quantized and knowledge-distilled models exhibit higher \%ASR values of 62.3\% and 87.5\%, respectively, indicating reduced robustness following compression. In contrast, the pruned model achieves a lower \%ASR of 34.6\%, suggesting greater robustness than the uncompressed model in this context. A similar trend is evident for the GraphCodeBERT model. The uncompressed model records a \%ASR of 54.9\%. Pruning increases this value to 61.2\%, and knowledge distillation further elevates it to 91.1\%, indicating substantial robustness degradation. Quantization, however, again improves robustness, reducing \%ASR to 39.2\%.  Collectively, these findings demonstrate that the robustness degradation observed in compressed models extends beyond identifier-renaming attacks to include the more advanced CODA attack, which combines identifier renaming with semantic-preserving transformations. This result reinforces the conclusion that model compression, particularly knowledge distillation, can substantially increase the vulnerability of language models for code in more challenging adversarial scenarios.

To further quantify the above observations, we perform a win-rate analysis by comparing each compressed model with its uncompressed counterpart across all experimental combinations, including model variants, tasks, and attack types. For each comparison pair, a \textbf{win} is recorded when the compressed model exhibits a lower attack success rate (lower \%ASR) (meaning higher robustness) than the uncompressed model. Table \ref{WinRateTable} summarizes the overall win rates for compressed models obtained through pruning, quantization, and knowledge distillation. Overall, compressed models rarely outperform their uncompressed counterparts, achieving wins in no more than approximately 31\% of the evaluated settings. Among the three strategies, quantized models achieve the highest overall win rate, suggesting that quantization better preserves robustness compared to pruning and knowledge distillation. In particular, the knowledge‑distilled models fail entirely to demonstrate robustness against the uncompressed models across all the evaluated settings, as reflected in the win‑rate analysis. These findings highlight the increased vulnerability of compressed models compared to their uncompressed counterparts in the context of adversarial attacks.

\begin{table}[htbp]
\centering
\caption{Win rate analysis comparing the adversarial robustness of compressed models against their corresponding uncompressed counterparts across all evaluated models, tasks, and attack combinations. A win is counted when the compressed model achieves a lower \%ASR than the uncompressed model. Reported values are rounded to enhance readability.}
\label{WinRateTable}
\begin{tabular}{l|c|c}
\toprule
\textbf{Compression Strategy} & \textbf{Total Comparisons} & \textbf{Win Rate (\%)} \\
\midrule
Pruning                       & 68                         & 9                   \\
Quantization                  & 68                         & 31          \\
Knowledge Distillation        & 20                         & \textbf{0}                   \\
\bottomrule
\end{tabular}
\end{table}

\begin{center}
\begin{tcolorbox}[
    enhanced,
    attach boxed title to top left={yshift=-3mm,yshifttext=-1mm}, 
    colback=mycolor_box,                 
    colframe=black,                
    colbacktitle= mycolor_title,            
    coltitle=black,                
    title=Result RQ2,            
    fonttitle=\bfseries,           
    boxed title style={size=small},
    width=0.95\textwidth,            
    boxsep=1mm,                    
]

Compression techniques generally reduce the robustness of language models for code under adversarial attacks, making them more vulnerable compared to their uncompressed counterparts for deployment in software analytics tasks. Additionally, the \textbf{win-rate} analysis shows that compressed models outperform their uncompressed counterparts in at most \textbf{31\%} of the experimental combinations, further underscoring their limited robustness advantage.

\end{tcolorbox}
\label{RQ1Result}
\end{center}

\subsection{Answering RQ3: Identifying Robust Compression Strategies} 
While compression strategies typically reduce model robustness to adversarial attacks (as discussed in RQ2), it is essential to identify which strategy yields the most robust models for practical deployment. To answer this question, we compare the robustness of models under adversarial attacks across three compression techniques: pruning, quantization, and knowledge distillation. Since the knowledge distillation technique (\textit{Compressor}) is applied only to the encoder-only architecture (e.g., CodeBERT and GraphCodeBERT models) and the program understanding tasks (e.g., clone detection and vulnerability detection), we focus on the pruned and quantized versions of CodeBERT and GraphCodeBERT models for these tasks to ensure consistency and enable a fair comparison in addressing RQ3. Table \ref{ASRAllComp} presents the results of adversarial robustness comparisons across the three compression techniques using the \%ASR metric.

\begin{table}[htbp]
\centering
\caption{Adversarial robustness (\%ASR) of compressed models under different attack techniques for clone detection (CD) and vulnerability detection (VD). Lower \%ASR indicates higher robustness. PR: Pruning, QT: Quantization, KD: Knowledge Distillation.}
\begin{tabular}{l|l|l|cccc}
\toprule
\textbf{Task} & \textbf{Model} & \textbf{Compression} 
& \textbf{ALERT} & \textbf{Beam} & \textbf{MHM} & \textbf{WIR-Random} \\
\midrule

\multirow{6}{*}{CD}
& \multirow{3}{*}{CodeBERT}
& PR & 16.8 & 21.1 & 35.5 & 37.5 \\
& & QT & 9.7  & 8.2  & 33.3 & 29.5 \\
& & KD & 97.1 & 95.3 & 95.7 & 95.7 \\
\cmidrule(lr){2-7}

& \multirow{3}{*}{GraphCodeBERT}
& PR & 17.9 & 19.6 & 39.4 & 66.3 \\
& & QT & 16.1 & 7.6  & 32.6 & 49.5 \\
& & KD & 18.6 & 73.6 & \textbf{25.4} & 81.2 \\

\midrule

\multirow{6}{*}{VD}
& \multirow{3}{*}{CodeBERT}
& PR & 17.7 & 15.8 & 48.7 & 58.2 \\
& & QT & 37.4 & 44.8 & 50.1 & 44.1 \\
& & KD & 56.9 & 55.1 & 83.8 & 72.3 \\
\cmidrule(lr){2-7}

& \multirow{3}{*}{GraphCodeBERT}
& PR & 48.1 & 41.2 & 45.7 & 39.3 \\
& & QT & 72.8 & 44.8 & 31.6 & 42.3 \\
& & KD & 70.9 & 76.3 & 82.2 & 77.9 \\

\bottomrule
\end{tabular}
\label{ASRAllComp}
\end{table}

From Table \ref{ASRAllComp}, we observe that for the clone detection and vulnerability detection tasks, the knowledge-distilled model consistently exhibits higher \%ASR values across all attacks, indicating lower robustness than the pruned and quantized variants. For instance, in clone detection and under the ALERT attack, the \%ASR for the distilled model reaches 97.1\%, while the pruned and quantized models show significantly lower \%ASR values of 16.8\% and 9.7\%, respectively. This trend holds across all attack types, suggesting that the distilled model is more susceptible to adversarial attacks. Between the quantized and pruned models, we observe that the quantized model is more robust in the clone detection task. However, in the vulnerability detection task, the pruned model is more robust than the quantized model. We also observe a very similar pattern with the GraphCodeBERT model. This pattern also holds for the more advanced attack technique CODA, as shown in Figure \ref{CODAAttack}. 

In terms of AMQ, as shown in Table \ref{AMQAllComp}, the knowledge-distilled model requires fewer queries to succeed in the clone detection task, indicating lower robustness. In contrast, although the quantized model shows higher AMQ values for the clone detection task, it requires fewer queries to succeed in the vulnerability detection task. Moreover, the pruned model consistently requires higher AMQ values to successfully attack in the vulnerability detection task, suggesting greater robustness. We also observe a similar pattern for the GraphCodeBERT model on the clone detection and vulnerability tasks. These observations suggest that, based on the AMQ metric, the impact of compression techniques on model robustness varies across tasks, with no single technique consistently outperforming the others.

\begin{table}[htbp]
\centering
\caption{Adversarial robustness in terms of Average Model Query (AMQ) for compressed models under different attack techniques on clone detection (CD) and vulnerability detection (VD). Higher AMQ indicates that more queries are required to generate successful adversarial examples, suggesting greater robustness. PR: Pruning, QT: Quantization, KD: Knowledge Distillation.}
\begin{tabular}{l|l|l|cccc}
\toprule
\textbf{Task} & \textbf{Model} & \textbf{Compression} 
& \textbf{ALERT} & \textbf{Beam} & \textbf{MHM} & \textbf{WIR-Random} \\
\midrule

\multirow{6}{*}{CD}
& \multirow{3}{*}{CodeBERT}
& PR & 519 & 867 & 534 & 174 \\
& & QT & 845 & 200 & 529 & 170 \\
& & KD & 118 & 181 & 87  & 90  \\
\cmidrule(lr){2-7}
& \multirow{3}{*}{GraphCodeBERT}
& PR & 486 & 832 & 571 & 188 \\
& & QT & 793 & 224 & 503 & 162 \\
& & KD & 131 & 205 & 94  & 101 \\
\midrule

\multirow{6}{*}{VD}
& \multirow{3}{*}{CodeBERT}
& PR & 413 & 799 & 670 & 171 \\
& & QT & 249 & 177 & 249 & 91  \\
& & KD & 275 & 531 & 282 & 119 \\
\cmidrule(lr){2-7}
& \multirow{3}{*}{GraphCodeBERT}
& PR & 438 & 756 & 641 & 184 \\
& & QT & 227 & 193 & 268 & 104 \\
& & KD & 294 & 497 & 301 & 128 \\
\bottomrule
\end{tabular}
\label{AMQAllComp}
\end{table}

To statistically validate whether the observed performance differences across the three compression techniques are significant, we apply the Friedman Test \cite{friedman1937use} to assess whether the differences among pruning, quantization, and knowledge distillation are statistically significant in terms of the \%ASR and AMQ values. A $p$-value less than 0.05 indicates that the differences in the performance of model compression techniques on the robustness of compressed models under adversarial attacks are statistically significant. For the CodeBERT model, the Friedman test yields $p$-values of 0.018 and 0.037 for the \%ASR metric, and 0.038 and 0.017 for the AMQ metric, in the clone detection and vulnerability detection tasks, respectively. For the GraphCodeBERT model, the Friedman test yields $p$-values of 0.011 and 0.043 for the \%ASR metric, and 0.041 and 0.025 for the AMQ metric, in the clone detection and vulnerability detection tasks, respectively. Since the $p$-values for both metrics and models are below the threshold (0.05), confirming a performance difference among the model compression strategies on the robustness of compressed models under adversarial attacks. These findings suggest that selecting compression strategies can significantly impact language models for practical deployment in software analytics tasks.

\begin{center}
\begin{tcolorbox}[
    enhanced,
    attach boxed title to top left={yshift=-3mm,yshifttext=-1mm}, 
    colback=mycolor_box,                 
    colframe=black,                
    colbacktitle= mycolor_title,            
    coltitle=black,                
    title=Result RQ3,            
    fonttitle=\bfseries,           
    boxed title style={size=small},
    width=0.95\textwidth,            
    boxsep=1mm,                    
]

No single compression strategy consistently produces the most robust models across different software analytics tasks. Knowledge distillation yields the least robust models, while quantization demonstrates higher robustness for clone detection, and pruning performs better for vulnerability detection. These trends hold across both \%ASR and AMQ metrics, highlighting task-dependent trade-offs in robustness.

\end{tcolorbox}
\label{RQ2Result}
\end{center}

\subsection{Answering RQ4: Comparing Compression Efficiency and Model Robustness}
To address the final research question (RQ4), we investigate whether reducing model size via compression techniques comes at the cost of adversarial robustness. Specifically, we analyze the relationship between model compactness (e.g., compression efficiency) and robustness under adversarial attacks to determine how robustness is influenced when models are compressed for efficient deployment. Since global unstructured pruning only sets weights to zero without altering the network structure, it does not reduce model size \cite{shi2022compressing}. Thus, we consider only quantization and the knowledge distillation technique \textit{Compressor} to address this research question. Furthermore, since \textit{Compressor} is applicable exclusively to encoder-only models and classification tasks, only the CodeBERT and GraphCodeBERT models are included in the analysis for RQ4.

\begin{figure}[htbp]
  \centering
  \subfigure[CodeBERT + Clone Detection]{\label{CBCDSize}\includegraphics[width=7.5cm, height=5.5cm]{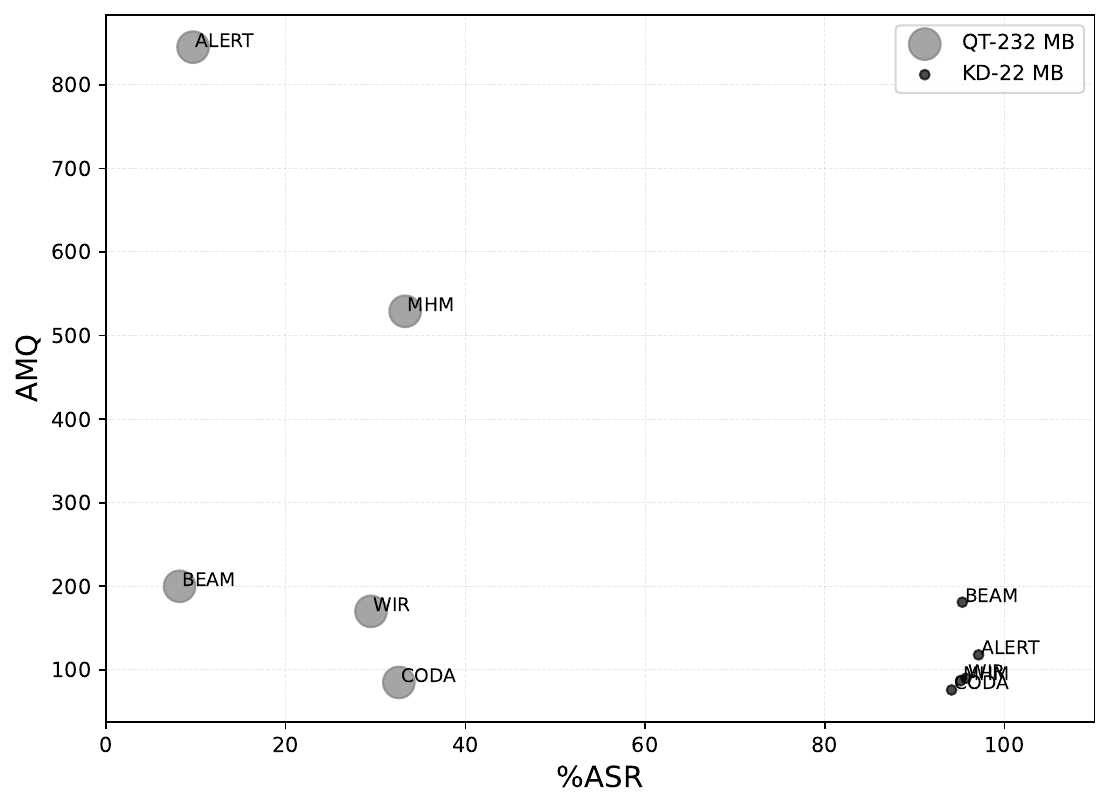}}
  \subfigure[CodeBERT + Vulnerability Detection]{\label{CBVDSize}\includegraphics[width=7.5cm, height=5.5cm]{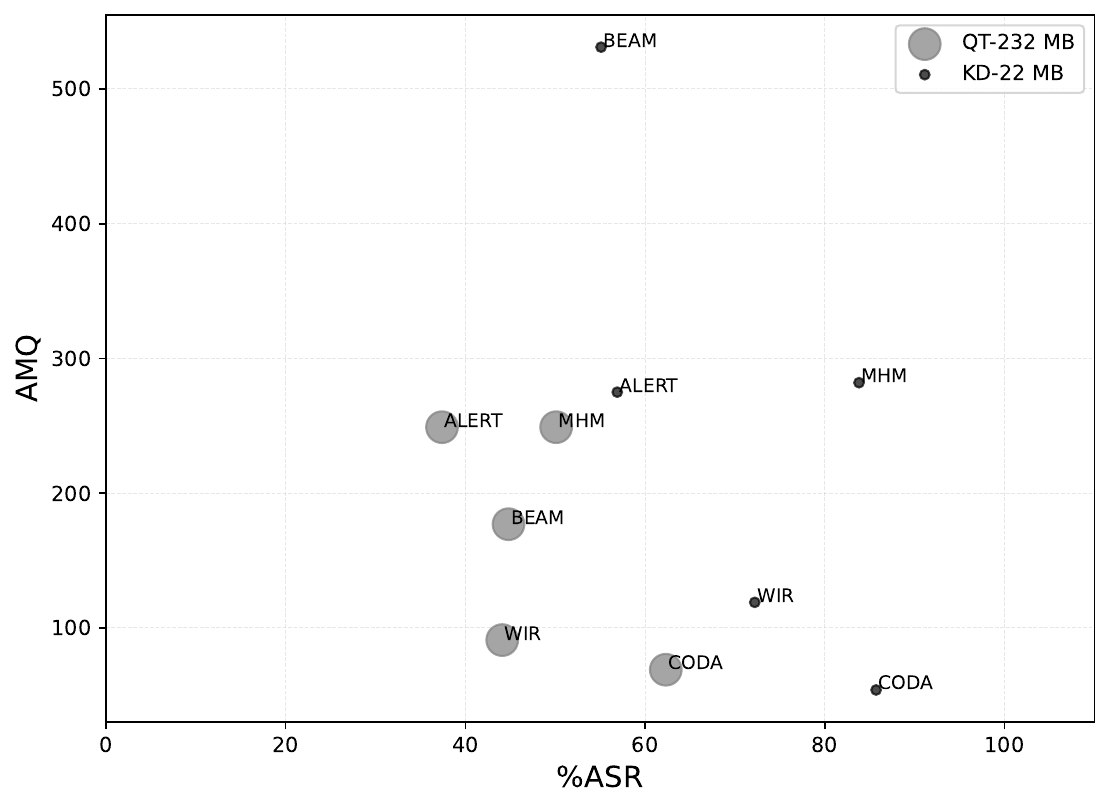}}
  \subfigure[GraphCodeBERT + Clone Detection]{\label{GCBCDSize}\includegraphics[width=7.5cm, height=5.5cm]{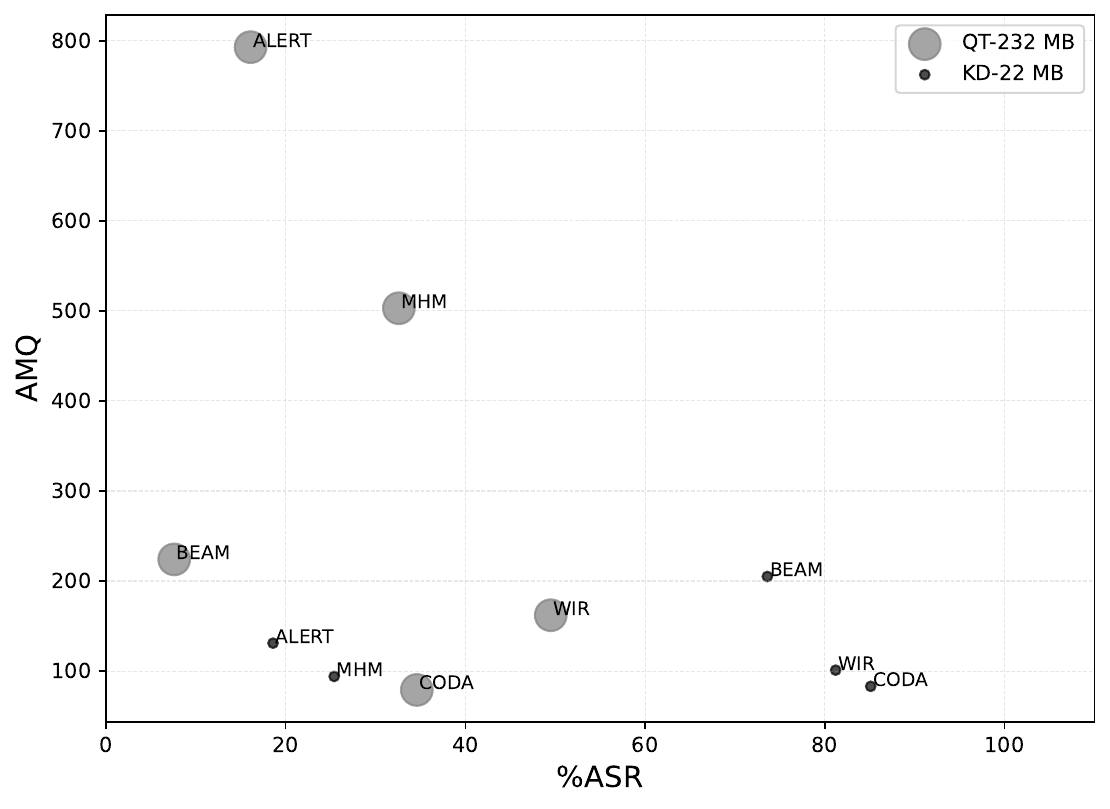}}
  \subfigure[GraphCodeBERT + Vulnerability Detection]{\label{GCBVDSize}\includegraphics[width=7.5cm, height=5.5cm]{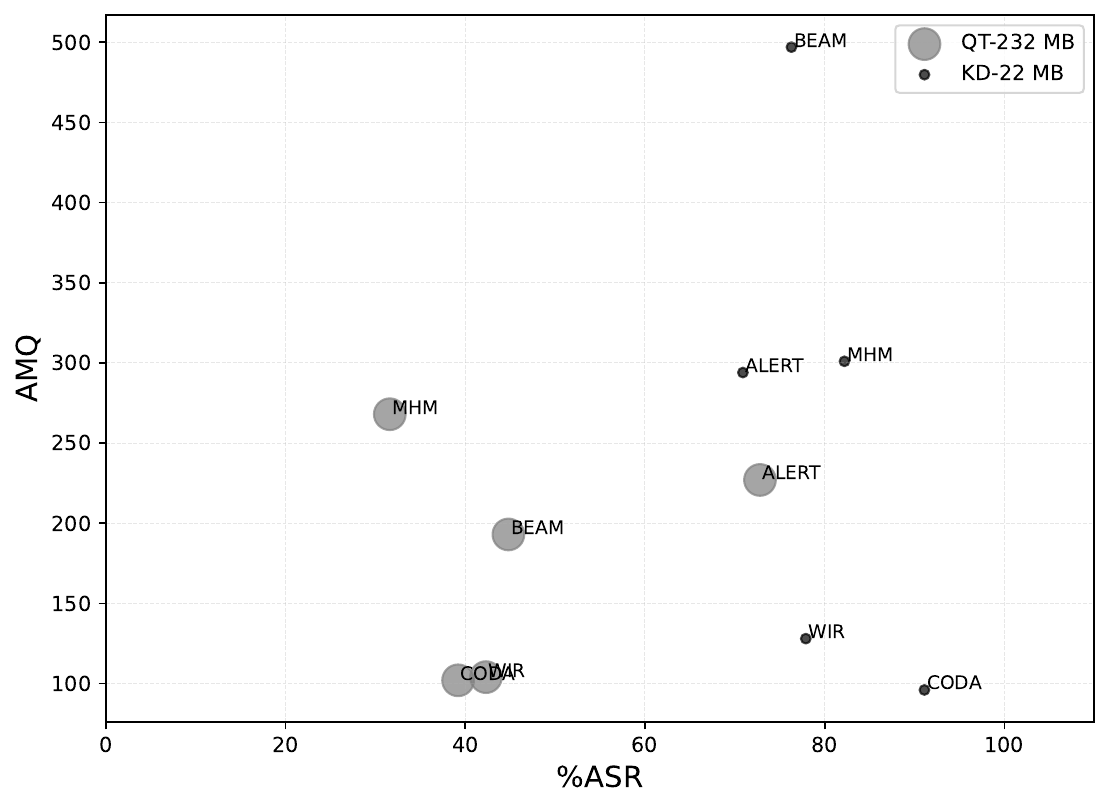}}
  \caption{Relationship between compressed model size and adversarial robustness for clone detection and vulnerability detection tasks. The horizontal axis denotes attack success rate (\%ASR), the vertical axis indicates the average number of model queries (AMQ), and bubble size represents the size of the compressed model.}
  \label{SizeRobust}
\end{figure}

Figure \ref{SizeRobust} shows the relationship between model size and adversarial robustness for compressed models across both clone detection and vulnerability detection tasks. Each point represents a compressed model produced by a given compression strategy and evaluated under various adversarial attacks. The horizontal axis indicates the attack success rate (\%ASR), and the vertical axis shows the average number of model queries (AMQ) necessary to generate successful adversarial examples. The size of each point reflects the size of the corresponding compressed model.

A consistent pattern is observed across both tasks: models with smaller sizes generally exhibit reduced robustness to adversarial attacks. Specifically, the knowledge-distilled model achieves the largest reduction in model size (22 MB), but also demonstrates substantially higher attack success rates across most attack techniques. In contrast, quantized models, which retain a larger size (232 MB), typically show lower attack success rates and require more queries to generate successful attacks. These findings suggest that models subjected to more aggressive compression are generally more vulnerable to adversarial attacks. The results also reveal differences in attack behavior among techniques. For example, WIR-Random and CODA frequently achieve high attack success rates with relatively few queries, indicating their efficiency as an attack strategy across both tasks. In contrast, ALERT and BEAM generally require more queries to achieve successful attacks in several settings.

Overall, these findings indicate that compression strategies can maintain model accuracy prior to adversarial attacks, as demonstrated in Tables \ref{TablePerformCDVD} and \ref{TablePerformCS}, provided that the compression level is carefully chosen to preserve original task performance. Nevertheless, even with accuracy-preserving configurations, the reduced size of compressed models substantially diminishes their robustness to adversarial attacks. Notably, models subjected to more aggressive compression exhibit higher attack success rates and require fewer queries for successful attacks. The results provide empirical evidence that improvements in compression efficiency do not necessarily preserve adversarial robustness, emphasizing the need for robustness-aware compression strategies when deploying compressed language models of code in real-world software engineering systems.

\begin{center}
\begin{tcolorbox}[
    enhanced,
    attach boxed title to top left={yshift=-3mm,yshifttext=-1mm}, 
    colback=mycolor_box,                 
    colframe=black,                
    colbacktitle= mycolor_title,            
    coltitle=black,                
    title=Result RQ4,            
    fonttitle=\bfseries,           
    boxed title style={size=small},
    width=0.95\textwidth,            
    boxsep=1mm,                    
]

There is a trade-off between model size reduction and robustness. As model size decreases due to compression, robustness against adversarial attacks typically declines, with smaller models exhibiting reduced robustness.

\end{tcolorbox}
\label{RQ4Result}
\end{center}

\section{Key Findings \& Guidelines}
\label{DSC}
This study investigates the impact of various compression strategies on the robustness of language models for code against adversarial attacks. Below, we present the key insights derived from our experimental results.

\subsection*{Implications for Researchers:}

\begin{itemize}

    \item textbf{Compressed models exhibit reduced adversarial robustness:} Our findings reveal that compressed models are less robust than their uncompressed counterparts in software analytics tasks under adversarial attacks. Across all compression strategies, pruning, quantization, and knowledge distillation, the \%ASR is higher for compressed models, indicating increased vulnerability. This suggests that while compression enhances efficiency and effectiveness \cite{shi2022compressing, shi2024greening, d2024compression}, it reduces the resilience of compressed models to adversarial attacks.

    \item \textbf{No compression technique is universally robust: } While pruning and quantization offer better robustness than knowledge distillation, models compressed using these techniques still exhibit vulnerabilities to certain adversarial attacks, indicating that no single compression method consistently outperforms the others across all models and software analytics tasks. To further investigate this behavior, we manually inspected 20 adversarial code examples from the vulnerability detection task. This analysis revealed heterogeneous behavior across compression techniques. In some cases, the pruned model preserved correct predictions under adversarial perturbations, while the quantized and knowledge-distilled models did not. In other cases, the quantized model maintained robustness, whereas the pruned and knowledge-distilled models misclassified the same modified code. These observations reinforce our central claim that robustness under compression is both task-dependent and instance-dependent.

    \item \textbf{Task-specific robustness differences are significant:} The effectiveness of compression techniques varies across tasks; for example, quantized models are more robust in clone detection, while pruned models perform better in vulnerability detection. Friedman Test results also confirm that the differences in model robustness among compression techniques are statistically significant, reinforcing the impact of compression choice on model behavior.

    \item \textbf{Clear trade off between compactness and robustness:} Our results confirm a trade-off between model compactness and adversarial robustness: more aggressive compression techniques, such as knowledge distillation, significantly reduce model size but at the cost of increased vulnerabilities to attacks. This supports the notion that smaller models, while efficient in terms of storage and computational cost, may be more susceptible to adversarial perturbations.

    \item \textbf{Vulnerability to stronger attacks is likely:} Since compressed models perform poorly against identifier-renaming adversarial attacks, we can hypothesize that they are even more likely to be vulnerable to sophisticated ones. Our empirical results using the more advanced attack techniques (e.g., CODA) also support this hypothesis.
    
\end{itemize}

\subsection*{Guidelines for Practitioners:} 

\begin{itemize}

    \item \textbf{Balance efficiency and robustness:} When deploying models in security-critical applications (e.g., vulnerability detection), it is crucial to balance compression efficiency and robustness. In adversarially prone environments, uncompressed or lightly compressed models may be preferable unless additional safeguards, such as adversarial defense, are implemented.
    
    \item \textbf{Evaluate compression per task:} Since no single compression strategy consistently outperforms the others across all software analytics tasks and models, practitioners should evaluate multiple strategies on their specific tasks before compressing models for deployment.

    \item \textbf{Apply knowledge distillation cautiously:} Although knowledge distillation achieves the highest compression rate, its reduced robustness suggests that it may not be suitable for security-sensitive applications without additional defense mechanisms.
    
\end{itemize}

In conclusion, our experimental findings underscore the need to develop compression strategies that balance computational efficiency with robustness against adversarial attacks. Additionally, we advocate for developing and integrating defense mechanisms within the compression pipeline, similar to those used in computer vision and natural language processing tasks \cite{ye2019adversarial, du2021robustness, gourtani2024improving, zhu2022safety, goldblum2020adversarially} to produce adversarially robust compressed models suitable for deployment in software analytics tasks.

\section{Threats to Validity}
\label{ThreatValidity}
This section provides a brief overview of the construct, conclusion, and the internal and external threats associated with our study.

\subsection{Construct Validity}
\label{ConsThreat}
Comparing compression techniques to evaluate their impact on the robustness of compressed models may threaten the construct validity of our study. To mitigate this threat, we applied consistent fine-tuning strategies across all models and tasks, ensuring a fair comparison. The performance (e.g., accuracy) of the compressed models before adversarial attacks could pose another potential construct threat to this study. However, we mitigated this threat by selecting compressed models with minimal performance degradation, ensuring that accuracy drops by no more than 5\%. This study evaluates knowledge distillation using the \textit{Compressor} framework, with a primary focus on encoder-only architectures and classification-oriented software analytics tasks. This design choice is consistent with the methodological assumptions of current SOTA distillation approaches for language models of code, which align teacher and student logits and are therefore well-suited to classification settings such as clone detection and vulnerability prediction. However, this focus may limit the generalizability of the findings to encoder-decoder or decoder-only models and to generation-oriented tasks such as code summarization. Recent approaches (e.g., SODA \cite{chen2025smaller}) targeting generative models remain difficult to reproduce reliably due to incomplete public implementations available at the time of this study. To address this limitation, the claims are explicitly scoped to the evaluated model families and tasks, and broader distillation coverage is identified as an important direction for future research. Our study uses the OWASP vulnerability dataset, which may be subject to known limitations, including potential label noise, duplicate samples, and reliance on function-level annotations that may not fully capture contextual vulnerabilities \cite{ding2024vulnerability, risse2025top}. Furthermore, the class distribution adopted for experimental consistency may not reflect real-world prevalence. To address these issues, we employ the standardized preprocessing and data splits established by the benchmark. Notably, since both compressed and uncompressed models are evaluated using the identical dataset and experimental pipeline, these limitations are unlikely to affect the validity of the relative robustness comparisons presented in this study. Lastly, it could be argued that the selected SOTA adversarial attacks do not encompass the full range of attacks applicable to compressed models. Nevertheless, we deliberately chose these attacks because they introduce both structural modifications and minimal identifier-renaming perturbations, while still demonstrating high effectiveness in misleading compressed models in our experiments. Therefore, we hypothesize that if compressed models exhibit poor performance against these attacks, they are likely to be even more susceptible to more advanced adversarial attacks.

\subsection{Internal Validity}
\label{IntThreat}
The robustness of the compressed models under adversarial attacks largely depends on the quality of the adversarial examples. If the examples are syntactically or semantically incorrect, the model's failure might not truly indicate a lack of robustness. Instead, it could be due to the model's incapacity to process unrealistic inputs. In addition, excessive changes to the original inputs during adversarial example generation can result in out-of-distribution data, further limiting the reliability of robustness evaluations. To mitigate these threats, we ensured that the adversarial examples preserved the syntax and semantic meaning of the original inputs, as described in Section \ref{Attack}. Furthermore, RQ1 confirms that the generated adversarial examples are generally of high quality, indicating that minimal necessary changes were made to effectively mislead the models. Therefore, the lack of robustness under adversarial attacks can be attributed to the compressed models themselves, as a consequence of the applied compression techniques. Potential errors in the implementation of fine-tuning and compressing models could affect the outcomes of our study. To address this concern, we utilized a well-established training pipeline provided by \textit{CodeXGLUE} for fine-tuning and testing models, and implemented compression strategies using actively maintained libraries.

\subsection{External Validity}
\label{ExtThreat}
The generalizability of our study findings may threaten the external validity of our empirical results. To mitigate this threat, we selected six language models for code representing different transformer-based architectures, evaluated them across three software analytics tasks, and applied five SOTA adversarial attack approaches along with three model compression techniques commonly used in software analytics. Therefore, our empirical results demonstrate a level of robustness that supports the potential generalizability of our study. The datasets used in our study contain only code snippets written in Java. Therefore, we encourage further analysis of other software analytics tasks that involve code snippets in programming languages other than Java (e.g., C/C++, Python). To support such analysis, we have made every effort to present our methodology clearly and have made our code and scripts publicly accessible.

\subsection{Conclusion Validity}
\label{ConThreat}
Across all evaluated models, tasks, and attack settings, the results demonstrate that compression strategies generally reduce the adversarial robustness of compressed models relative to their uncompressed counterparts. While very large-scale code language models, such as CodeLlama \cite{roziere2023code}, Deepseek v3 \cite{liu2024deepseek}, Qwen3 \cite{yang2025qwen3}, ChatGPT \cite{achiam2023gpt}, CodeGen \cite{nijkamp2022codegen}, and instruction-tuned variants, are not included, the experimental design intentionally spans multiple architectures and model types to capture adversarial robustness behaviors not limited to a single model or configuration. The consistency of observed trends across diverse models and settings provides strong evidence that the impact of compression on robustness is systematic rather than model-specific. Although very large code language models are more sophisticated, they still rely on transformer architectures as their core building blocks, as do models such as CodeBERT and CodeGPT used in our study. Therefore, the applicability of our study remains valid. Additionally, incorporating these models requires substantial computational resources and is currently constrained by the absence of compatible compression and adversarial evaluation pipelines. Future research should extend this analysis to include larger and more recent code language models to further validate the generalizability of these findings in very large-model settings.

\section{Related Work}
\label{RW}

A substantial body of research has focused on compressing large pre-trained models in computer vision and natural language processing. \cite{xu2023survey, zhu2024survey, xu2025resource, xu2021survey, gordon2020compressing, xu2021beyond, ye2019adversarial, du2021robustness, gourtani2024improving, zhu2022safety, sanh2019distilbert, sun2019patient, jiao2019tinybert, buciluǎ2006model, jiang2023lion, zhang2018structadmm, tang2019distilling, xu2020bert}. Zafrir et al. \cite{zafrir2019q8bert} quantized the network parameters and activation functions of the BERT model into 8-bit integers, resulting in a 25\% reduction in model size compared to the original. Gordon et al. \cite{gordon2020compressing} demonstrated that pruning 30\%–40\% of BERT’s parameters has little to no impact on its performance in downstream tasks. Fan et al. \cite{fan2019reducing} introduced a training approach that randomly omits layers, allowing for the removal of specific layers during inference with minimal loss in accuracy. Similarly, Michel et al. \cite{michel2019sixteen} found that many attention heads in pre-trained models can be pruned without substantially affecting performance. Several knowledge distillation techniques have been developed to compress models to sizes between 100 and 200 MB \cite{sanh2019distilbert, sun2019patient, jiao2019tinybert, tang2019distilling, xu2020bert}.

Model compression techniques have recently attracted growing interest in the field of software engineering. Shi et al. proposed two knowledge distillation-based approaches, \textit{Compressor} \cite{shi2022compressing} and \textit{AVATAR} \cite{shi2024greening}, to compress CodeBERT and GraphCodeBERT models to a size of just 3 MB, with little to no loss in accuracy. Their findings showed that both \textit{Compressor} and \textit{AVATAR} significantly reduce inference time and energy usage while maintaining model performance. MORPH \cite{panichella2025metamorphic} extends the AVATAR framework by integrating metamorphic testing with multi-objective optimization in a knowledge distillation pipeline. This approach aims to enhance robustness on metamorphic code while jointly optimizing for accuracy, model size, and efficiency. Empirical evaluations on CodeBERT and GraphCodeBERT indicate significant improvements in robustness and efficiency compared to AVATAR, without compromising predictive performance. Wei et al. \cite{wei2023towards} evaluated quantized models for code generation tasks, analyzing resource consumption, carbon footprint, and accuracy. They concluded that quantization, when applied under certain conditions, significantly improves model efficiency with minimal loss in accuracy or robustness. Sun et al. \cite{sun2024neural} investigated dynamic inference as a technique to accelerate code completion. Zhang et al. \cite{zhang2022diet} proposed a method to streamline input programs for CodeBERT, substantially reducing computational costs while maintaining model performance.

Saad et al. \cite{saad2024alpine} developed \textit{ALPINE}, a programming language-agnostic pruning technique that significantly lowers the computational overhead of language models for code. Furthermore, Aloisio et al. \cite{d2024compression} empirically investigated the impact of different compression strategies on the efficiency and effectiveness of the CodeBERT model across three software analytics tasks: code summarization, vulnerability detection, and code search. More recently, Chen et al. \cite{chen2025smaller} proposed SODA, a self-paced knowledge distillation framework that adaptively transfers programming knowledge from teacher to student models using correctness-aware and fault-aware learning. SODA iteratively refines student models through feedback-driven updates, achieving a 65.96\% gain in average Pass@1 and outperforming existing KD baselines. Notably, their distilled model SodaCoder-DS-6.7B surpasses ChatGPT across seven programming languages. All of the research mentioned above focuses on compressing models or inputs to improve efficiency, effectiveness, memory requirement, and energy consumption while overlooking the robustness of these compressed models under adversarial attacks.

While some research in computer vision and natural language processing has evaluated and enhanced the robustness of compressed models under adversarial attacks and out-of-distribution (OOD) data \cite{xu2021beyond, ye2019adversarial, du2021robustness, gourtani2024improving, zhu2022safety, goldblum2020adversarially, dong2024robust, bai2023guided, kuang2023improving}, this area remains largely unexplored in software engineering despite an initial work by Wei et al. \cite{wei2023towards}. They demonstrated that the quantized models perform similarly to the uncompressed model, even after perturbations were added to the natural text during code search. However, a significant gap remains in comprehensive studies that systematically investigate the impact of different model compression strategies on the robustness of language models for code under adversarial attacks across various software analytics tasks. To address this gap, we conduct an extensive study evaluating the effects of three established compression techniques, such as pruning, quantization, and knowledge distillation, on the adversarial robustness of CodeBERT, CodeGPT, and PLBART across three tasks: clone detection, code summarization, and vulnerability detection.

\section{Conclusion}
\label{end}
In this paper, we conduct an extensive empirical study to investigate how model compression affects the robustness of language models for code under adversarial attacks. Our experimental results demonstrate that, before applying adversarial attacks, compressed models perform similarly to their uncompressed counterparts across all tasks. However, when subjected to adversarial attacks, compressed models exhibit reduced robustness, highlighting a trade-off between compression efficiency and robustness. This degradation is observed across various common model compression strategies, including pruning, quantization, and knowledge distillation, with knowledge-distilled models exhibiting the most significant drop in robustness. These findings underscore the importance of considering adversarial robustness when applying model compression techniques to models in security-critical software analytics applications. Future research should focus on developing novel compression strategies for models that preserve adversarial robustness while maintaining their computational efficiency in software analytics tasks.

\section*{Data Availability}
Our code and the corresponding dataset are publicly available to enhance further research \footnote{\href{https://doi.org/10.5281/zenodo.15272413}{Replication-packages}}.

\section*{Funding}
This research is supported in part by the Natural Sciences and Engineering Research Council of Canada (NSERC) Discovery Grants program, the Canada Foundation for Innovation's John R. Evans Leaders Fund (CFI-JELF), and by the industry-stream NSERC CREATE in Software Analytics Research (SOAR).


\section*{Declaration of generative AI and AI-assisted technologies in the writing process}
During the preparation of this work, the author(s) used Grammarly \footnote{https://app.grammarly.com/} and ChatGPT\footnote{https://chat.openai.com/} to find grammatical mistakes and improve sentence clarity/ presentation. After using these tools/services, the author(s) reviewed and edited the content as needed and take(s) full responsibility for the content of the publication.



\bibliography{sn-bibliography.bib}

@String{BIT = "{BIT}" }

@String{Computing = "Computing" }

@String{Computer = "{IEEE} Computer" }

@String{Springer = "Springer-Verlag" }

@inproceedings{buciluǎ2006model,
  title={Model compression},
  author={Buciluǎ, Cristian and Caruana, Rich and Niculescu-Mizil, Alexandru},
  booktitle={Proceedings of the 12th ACM SIGKDD international conference on Knowledge discovery and data mining},
  pages={535--541},
  year={2006}
}

@article{jiang2023lion,
  title={Lion: Adversarial distillation of proprietary large language models},
  author={Jiang, Yuxin and Chan, Chunkit and Chen, Mingyang and Wang, Wei},
  journal={arXiv preprint arXiv:2305.12870},
  year={2023}
}

@article{vaswani2017attention,
  title={Attention is all you need},
  author={Vaswani, Ashish and Shazeer, Noam and Parmar, Niki and Uszkoreit, Jakob and Jones, Llion and Gomez, Aidan N and Kaiser, {\L}ukasz and Polosukhin, Illia},
  journal={Advances in neural information processing systems},
  volume={30},
  year={2017}
}

@inproceedings{zhu2022safety,
  title={Safety and performance, why not both? bi-objective optimized model compression toward ai software deployment},
  author={Zhu, Jie and Wang, Leye and Han, Xiao},
  booktitle={Proceedings of the 37th IEEE/ACM International Conference on Automated Software Engineering},
  pages={1--13},
  year={2022}
}

@inproceedings{shi2022compressing,
  author = {Shi, Jieke and Yang, Zhou and Xu, Bowen and Kang, Hong Jin and Lo, David},
  title = {Compressing Pre-Trained Models of Code into 3 MB},
  year = {2023},
  isbn = {9781450394758},
  publisher = {Association for Computing Machinery},
  address = {New York, NY, USA},
  url = {https://doi.org/10.1145/3551349.3556964},
  doi = {10.1145/3551349.3556964},
  booktitle = {Proceedings of the 37th IEEE/ACM International Conference on Automated Software Engineering},
  articleno = {24},
  numpages = {12},
  location = {Rochester, MI, USA},
  series = {ASE '22}
}

@inproceedings{panichella2025metamorphic,
  title={Metamorphic-Based Many-Objective Distillation of LLMs for Code-related Tasks},
  author={Panichella, Annibale},
  booktitle={2025 IEEE/ACM 47th International Conference on Software Engineering (ICSE)},
  pages={766--766},
  year={2025},
  organization={IEEE Computer Society}
}

@inproceedings{shi2024greening,
  author = {Shi, Jieke and Yang, Zhou and Kang, Hong Jin and Xu, Bowen and He, Junda and Lo, David},
  title = {Greening Large Language Models of Code},
  year = {2024},
  isbn = {9798400704994},
  publisher = {Association for Computing Machinery},
  address = {New York, NY, USA},
  url = {https://doi-org.libproxy.smu.edu.sg/10.1145/3639475.3640097},
  doi = {10.1145/3639475.3640097},
  booktitle = {Proceedings of the 46th International Conference on Software Engineering: Software Engineering in Society},
  pages = {142–153},
  numpages = {12},
  location = {Lisbon, Portugal},
  series = {ICSE-SEIS'24}
}

@article{shi2024efficient,
  title={Efficient and green large language models for software engineering: Vision and the road ahead},
  author={Shi, Jieke and Yang, Zhou and Lo, David},
  journal={ACM Transactions on Software Engineering and Methodology},
  year={2024},
  publisher={ACM New York, NY}
}

@inproceedings{tian2023code,
  title={Code difference guided adversarial example generation for deep code models},
  author={Tian, Zhao and Chen, Junjie and Jin, Zhi},
  booktitle={2023 38th IEEE/ACM International Conference on Automated Software Engineering (ASE)},
  pages={850--862},
  year={2023},
  organization={IEEE}
}

@article{kullback1951information,
  title={On information and sufficiency},
  author={Kullback, Solomon and Leibler, Richard A},
  journal={The annals of mathematical statistics},
  volume={22},
  number={1},
  pages={79--86},
  year={1951},
  publisher={JSTOR}
}

@article{dong2025can,
  title={Can Compressed LLMs Truly Act? An Empirical Evaluation of Agentic Capabilities in LLM Compression},
  author={Dong, Peijie and Tang, Zhenheng and Liu, Xiang and Li, Lujun and Chu, Xiaowen and Li, Bo},
  journal={arXiv preprint arXiv:2505.19433},
  year={2025}
}

@article{woolson2007wilcoxon,
  title={Wilcoxon signed-rank test},
  author={Woolson, Robert F},
  journal={Wiley encyclopedia of clinical trials},
  pages={1--3},
  year={2007},
  publisher={Wiley Online Library}
}

@article{awal2024large,
  title={Large Language Models as Robust Data Generators in Software Analytics: Are We There Yet?},
  author={Awal, Md Abdul and Rochan, Mrigank and Roy, Chanchal K},
  journal={arXiv preprint arXiv:2411.10565},
  year={2024}
}

@inproceedings{wei2023towards,
  title={Towards greener yet powerful code generation via quantization: An empirical study},
  author={Wei, Xiaokai and Gonugondla, Sujan Kumar and Wang, Shiqi and Ahmad, Wasi and Ray, Baishakhi and Qian, Haifeng and Li, Xiaopeng and Kumar, Varun and Wang, Zijian and Tian, Yuchen and others},
  booktitle={Proceedings of the 31st ACM Joint European Software Engineering Conference and Symposium on the Foundations of Software Engineering},
  pages={224--236},
  year={2023}
}

@article{chen2025smaller,
  title={Smaller but Better: Self-Paced Knowledge Distillation for Lightweight yet Effective LCMs},
  author={Chen, Yujia and Ye, Yang and Li, Zhongqi and Ma, Yuchi and Gao, Cuiyun},
  journal={Proceedings of the ACM on Software Engineering},
  volume={2},
  number={FSE},
  pages={3057--3080},
  year={2025},
  publisher={ACM New York, NY, USA}
}

@article{saad2024alpine,
  title={ALPINE: An adaptive language-agnostic pruning method for language models for code},
  author={Saad, Mootez and L{\'o}pez, Jos{\'e} Antonio Hern{\'a}ndez and Chen, Boqi and Varr{\'o}, D{\'a}niel and Sharma, Tushar},
  journal={arXiv preprint arXiv:2407.04147},
  year={2024}
}

@article{xu2021survey,
  title={A survey on green deep learning},
  author={Xu, Jingjing and Zhou, Wangchunshu and Fu, Zhiyi and Zhou, Hao and Li, Lei},
  journal={arXiv preprint arXiv:2111.05193},
  year={2021}
}

@article{d2024compression,
  title={On the Compression of Language Models for Code: An Empirical Study on CodeBERT},
  author={d'Aloisio, Giordano and Traini, Luca and Sarro, Federica and Di Marco, Antinisca},
  journal={arXiv preprint arXiv:2412.13737},
  year={2024}
}

@inproceedings{sun2024neural,
  title={When neural code completion models size up the situation: Attaining cheaper and faster completion through dynamic model inference},
  author={Sun, Zhensu and Du, Xiaoning and Song, Fu and Wang, Shangwen and Li, Li},
  booktitle={Proceedings of the IEEE/ACM 46th International Conference on Software Engineering},
  pages={1--12},
  year={2024}
}

@inproceedings{zhang2022diet,
  title={Diet code is healthy: Simplifying programs for pre-trained models of code},
  author={Zhang, Zhaowei and Zhang, Hongyu and Shen, Beijun and Gu, Xiaodong},
  booktitle={Proceedings of the 30th ACM Joint European Software Engineering Conference and Symposium on the Foundations of Software Engineering},
  pages={1073--1084},
  year={2022}
}

@article{xu2025resource,
  title={Resource-efficient algorithms and systems of foundation models: A survey},
  author={Xu, Mengwei and Cai, Dongqi and Yin, Wangsong and Wang, Shangguang and Jin, Xin and Liu, Xuanzhe},
  journal={ACM Computing Surveys},
  volume={57},
  number={5},
  pages={1--39},
  year={2025},
  publisher={ACM New York, NY}
}

@article{zhu2024survey,
  title={A survey on model compression for large language models},
  author={Zhu, Xunyu and Li, Jian and Liu, Yong and Ma, Can and Wang, Weiping},
  journal={Transactions of the Association for Computational Linguistics},
  volume={12},
  pages={1556--1577},
  year={2024},
  publisher={MIT Press 255 Main Street, 9th Floor, Cambridge, Massachusetts 02142, USA~…}
}

@inproceedings{hort2023exploratory,
  title={An exploratory literature study on sharing and energy use of language models for source code},
  author={Hort, Max and Grishina, Anastasiia and Moonen, Leon},
  booktitle={2023 ACM/IEEE International Symposium on Empirical Software Engineering and Measurement (ESEM)},
  pages={1--12},
  year={2023},
  organization={IEEE}
}

@inproceedings{castano2023exploring,
  title={Exploring the carbon footprint of hugging face's ml models: A repository mining study},
  author={Casta{\~n}o, Joel and Mart{\'\i}nez-Fern{\'a}ndez, Silverio and Franch, Xavier and Bogner, Justus},
  booktitle={2023 ACM/IEEE International Symposium on Empirical Software Engineering and Measurement (ESEM)},
  pages={1--12},
  year={2023},
  organization={IEEE}
}

@inproceedings{zafrir2019q8bert,
  title={Q8bert: Quantized 8bit bert},
  author={Zafrir, Ofir and Boudoukh, Guy and Izsak, Peter and Wasserblat, Moshe},
  booktitle={2019 Fifth Workshop on Energy Efficient Machine Learning and Cognitive Computing-NeurIPS Edition (EMC2-NIPS)},
  pages={36--39},
  year={2019},
  organization={IEEE}
}

@article{sanh2020movement,
  title={Movement pruning: Adaptive sparsity by fine-tuning},
  author={Sanh, Victor and Wolf, Thomas and Rush, Alexander},
  journal={Advances in neural information processing systems},
  volume={33},
  pages={20378--20389},
  year={2020}
}

@article{hinton2015distilling,
  title={Distilling the knowledge in a neural network},
  author={Hinton, Geoffrey and Vinyals, Oriol and Dean, Jeff},
  journal={arXiv preprint arXiv:1503.02531},
  year={2015}
}

@inproceedings{ye2019adversarial,
  title={Adversarial robustness vs. model compression, or both?},
  author={Ye, Shaokai and Xu, Kaidi and Liu, Sijia and Cheng, Hao and Lambrechts, Jan-Henrik and Zhang, Huan and Zhou, Aojun and Ma, Kaisheng and Wang, Yanzhi and Lin, Xue},
  booktitle={Proceedings of the IEEE/CVF international conference on computer vision},
  pages={111--120},
  year={2019}
}

@article{xu2024survey,
  title={A survey of resource-efficient llm and multimodal foundation models},
  author={Xu, Mengwei and Yin, Wangsong and Cai, Dongqi and Yi, Rongjie and Xu, Daliang and Wang, Qipeng and Wu, Bingyang and Zhao, Yihao and Yang, Chen and Wang, Shihe and others},
  journal={arXiv preprint arXiv:2401.08092},
  year={2024}
}

@article{xu2024surveyKD,
  title={A survey on knowledge distillation of large language models},
  author={Xu, Xiaohan and Li, Ming and Tao, Chongyang and Shen, Tao and Cheng, Reynold and Li, Jinyang and Xu, Can and Tao, Dacheng and Zhou, Tianyi},
  journal={arXiv preprint arXiv:2402.13116},
  year={2024}
}

@inproceedings{xu2023survey,
  title={A survey on model compression and acceleration for pretrained language models},
  author={Xu, Canwen and McAuley, Julian},
  booktitle={Proceedings of the AAAI Conference on Artificial Intelligence},
  volume={37},
  number={9},
  pages={10566--10575},
  year={2023}
}

@article{wang2024model,
  title={Model compression and efficient inference for large language models: A survey},
  author={Wang, Wenxiao and Chen, Wei and Luo, Yicong and Long, Yongliu and Lin, Zhengkai and Zhang, Liye and Lin, Binbin and Cai, Deng and He, Xiaofei},
  journal={arXiv preprint arXiv:2402.09748},
  year={2024}
}

@article{xu2021beyond,
  title={Beyond preserved accuracy: Evaluating loyalty and robustness of BERT compression},
  author={Xu, Canwen and Zhou, Wangchunshu and Ge, Tao and Xu, Ke and McAuley, Julian and Wei, Furu},
  journal={arXiv preprint arXiv:2109.03228},
  year={2021}
}

@article{zhang2022towards,
  title={Towards robustness of deep program processing models—detection, estimation, and enhancement},
  author={Zhang, Huangzhao and Fu, Zhiyi and Li, Ge and Ma, Lei and Zhao, Zhehao and Yang, Hua’an and Sun, Yizhe and Liu, Yang and Jin, Zhi},
  journal={ACM Transactions on Software Engineering and Methodology (TOSEM)},
  volume={31},
  number={3},
  pages={1--40},
  year={2022},
  publisher={ACM New York, NY}
}

@article{schwartz2020green,
  title={Green ai},
  author={Schwartz, Roy and Dodge, Jesse and Smith, Noah A and Etzioni, Oren},
  journal={Communications of the ACM},
  volume={63},
  number={12},
  pages={54--63},
  year={2020},
  publisher={ACM New York, NY, USA}
}

@article{lecun1989optimal,
  title={Optimal brain damage},
  author={LeCun, Yann and Denker, John and Solla, Sara},
  journal={Advances in neural information processing systems},
  volume={2},
  year={1989}
}

@article{gordon2020compressing,
  title={Compressing bert: Studying the effects of weight pruning on transfer learning},
  author={Gordon, Mitchell A and Duh, Kevin and Andrews, Nicholas},
  journal={arXiv preprint arXiv:2002.08307},
  year={2020}
}

@article{gray1998quantization,
  title={Quantization},
  author={Gray, Robert M. and Neuhoff, David L.},
  journal={IEEE transactions on information theory},
  volume={44},
  number={6},
  pages={2325--2383},
  year={1998},
  publisher={IEEE}
}

@inproceedings{svajlenko2014towards,
  title={Towards a big data curated benchmark of inter-project code clones},
  author={Svajlenko, Jeffrey and Islam, Judith F and Keivanloo, Iman and Roy, Chanchal K and Mia, Mohammad Mamun},
  booktitle={2014 IEEE International Conference on Software Maintenance and Evolution},
  pages={476--480},
  year={2014},
  organization={IEEE}
}

@inproceedings{wang2020detecting,
  title={Detecting code clones with graph neural network and flow-augmented abstract syntax tree},
  author={Wang, Wenhan and Li, Ge and Ma, Bo and Xia, Xin and Jin, Zhi},
  booktitle={2020 IEEE 27th International Conference on Software Analysis, Evolution and Reengineering (SANER)},
  pages={261--271},
  year={2020},
  organization={IEEE}
}

@inproceedings{hough2020revealing,
  title={Revealing injection vulnerabilities by leveraging existing tests},
  author={Hough, Katherine and Welearegai, Gebrehiwet and Hammer, Christian and Bell, Jonathan},
  booktitle={Proceedings of the ACM/IEEE 42nd International Conference on Software Engineering},
  pages={284--296},
  year={2020}
}

@article{sayar2023depth,
  title={An in-depth study of java deserialization remote-code execution exploits and vulnerabilities},
  author={Sayar, Imen and Bartel, Alexandre and Bodden, Eric and Le Traon, Yves},
  journal={ACM Transactions on Software Engineering and Methodology},
  volume={32},
  number={1},
  pages={1--45},
  year={2023},
  publisher={ACM New York, NY}
}

@article{du2021robustness,
  title={Robustness challenges in model distillation and pruning for natural language understanding},
  author={Du, Mengnan and Mukherjee, Subhabrata and Cheng, Yu and Shokouhi, Milad and Hu, Xia and Awadallah, Ahmed Hassan},
  journal={arXiv preprint arXiv:2110.08419},
  year={2021}
}

@inproceedings{gourtani2024improving,
  title={Improving Robustness of Compressed Models with Weight Sharing through Knowledge Distillation},
  author={Gourtani, Saeed Khalilian and Meratnia, Nirvana},
  booktitle={2024 IEEE 10th International Conference on Edge Computing and Scalable Cloud (EdgeCom)},
  pages={13--21},
  year={2024},
  organization={IEEE}
}

@article{zhang2018structadmm,
  title={StructADMM: A systematic, high-efficiency framework of structured weight pruning for DNNs},
  author={Zhang, Tianyun and Ye, Shaokai and Zhang, Kaiqi and Ma, Xiaolong and Liu, Ning and Zhang, Linfeng and Tang, Jian and Ma, Kaisheng and Lin, Xue and Fardad, Makan and others},
  journal={arXiv preprint arXiv:1807.11091},
  year={2018}
}

@article{fan2019reducing,
  title={Reducing transformer depth on demand with structured dropout},
  author={Fan, Angela and Grave, Edouard and Joulin, Armand},
  journal={arXiv preprint arXiv:1909.11556},
  year={2019}
}

@article{michel2019sixteen,
  title={Are sixteen heads really better than one?},
  author={Michel, Paul and Levy, Omer and Neubig, Graham},
  journal={Advances in neural information processing systems},
  volume={32},
  year={2019}
}

@article{sun2019patient,
  title={Patient knowledge distillation for bert model compression},
  author={Sun, Siqi and Cheng, Yu and Gan, Zhe and Liu, Jingjing},
  journal={arXiv preprint arXiv:1908.09355},
  year={2019}
}

@article{jiao2019tinybert,
  title={Tinybert: Distilling bert for natural language understanding},
  author={Jiao, Xiaoqi and Yin, Yichun and Shang, Lifeng and Jiang, Xin and Chen, Xiao and Li, Linlin and Wang, Fang and Liu, Qun},
  journal={arXiv preprint arXiv:1909.10351},
  year={2019}
}

@article{tang2019distilling,
  title={Distilling task-specific knowledge from bert into simple neural networks},
  author={Tang, Raphael and Lu, Yao and Liu, Linqing and Mou, Lili and Vechtomova, Olga and Lin, Jimmy},
  journal={arXiv preprint arXiv:1903.12136},
  year={2019}
}

@article{xu2020bert,
  title={Bert-of-theseus: Compressing bert by progressive module replacing},
  author={Xu, Canwen and Zhou, Wangchunshu and Ge, Tao and Wei, Furu and Zhou, Ming},
  journal={arXiv preprint arXiv:2002.02925},
  year={2020}
}

@inproceedings{goldblum2020adversarially,
  title={Adversarially robust distillation},
  author={Goldblum, Micah and Fowl, Liam and Feizi, Soheil and Goldstein, Tom},
  booktitle={Proceedings of the AAAI conference on artificial intelligence},
  volume={34},
  number={04},
  pages={3996--4003},
  year={2020}
}

@inproceedings{dong2024robust,
  title={Robust distillation via untargeted and targeted intermediate adversarial samples},
  author={Dong, Junhao and Koniusz, Piotr and Chen, Junxi and Wang, Z Jane and Ong, Yew-Soon},
  booktitle={Proceedings of the IEEE/CVF Conference on Computer Vision and Pattern Recognition},
  pages={28432--28442},
  year={2024}
}

@article{bai2023guided,
  title={Guided adversarial contrastive distillation for robust students},
  author={Bai, Tao and Zhao, Jun and Wen, Bihan},
  journal={IEEE Transactions on Information Forensics and Security},
  year={2023},
  publisher={IEEE}
}

@article{kuang2023improving,
  title={Improving adversarial robustness via information bottleneck distillation},
  author={Kuang, Huafeng and Liu, Hong and Wu, Yongjian and Satoh, Shin'ichi and Ji, Rongrong},
  journal={Advances in Neural Information Processing Systems},
  volume={36},
  pages={10796--10813},
  year={2023}
}

@inproceedings{hellendoorn2019code,
  title={When code completion fails: A case study on real-world completions},
  author={Hellendoorn, Vincent J and Proksch, Sebastian and Gall, Harald C and Bacchelli, Alberto},
  booktitle={2019 IEEE/ACM 41st International Conference on Software Engineering (ICSE)},
  pages={960--970},
  year={2019},
  organization={IEEE}
}

@article{friedman1937use,
  title={The use of ranks to avoid the assumption of normality implicit in the analysis of variance},
  author={Friedman, Milton},
  journal={Journal of the american statistical association},
  volume={32},
  number={200},
  pages={675--701},
  year={1937},
  publisher={Taylor \& Francis}
}

@misc{awal2025largelanguagemodelsrobust,
      title={Large Language Models as Robust Data Generators in Software Analytics: Are We There Yet?}, 
      author={Md. Abdul Awal and Mrigank Rochan and Chanchal K. Roy},
      year={2025},
      eprint={2411.10565},
      archivePrefix={arXiv},
      primaryClass={cs.SE},
      url={https://arxiv.org/abs/2411.10565}, 
}

@inproceedings{sainath2013low,
  title={Low-rank matrix factorization for deep neural network training with high-dimensional output targets},
  author={Sainath, Tara N and Kingsbury, Brian and Sindhwani, Vikas and Arisoy, Ebru and Ramabhadran, Bhuvana},
  booktitle={2013 IEEE international conference on acoustics, speech and signal processing},
  pages={6655--6659},
  year={2013},
  organization={IEEE}
}

@article{lan2019albert,
  title={Albert: A lite bert for self-supervised learning of language representations},
  author={Lan, Zhenzhong and Chen, Mingda and Goodman, Sebastian and Gimpel, Kevin and Sharma, Piyush and Soricut, Radu},
  journal={arXiv preprint arXiv:1909.11942},
  year={2019}
}

@article{devlin2018bert,
  title={Bert: Pre-training of deep bidirectional transformers for language understanding},
  author={Devlin, Jacob},
  journal={arXiv:1810.04805},
  year={2018}
}

@article{sanh2019distilbert,
  title={DistilBERT, a distilled version of BERT: smaller, faster, cheaper and lighter},
  author={Sanh, V},
  journal={arXiv:1910.01108},
  year={2019}
}

@article{feng2020codebert,
  title={Codebert: A pre-trained model for programming and natural languages},
  author={Feng, Z. and Guo, D. and Tang, D. and Duan, N. and Feng, X. and Gong, M. and Shou, L. and Qin, Bing and Liu, T. and Jiang, D. and others},
  journal={arXiv:2002.08155},
  year={2020}
}

@article{guo2022unixcoder,
  title={Unixcoder: Unified cross-modal pre-training for code representation},
  author={Guo, Daya and Lu, Shuai and Duan, Nan and Wang, Yanlin and Zhou, Ming and Yin, Jian},
  journal={arXiv:2203.03850},
  year={2022}
}

@article{guo2020graphcodebert,
  title={Graphcodebert: Pre-training code representations with data flow},
  author={Guo, Daya and Ren, Shuo and Lu, Shuai and Feng, Zhangyin and Tang, Duyu and Liu, Shujie and Zhou, Long and Duan, Nan and Svyatkovskiy, Alexey and Fu, Shengyu and others},
  journal={arXiv:2009.08366},
  year={2020}
}

@article{wang2021codet5,
  title={Codet5: Identifier-aware unified pre-trained encoder-decoder models for code understanding and generation},
  author={Wang, Yue and Wang, Weishi and Joty, Shafiq and Hoi, Steven CH},
  journal={arXiv:2109.00859},
  year={2021}
}

@article{yang2025qwen3,
  title={Qwen3 technical report},
  author={Yang, An and Li, Anfeng and Yang, Baosong and Zhang, Beichen and Hui, Binyuan and Zheng, Bo and Yu, Bowen and Gao, Chang and Huang, Chengen and Lv, Chenxu and others},
  journal={arXiv preprint arXiv:2505.09388},
  year={2025}
}

@inproceedings{du2023extensive,
  title={An Extensive Study on Adversarial Attack against Pre-trained Models of Code},
  author={Du, Xiaohu and Wen, Ming and Wei, Zichao and Wang, Shangwen and Jin, Hai},
  booktitle={31st FSE},
  pages={489--501},
  year={2023}
}

@article{zhou2019devign,
  title={Devign: Effective vulnerability identification by learning comprehensive program semantics via graph neural networks},
  author={Zhou, Yaqin and Liu, Shangqing and Siow, Jingkai and Du, Xiaoning and Liu, Yang},
  journal={Advances in NIPS},
  volume={32},
  year={2019}
}

@article{ding2024vulnerability,
  title={Vulnerability detection with code language models: How far are we?},
  author={Ding, Yangruibo and Fu, Yanjun and Ibrahim, Omniyyah and Sitawarin, Chawin and Chen, Xinyun and Alomair, Basel and Wagner, David and Ray, Baishakhi and Chen, Yizheng},
  journal={arXiv preprint arXiv:2403.18624},
  year={2024}
}

@article{risse2025top,
  title={Top score on the wrong exam: On benchmarking in machine learning for vulnerability detection},
  author={Risse, Niklas and Liu, Jing and B{\"o}hme, Marcel},
  journal={Proceedings of the ACM on Software Engineering},
  volume={2},
  number={ISSTA},
  pages={388--410},
  year={2025},
  publisher={ACM New York, NY, USA}
}

@article{husain2019codesearchnet,
  title={Codesearchnet challenge: Evaluating the state of semantic code search},
  author={Husain, Hamel and Wu, Ho-Hsiang and Gazit, Tiferet and Allamanis, Miltiadis and Brockschmidt, Marc},
  journal={arXiv:1909.09436},
  year={2019}
}

@article{ahmad2021unified,
  title={Unified pre-training for program understanding and generation},
  author={Ahmad, Wasi Uddin and Chakraborty, Saikat and Ray, Baishakhi and Chang, Kai-Wei},
  journal={arXiv:2103.06333},
  year={2021}
}

@inproceedings{ilyas2018black,
  title={Black-box adversarial attacks with limited queries and information},
  author={Ilyas, Andrew and Engstrom, Logan and Athalye, Anish and Lin, Jessy},
  booktitle={ICML},
  pages={2137--2146},
  year={2018},
  organization={PMLR}
}

@article{ebrahimi2017hotflip,
  title={Hotflip: White-box adversarial examples for text classification},
  author={Ebrahimi, Javid and Rao, Anyi and Lowd, Daniel and Dou, Dejing},
  journal={arXiv arXiv:1712.06751},
  year={2017}
}

@inproceedings{zhang2020generating,
  title={Generating adversarial examples for holding robustness of source code processing models},
  author={Zhang, Huangzhao and Li, Zhuo and Li, Ge and Ma, Lei and Liu, Yang and Jin, Zhi},
  booktitle={Proceedings of the AAAI Conference on AI},
  volume={34},
  number={01},
  pages={1169--1176},
  year={2020}
}

@inproceedings{yang2022natural,
  title={Natural attack for pre-trained models of code},
  author={Yang, Zhou and Shi, Jieke and He, Junda and Lo, David},
  booktitle={Proceedings of the 44th ICSE},
  pages={1482--1493},
  year={2022}
}

@article{nijkamp2022codegen,
  title={Codegen: An open large language model for code with multi-turn program synthesis},
  author={Nijkamp, Erik and Pang, Bo and Hayashi, Hiroaki and Tu, Lifu and Wang, Huan and Zhou, Yingbo and Savarese, Silvio and Xiong, Caiming},
  journal={arXiv preprint arXiv:2203.13474},
  year={2022}
}

@article{li2022codereviewer,
  title={Codereviewer: Pre-training for automating code review activities},
  author={Li, Zhiyu and Lu, Shuai and Guo, Daya and Duan, Nan and Jannu, Shailesh and Jenks, Grant and Majumder, Deep and Green, Jared and Svyatkovskiy, Alexey and Fu, Shengyu and others},
  journal={arXiv preprint arXiv:2203.09095},
  year={2022}
}

@inproceedings{shi2019automatic,
  title={Automatic code review by learning the revision of source code},
  author={Shi, Shu-Ting and Li, Ming and Lo, David and Thung, Ferdian and Huo, Xuan},
  booktitle={Proceedings of the AAAI conference on artificial intelligence},
  volume={33},
  number={01},
  pages={4910--4917},
  year={2019}
}

@article{achiam2023gpt,
  title={Gpt-4 technical report},
  author={Achiam, Josh and Adler, Steven and Agarwal, Sandhini and Ahmad, Lama and Akkaya, Ilge and Aleman, Florencia Leoni and Almeida, Diogo and Altenschmidt, Janko and Altman, Sam and Anadkat, Shyamal and others},
  journal={arXiv:2303.08774},
  year={2023}
}

@article{roziere2023code,
  title={Code llama: Open foundation models for code},
  author={Roziere, Baptiste and Gehring, Jonas and Gloeckle, Fabian and Sootla, Sten and Gat, Itai and Tan, Xiaoqing Ellen and Adi, Yossi and Liu, Jingyu and Remez, Tal and Rapin, J{\'e}r{\'e}my and others},
  journal={arXiv:2308.12950},
  year={2023}
}

@article{liu2024deepseek,
  title={Deepseek-v3 technical report},
  author={Liu, Aixin and Feng, Bei and Xue, Bing and Wang, Bingxuan and Wu, Bochao and Lu, Chengda and Zhao, Chenggang and Deng, Chengqi and Zhang, Chenyu and Ruan, Chong and others},
  journal={arXiv:2412.19437},
  year={2024}
}

@article{mondal2019empirical,
  title={An empirical study on bug propagation through code cloning},
  author={Mondal, Manishankar and Roy, Banani and Roy, Chanchal K and Schneider, Kevin A},
  journal={Journal of Systems and Software},
  volume={158},
  pages={110407},
  year={2019},
  publisher={Elsevier}
}

@misc
{technicaldebtcost,
author = {McGraw-Hill Book Co, New York},
title={Cast worldwide application software quality study: summary of key findings.},
note = {Cast report Charette RN (1989) Software engineering, risk analysis and management Intertext publications},
Year = {2012}
}

@article{roy2007survey,
  title={A survey on software clone detection research},
  author={Roy, Chanchal Kumar and Cordy, James R},
  journal={Queen’s School of computing TR},
  volume={541},
  number={115},
  pages={64--68},
  year={2007}
}

@article{szegedy2013intriguing,
  title={Intriguing properties of neural networks},
  author={Szegedy, Christian and Zaremba, Wojciech and Sutskever, Ilya and Bruna, Joan and Erhan, Dumitru and Goodfellow, Ian and Fergus, Rob},
  journal={arXiv:1312.6199},
  year={2013}
}

@inproceedings{carlini2017adversarial,
  title={Adversarial examples are not easily detected: Bypassing ten detection methods},
  author={Carlini, Nicholas and Wagner, David},
  booktitle={Proceedings of the 10th ACM workshop on artificial intelligence and security},
  pages={3--14},
  year={2017}
}

@article{guo2016exploring,
  title={Exploring the costs of technical debt management--a case study},
  author={Guo, Yuepu and Sp{\'\i}nola, Rodrigo Oliveira and Seaman, Carolyn},
  journal={Empirical Software Engineering},
  volume={21},
  pages={159--182},
  year={2016},
  publisher={Springer}
}

@inproceedings{ahmed2024automatic,
  title={Automatic semantic augmentation of language model prompts (for code summarization)},
  author={Ahmed, Toufique and Pai, Kunal Suresh and Devanbu, Premkumar and Barr, Earl},
  booktitle={Proceedings of the IEEE/ACM 46th International Conference on Software Engineering},
  pages={1--13},
  year={2024}
}

@article{lu2021codexglue,
  title={Codexglue: A machine learning benchmark dataset for code understanding and generation},
  author={Lu, Shuai and Guo, Daya and Ren, Shuo and Huang, Junjie and Svyatkovskiy, Alexey and Blanco, Ambrosio and Clement, Colin and Drain, Dawn and Jiang, Daxin and Tang, Duyu and others},
  journal={arXiv:2102.04664},
  year={2021}
}

@inproceedings{zeng2022extensive,
  title={An extensive study on pre-trained models for program understanding and generation},
  author={Zeng, Zhengran and Tan, Hanzhuo and Zhang, Haotian and Li, Jing and Zhang, Yuqun and Zhang, Lingming},
  booktitle={31st ACM SIGSOFT ISSTA},
  pages={39--51},
  year={2022}
}

\end{document}